\documentclass{article}
\usepackage{graphicx}
\usepackage{amsmath,amssymb}
\usepackage[round]{natbib}
\usepackage{color}
\usepackage{fullpage}
\usepackage{multirow}
\usepackage{hyperref}

\newcommand{\wikiincludegraphics}[1]{\includegraphics[width=6.5in]{#1.pdf}}

\newcommand{\calL}{\mathcal{L}}
\newcommand{\E}{\mathbb{E}}

\begin{document}

\title{The Geography of Recent Genetic Ancestry across Europe.}
\author{Peter Ralph$^{1}$ and Graham Coop$^{2}$ \\
\small $^1$ Department of Evolution and Ecology \& Center for Population Biology,\\
\small University of California, Davis.\\
\small $^2$ Department of Molecular and Computational Biology,\\
\small University of Southern California.\\
\small To whom correspondence should be addressed: \texttt{plralph@ucdavis.edu,gmcoop@ucdavis.edu}\\
}
\date{}
\maketitle

\section*{Abstract}

The recent genealogical history of human populations 
is a complex mosaic formed by individual migration, 
large-scale population movements, 
and other demographic events.
Population genomics datasets can provide a window into this recent history,
as rare traces of recent shared genetic ancestry
are detectable due to long segments of shared genomic material.
We make use of genomic data for 2257 Europeans (in the POPRES dataset)
to conduct one of the first surveys of recent genealogical ancestry over the past three thousand years at a continental scale.  

We detected 1.9 million shared long genomic segments,
and used the lengths of these to infer the distribution of shared ancestors across time and geography.
We find that a pair of modern Europeans living in neighboring populations 
share around 2--12 genetic common ancestors from the last 1500 years,
and upwards of 100 genetic ancestors from the previous 1000 years.
These numbers drop off exponentially with geographic distance,
but since these genetic ancestors are a tiny fraction of common genealogical ancestors,  
individuals from opposite ends of Europe are still expected to share 
millions of common genealogical ancestors over the last 1000 years.
There is also substantial regional variation in the number of shared genetic ancestors.
For example, there are especially high numbers of common ancestors shared between many eastern populations
which date roughly to the migration period (which includes the Slavic and Hunnic expansions into that region).
Some of the lowest levels of common ancestry are seen in the Italian and Iberian peninsulas,
which may indicate different effects of historical population expansions in these areas
and/or more stably structured populations.

Population genomic datasets have considerable power to uncover recent demographic history,
and will allow a much fuller picture of the close genealogical kinship of individuals across the world.

\vskip 2em

\hrule

\vskip 2em

\section*{Author Summary}

Few of us know our family histories more than a few generations back.
Therefore, it is easy to overlook the fact that we are all distant
cousins, related to one another via a vast network of relationships.
Here we use genome-wide data from European individuals to investigate
these relationships over the past three thousand years,
by looking for long stretches of shared genome between pairs of
individuals inherited from common genetic ancestors.
We quantify this ubiquitous recent common ancestry,
showing that for instance even pairs of individuals on opposite ends of Europe share
hundreds of genetic common ancestors over this time period.
Despite this degree of commonality,
there are also striking regional differences.
For instance, southeastern Europeans share large numbers of common ancestors
which date to the era of the Slavic and Hunnic expansions around 1500
years ago,
while most common ancestors that Italians share with other populations lived
longer ago than 2500 years.
The study of long stretches of shared genetic material holds the promise of rich information
about many aspects of recent population history.

\pagebreak

\section{Introduction}

Even seemingly unrelated humans are distant cousins to each other,
as all members of a species are related to each other through a vastly ramified family tree (their pedigree).
We can see traces of these relationships in genetic data 
when individuals inherit shared genetic material from a common ancestor.
Traditionally, population genetics has studied the distant bulk of these genetic relationships,
which in humans typically date from hundreds of thousands of years ago \citep[e.g.][]{cann1987mitochondrial,takahata1993allelic}.
Such studies have provided deep insights into the origins of modern humans \citep[e.g.][]{li2011inference},
and into recent admixture between diverged populations  \citep[e.g.][]{moorjani2011history,henn2012genomic}.

Although most such genetic relationships among individuals are very old, some individuals are related on far shorter time scales.
Indeed, given that each individual has $2^n$ ancestors from $n$ generations ago,
theoretical considerations suggest that all humans are related genealogically to each other
over surprisingly short time scales \citep{chang1999,rohde2004modelling}.
We are usually unaware of these close genealogical ties, 
as few of us have knowledge of family histories more than a few
generations back, 
and these ancestors often do not contribute any genetic material to us \citep{donnelly1983probability}.
However, in large samples we can hope to identify genetic evidence of more recent relatedness,
and so obtain insight into the population history of the past tens of generations. 
Here we investigate such patterns of recent relatedness in a large European dataset.

The past several thousand years are replete with events that may have had significant impact on modern European relatedness,
such as the Neolithic expansion of farming,
the Roman empire, or the more recent expansions of the Slavs and the Vikings.

Our current understanding of these events is deduced from archaeological, linguistic, cultural, historical, and genetic evidence,
with widely varying degrees of certainty.
However, the demographic and genealogical impact of these events is still uncertain \citep[e.g.][]{gillett2006ethnogenesis}.
Genetic data describing the breadth of genealogical relationships
can therefore add another dimension to our understanding of these historical events.

Work from uniparentally inherited markers (mtDNA and Y chromosomes)
has improved our understanding of human demographic history \citep[e.g.][]{soares2010archaeogenetics}.
However, interpretation of these markers is difficult
since they only record a single lineage of each individual
(the maternal and paternal lineages, respectively),
rather than the entire distribution of ancestors.
Genome-wide genotyping and sequencing datasets have the potential to
provide a much richer picture of human history, 
as we can learn simultaneously about the diversity of ancestors 
that contributed to each individual's genome.

A number of genome-wide studies have begun to reveal quantitative insights into recent human history \citep{novembre2011perspectives}. 
Within Europe,
the first two principal axes of variation of the matrix of genotypes
are closely related to a rotation of latitude and longitude
\citep{menozzi1978synthetic,novembre2008europe,lao2008correlation}, 
as would be expected if patterns of ancestry are mostly shaped by local migration \citep{novembre2008interpreting}.
Other work has revealed a slight decrease in diversity running from south-to-north in Europe, 
with the highest haplotype and allelic diversity in the Iberian peninsula
\citep[e.g.][]{lao2008correlation,auton2009global,nelson2012abundance}, 
and the lowest haplotype diversity in England and Ireland \citep{odushlaine2010population}.
Recently, progress has also been made using genotypes of ancient individuals to understand the prehistory of Europe 
\citep{patterson2012ancient,skoglund2012origins,keller2012insights}.
However, we currently have little sense of the 
time scale of the historical events underlying modern geographic patterns of relatedness,
nor the degrees of genealogical relatedness they imply.

In this paper, we analyze those rare long chunks of genome
that are shared between pairs of individuals due to inheritance from recent common ancestors,
to obtain a detailed view of the geographic structure of recent relatedness.
To determine the time scale of these relationships,
we develop methodology that uses the lengths of shared genomic segments 
to infer the distribution of the ages of these recent common ancestors.
We find that even geographically distant Europeans share
ubiquitous common ancestry even within the past 1000 years,
and show that common ancestry from the past 3000 years 
is a result of both local migration and large-scale historical events.
We find considerable structure below the country level in sharing of recent ancestry,
lending further support to the idea that looking at runs of shared ancestry can identify very subtle population structure \citep[e.g.][]{lawson2012inference}.

Our method for inferring ages of common ancestors is conceptually similar to the work of \citet{palamara2012length},
who use total amount of long runs of shared genome to fit simple parametric models of recent history,
as well as to \citet{li2011inference} and \citet{harris2012inferring},
who use information from short runs of shared genome
to infer demographic history over much longer time scales.
Other conceptually similar work includes \citet{pool2009inference} and \citet{gravel2012population},
who used the length distribution of admixture tracts to fit parametric models of historical admixture.
We rely less on discrete, idealized populations or parametric demographic models than these other works,
and describe continuous geographic structure by obtaining average numbers of common ancestors shared 
by many populations across time in a relatively non-parametric fashion.

\subsection{Definitions: Genetic ancestry and identity by descent}
\label{ss:finding_ibd}

\begin{figure}[!htp]
  \begin{center}
    \includegraphics[width=.3\textwidth]{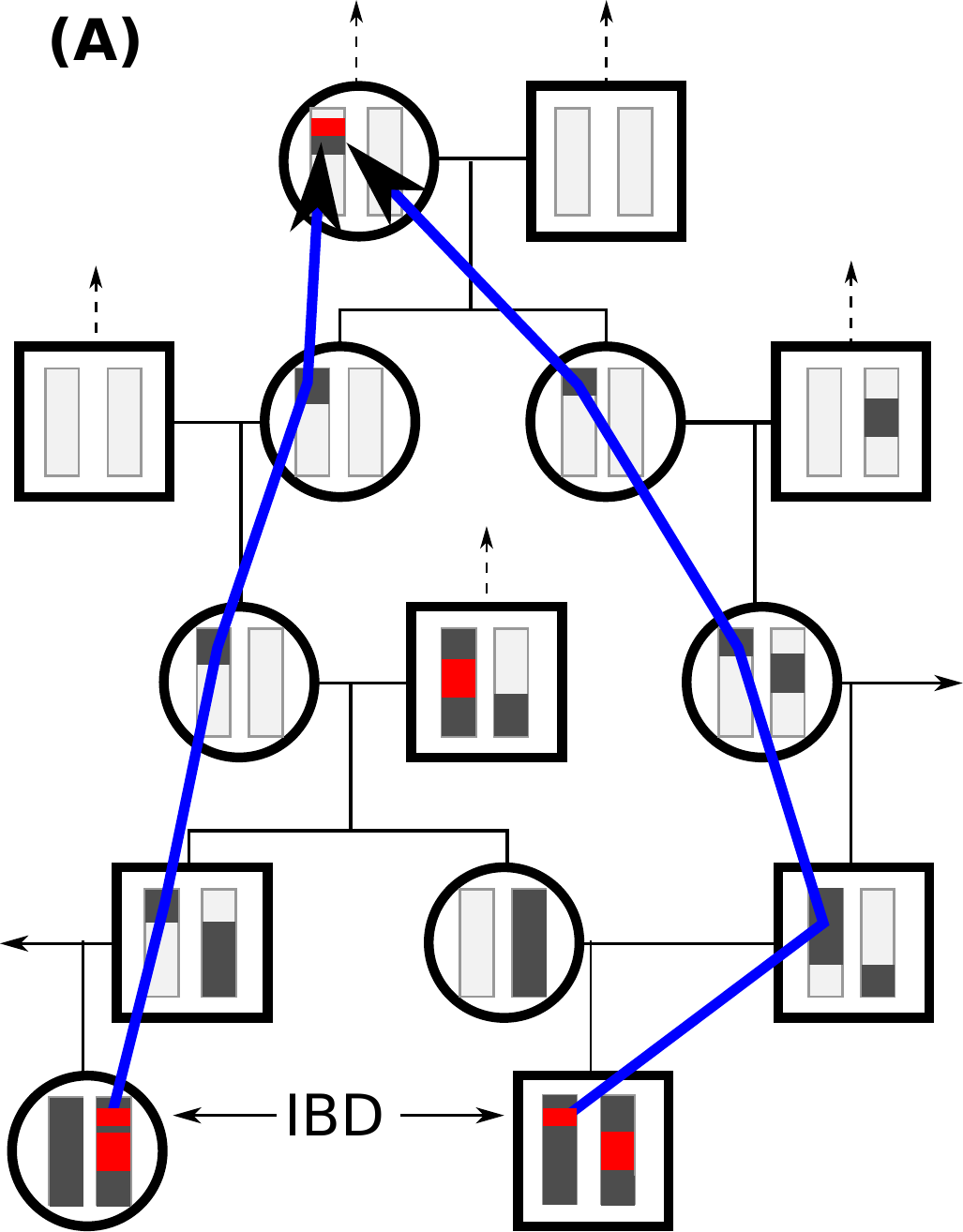}
    \hspace{2em}
    \includegraphics[width=.2\textwidth]{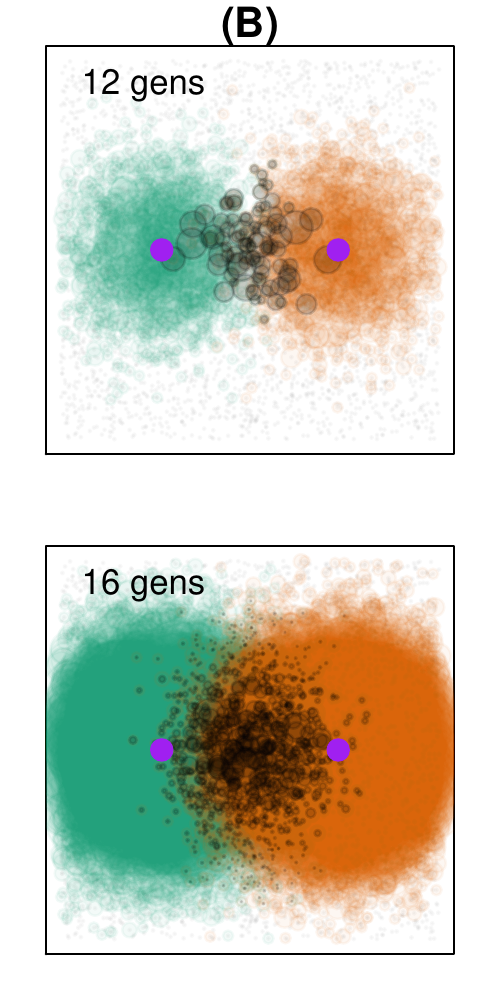}
    \caption{
      {\bf (A)} A hypothetical portion of the pedigree relating two sampled individuals, which shows six of their genealogical common ancestors,
      with the portions of ancestral chromosomes from which the sampled individuals have inherited shaded grey.
      The IBD blocks they have inherited from the two genetic common ancestors are colored red,
      and the blue arrow denotes the path through the pedigree along which one of these IBD blocks was inherited.
      {\bf (B)} Cartoon of the spatial locations of ancestors of two individuals --
      circle size is proportional to likelihood of genetic contribution,
      and shared ancestors are marked in grey.
      Note that common ancestors are likely located between the two,
      and their distribution becomes more diffuse further back in time.
    \label{fig:conceptual}
    }
  \end{center}
\end{figure}

We can only hope to learn from genetic data about those common ancestors from whom two individuals have both inherited the same genomic region.
If a pair of individuals have both inherited some genomic region from a common ancestor,
that ancestor is called a ``genetic common ancestor'', 
and the genomic region is shared ``identical by descent'' (IBD) by the two. 
Here we define an ``IBD block'' to be a contiguous segment of genome inherited (on at least one chromosome)
from a shared common ancestor without intervening recombination
(see figure~\ref{fig:conceptual}A).
A more usual definition of IBD 
restricts to those segments inherited from some prespecified set of ``founder'' individuals
\citep[e.g.][]{fisher1954fuller,donnelly1983probability,chapman2002effect},
but we allow ancestors to be arbitrarily far back in time.
Under our definition, everyone is IBD everywhere,
but mostly on very short, old segments \citep{powell2010reconciling}.
We measure lengths of IBD segments in units of Morgans (M) or centiMorgans (cM),
where 1 Morgan is defined to be the distance over which an average of one recombination (i.e.\ a crossover) occurs per meiosis.
Segments of IBD are broken up over time by recombination, 
which implies that older shared ancestry tends to result in shorter shared IBD blocks.

Sufficiently long segments of IBD can be identified
as long, contiguous regions over which the two individuals are identical 
(or nearly identical) at a set of Single Nucleotide Polymorphisms (SNPs) 
which segregate in the population. 
Formal, model-based methods to infer IBD are only computationally feasible 
for very recent ancestry \citep[e.g.][]{brown2012inferring},
but recently, fast heuristic algorithms have been developed
that can be applied to thousands of samples typed on genotyping chips 
\citep[e.g.][]{browning2011powerful,gusev2009whole}.

The relationship between 
numbers of long, shared segments of genome,
numbers of genetic common ancestors,
and numbers of genealogical common ancestors
can be difficult to envision.
Since everyone has exactly two biological parents,
every individual has exactly $2^n$ paths of length $n$ meioses leading back through their pedigree,
each such path ending in a grand$^{n-1}$parent.
However, due to Mendelian segregation and limited recombination, 
genetic material will only be passed down along a small subset of
these paths \citep{donnelly1983probability}. 
As $n$ grows, these paths proliferate rapidly and so the genealogical
paths of two individuals soon overlap significantly.
(These points are illustrated in in figure~\ref{fig:conceptual}.)
By observing the number of shared genomic blocks, we learn about 
the degree to which their genealogies overlap,
or the number of common ancestors from which both individuals have inherited genetic material.

At least one parent of each genetic common ancestor of two individuals is also a genetic common ancestor, 
so the number of genetic common ancestors at each point back in time is strictly increasing.
A more relevant quantity is the rate of appearance of {\em most recent} common genetic ancestors.
This quantity can be much more intuitive, and is closely related to the coalescent rate \citep{hudson1990gene},
as we demonstrate later.
For this reason, when we say ``genetic common ancestor'' or ``rate of genetic common ancestry'',
we are referring to only the {\em most recent} genetic common ancestors from which the individuals in question
inherited their shared segments of genome.

\section{Results}

We applied the {\tt fastIBD} method, implemented in {\tt BEAGLE} v3.3 \citep{browning2011powerful},
to the European subset of the POPRES dataset \citep[dbgap accession phs000145.v1.p1,][]{nelson2008population},
which includes language and country-of-origin data for several thousand Europeans genotyped at 500000 SNPs.
Our simulations showed that we have good power to detect long IBD blocks
(probability of detection 50\% for blocks longer than 2cM, rising to 98\% for blocks longer than 4cM),
and a low false positive rate (discussed further in section \ref{ss:error_model} of the methods).
We excluded from our analyses individuals who reported grandparents originating from non-European countries or more than one distinct country
(and refer to the remainder as ``Europeans'').
After removing obvious outlier individuals and close relatives, 
we were left with 2257 individuals
who we grouped using reported country of origin and language into 40 populations,
listed with sample sizes and average IBD levels in table~\ref{tab:ibd_summaries}.
For geographic analyses, we located each population at the largest population city in the appropriate region.
Pairs of individuals in this dataset were found to share a total of 1.9 million segments of IBD, 
an average of 0.74 per pair of individuals,
or 831 per individual.
The mean length of these blocks was 2.5cM, the median was 2.1cM
and the 25$^\mathrm{th}$ and 75$^\mathrm{th}$ quantiles are 1.5cM and 2.9cM respectively.
The majority of pairs sharing some IBD shared only a single block of IBD (94\%).
The total length of IBD blocks an individual shares with all others ranged between 30\% and 250\% (average 128\%) of the length of the genome
(greater than 100\% is possible as individuals may share IBD blocks with more than one other at the same genomic location).

\begin{table}[!htp]
\begin{center}
\begin{tabular}{|rlrrr||rlrrr|}
  \hline
\multicolumn{2}{|l}{\bf E group}  & $n$ & self & other & \multicolumn{2}{|l}{\bf N group}  &  $n$& self & other \\ \cline{1-10}
Albania                 &   AL    &   9 & 14.5 & 1.7   & Denmark       &   DK              &   1 &   -- & 0.9   \\
Austria                 &   AT    &  14 &  1.3 & 0.9   & Finland       &   FI              &   1 &   -- & 1.2   \\
Bosnia                  &   BO    &   9 &  4.1 & 1.6   & Latvia        &   LV              &   1 &   -- & 1.6   \\
Bulgaria                &   BG    &   1 &   -- & 1.3   & Norway        &   NO              &   2 &  2.0 & 0.8   \\
Croatia                 &   HR    &   9 &  2.8 & 1.6   & Sweden        &   SE              &  10 &  3.4 & 1.0   \\
Czech Republic          &   CZ    &   9 &  2.1 & 1.3   &               &                   &     &      &       \\ \cline{6-10}
Greece                  &   EL    &   5 &  1.8 & 0.9   & \multicolumn{2}{|l}{\bf W group}  &  $n$& self & other \\ \cline{6-10}
Hungary                 &   HU    &  19 &  1.9 & 1.2   & Belgium       &   BE              &  37 &  1.1 & 0.6   \\
Kosovo                  &   KO    &  15 &  9.9 & 1.7   & England       &   EN              &  22 &  1.3 & 0.7   \\
Montenegro              &   ME    &   1 &   -- & 1.8   & France        &   FR              &  86 &  0.7 & 0.5   \\
Macedonia               &   MA    &   4 &  2.5 & 1.4   & Germany       &   DE              &  71 &  1.1 & 0.9   \\
Poland                  &   PL    &  22 &  3.8 & 1.5   & Ireland       &   IE              &  60 &  2.6 & 0.6   \\
Romania                 &   RO    &  14 &  2.1 & 1.2   & Netherlands   &   NL              &  17 &  1.9 & 0.7   \\
Russia                  &   RU    &   6 &  4.3 & 1.4   & Scotland      &   SC              &   5 &  2.2 & 0.7   \\
Slovenia                &   SI    &   2 &  5.0 & 1.3   & Swiss French  &  CHf              & 839 &  1.3 & 0.6   \\
Serbia                  &   RS    &  11 &  2.7 & 1.5   & Swiss German  &  CHd              & 103 &  1.6 & 0.6   \\
Slovakia                &   SK    &   1 &   -- & 0.7   & Switzerland   &   CH              &  17 &  1.1 & 0.5   \\
Ukraine                 &   UA    &   1 &   -- & 1.5   & United Kingdom&   UK              & 358 &  1.2 & 0.7   \\
Yugoslavia              &   YU    &  10 &  3.4 & 1.5   &               &                   &     &      &       \\ \cline{6-10}
                        &         &     &      &       & \multicolumn{2}{|l}{\bf I group}  &  $n$& self & other \\ \cline{1-10}
\multicolumn{2}{|l}{\bf TC group} &  $n$& self & other & Italy         &   IT              & 213 &  0.6 & 0.5   \\ \cline{1-5}
Cyprus                  &   CY    &   3 &  2.7 & 0.4   & Portugal      &   PT              & 115 &  1.9 & 0.5   \\
Turkey                  &   TR    &   4 &  2.2 & 0.5   & Spain         &   ES              & 130 &  1.5 & 0.4   \\
\hline
\end{tabular}
\caption{
Populations, abbreviations, sample sizes ($n$), 
mean number of IBD blocks shared by a pair of individuals from that population (``self''),
and mean IBD rate averaged across all other populations (``other'');
sorted by regional groupings described in the text.
\label{tab:ibd_summaries}
}
\end{center}
\end{table}

The observed genomic density of long IBD blocks (per cM) 
can be affected by recent selection \citep{albrechtsen2010natural} and
by cis-acting recombination modifiers.
We find that the local density of IBD blocks of all lengths is relatively constant across the genome,
but in certain regions the length distribution is systematically perturbed
(see supplemental figure \ref{sfig:overlap_all}),
including around certain centromeres
and the large inversion on chromosome 8 \citep{giglio2001olfactory},
also seen by \citet{albrechtsen2010natural}.
Somewhat surprisingly, the MHC does not show an unusual pattern of IBD, 
despite having shown up in other genomic scans for IBD \citep{albrechtsen2010natural,gusev2012homologous}. 
However, there are a few other regions where differences in IBD rate are not predicted by differences in SNP density. 
Notably, there are two regions, on chromosomes 15 and 16, 
which are nearly as extreme in their deviations in IBD as the inversion on chromosome 8, 
and may also correspond to large inversions segregating in the sample.
These only make up a small portion of the genome, and do not significantly affect our other analyses (and so are not removed);
we leave further analysis for future work.

\subsection{Substructure and recent migrants}

We should expect significant within-population variability, 
as modern countries are relatively recent constructions of diverse assemblages of languages and heritages. 
To assess the uniformity of ancestry within populations,
we used a permutation test to measure, for each pair of populations $x$ and $y$,
the uniformity with which relationships with $x$ are distributed across individuals from $y$. 
Most comparisons show statistically significant heterogeneity (supplemental figure~\ref{sfig:substructure_summaries}),
which is probably due to population substructure (as well as correlations introduced by the pedigree). 
A notable exception is that nearly all populations showed no
significant heterogeneity of numbers of common ancestors with Italian samples,
suggesting that most common ancestors shared with Italy lived longer ago than the time that structure within modern-day countries formed.

\begin{figure}[!htp]
  \begin{center}
    \wikiincludegraphics{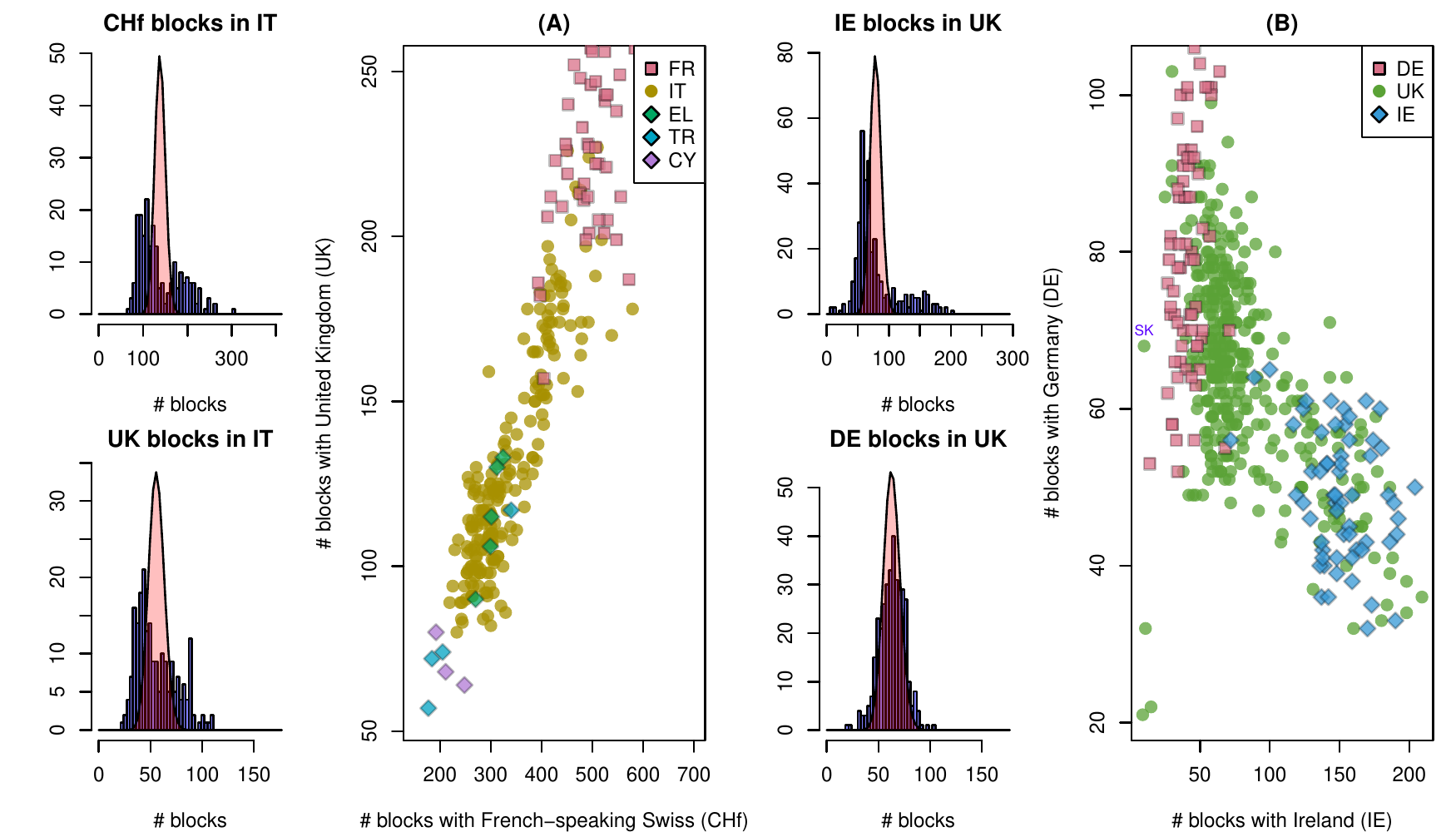}
    \caption{
    Substructure, in {\bf (A)} Italian, and {\bf (B)} UK samples.
    The leftmost plots of {\bf (A)} show histograms of the numbers of IBD blocks 
    that each Italian sample shares with any French-speaking Swiss (top) and anyone from the UK (bottom),
    overlaid with the expected distribution (Poisson) if there was no dependence between blocks. 
    Next is shown a scatterplot of numbers of blocks shared with French-speaking Swiss and UK samples,
    for all samples from France, Italy, Greece, Turkey, and Cyprus.
    We see that the numbers of recent ancestors each Italian shares with the French-speaking Swiss and with the United Kingdom are both bimodal,
    and that these two are positively correlated, ranging continuously between values typical for Turkey/Cyprus and for France.
    Figure {\bf (B)} is similar, showing that
    the substructure within the UK is part of a continuous trend ranging from Germany to Ireland.
    The outliers visible in the scatterplot of figure \ref{fig:substructure}B are easily explained as individuals with immigrant recent ancestors
    -- the three outlying UK individuals in the lower left share many more blocks with Italians than all other UK samples,
    and the individual labeled ``SK'' is a clear outlier for the number of blocks shared with the Slovakian sample.
    \label{fig:substructure}
    }
  \end{center}
\end{figure}

Two of the more striking examples of substructure are illustrated in figure \ref{fig:substructure}.
Here, we see that variation within countries can be reflective of continuous variation in ancestry that spans a broader geographic region,
crossing geographic, political, and linguistic boundaries.
Figure \ref{fig:substructure}A shows the distinctly bimodal distribution of numbers of IBD blocks
that each Italian shares with both French-speaking Swiss and the UK,
and that these numbers are strongly correlated.
Furthermore, the amount that Italians share with these two populations varies continuously
from values typical for Turkey and Cyprus, to values typical for France and Switzerland.
Interestingly, the Greek samples (EL) place near the middle of the Italian gradient.
It is natural to guess that there is a north-south gradient of recency of common ancestry
along the length of Italy, and that southern Italy has been historically more closely connected to the eastern Mediterranean.

In contrast, within samples from the UK and nearby regions
we see negative correlation between numbers of blocks shared with Irish and numbers of blocks shared with Germans.
From our data, we do not know if this substructure is also geographically arranged within the UK (our sample of which which may include individuals from Northern Ireland).
However, an obvious explanation of this pattern is that individuals within the UK differ in the number of recent ancestors shared with Irish, 
and that individuals with less Irish ancestry have a larger portion of their recent ancestry shared with Germans.
This suggests that there is variation across the UK 
-- perhaps a geographic gradient -- 
in terms of the amount of Celtic versus Germanic ancestry.

The first two principal components of the matrix of genotypes,
after suitable manipulations, can reproduce the geographic positions
of European populations
\citep[e.g.][]{menozzi1978synthetic,novembre2008europe,lao2008correlation}.
Therefore, it is natural to compare the structure we see within populations in terms of IBD sharing 
to the positions on the principal components map.
(A PCA map of these populations \citep[produced by EIGENSTRAT,][]{price2006eigenstrat} is shown in supplemental figure \ref{sfig:pca_map}.)
It is not known what the geographic resolution of the principal components map is,
but if relative positions within populations is meaningful,
then comparison of IBD to PCA can stand in for comparison to geography.
Indeed, as seen in supplemental figures \ref{sfig:pca_ibd_uk} and \ref{sfig:pca_ibd_it},
the substructure of figure \ref{fig:substructure} correlates well with the position on certain principal components,
further suggesting that the structure is geographically meaningful.
Conversely, since the substructure we see is highly statistically significant,
this demonstrates that the scatter of positions within populations on
the European PCA map is at least in part signal, rather than noise.

\subsection{Europe-wide patterns of relatedness}

\begin{figure}[!htp]
  \begin{center}
    \wikiincludegraphics{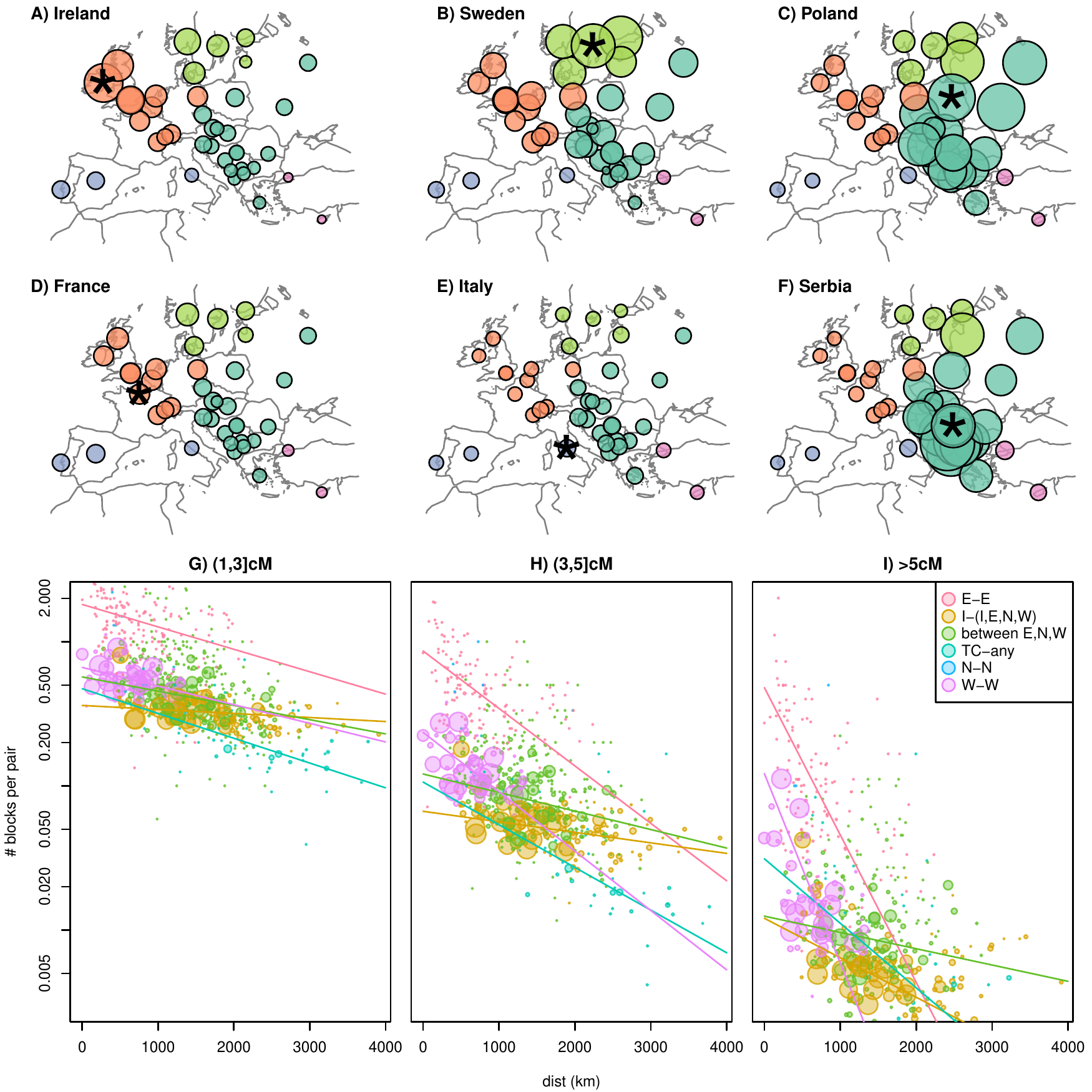}
    \caption{
    In all figures, colors give categories based on the regional groupings of table~\ref{tab:ibd_summaries}.
    {\bf (A--F)} The area of the circle located on a particular population
    is proportional to the mean number of IBD blocks of length at least 1cM shared 
    between random individuals chosen from that population and the population named in the label (also marked with a star).
    Both regional variation of overall IBD rates
    and gradual geographic decay are apparent.
    {\bf (G--I)} Mean number of IBD blocks of lengths 1--3cM (oldest), 3--5cM, and $>$5cM (youngest), 
    respectively, shared by a pair of individuals across all pairs of populations;
    the area of the point is proportional to sample size (number of distinct pairs), capped at a reasonable value;
    and lines show an exponential decay fit to each category (using a Poisson GLM weighted by sample size).
    Comparisons with no shared IBD are used in the fit but not shown in the figure (due to the log scale).
    ``E--E'', ``N--N'', and ``W--W'' denote any two populations both in the E, N, or W grouping, respectively;
    ``TC-any'' denotes any population paired with Turkey or Cyprus;
    ``I-(I,E,N,W)'' denotes Italy, Spain, or Portugal paired with any population except Turkey or Cyprus;
    and ``between E,N,W'' denotes the remaining pairs
    (when both populations are in E, N, or W, but the two are in different groups).
    The exponential fit for the N--N points is not shown due to the very small sample size.
    See supplemental figure \ref{sfig:sharing_rates_svg} for an SVG version of these plots where it is possible to identify individual points.
    \label{fig:sharing_and_maps}
    }
  \end{center}
\end{figure}

Individuals usually share the highest number of IBD blocks with others from the same population,
with some exceptions. 
For example, individuals in the UK share more IBD blocks on average, 
and hence more close genetic ancestors,
with individuals from Ireland than with other individuals from the UK (1.26 versus 1.09 blocks at least 1cM per pair, Mann-Whitney $p<10^{-10}$),
and Germans share similarly more with Polish than with other Germans (1.24 versus 1.05, $p=5.7\times10^{-6}$),
a pattern which could be due to recent asymmetric migration from a
smaller population into a larger population. 
In figure \ref{fig:sharing_and_maps} we depict the geography of rates of IBD sharing between populations,  
i.e.\ the average number of IBD blocks shared by a randomly chosen pair of individuals.
Above, maps show the IBD rate relative to certain chosen populations,
and below, all pairwise sharing rates are plotted against the geographic distance separating the populations.
It is evident that geographic proximity is a major determinant of IBD sharing (and hence recent relatedness), 
with the rate of pairwise IBD decreasing relatively smoothly as the
geographic separation of the pair of populations increases. 
Note that even populations represented by only a single sample are included,
as these showed surprisingly consistent signal despite the small sample size.

Superimposed on this geographic decay there is striking regional variation in rates of IBD.
To further explore this variation, we divided the populations into 
the four groups listed in table~\ref{tab:ibd_summaries},
using geographic location and correlations in the pattern of IBD sharing with other populations
(shown in supplemental figure~\ref{sfig:correlations}).
These groupings are defined as:
Europe ``E'', lying to the east of Germany and Austria;
Europe ``N'', lying to the north of Germany and Poland;
Europe ``W'', to the west of Germany and Austria and including these;
the Iberian and Italian peninsulas ``I''; and Turkey/Cyprus ``TC''.
Although the general pattern of regional IBD variation is strong,
none of these groups have sharp boundaries -- for instance, Germany, Austria, and Slovakia are intermediate between E and W. 
Furthermore, we suspect that the Italian and Iberian peninsulas likely
do not group together because of higher shared ancestry with each other,
but rather because of similarly low rates of IBD with other European populations. 
The overall mean IBD rates between these regions are shown in table~\ref{tab:group_rates},
and comparisons between different groupings are colored differently in figure~\ref{fig:sharing_and_maps}G--I,
showing that rates of IBD sharing between E populations and between N populations average
a factor of about three higher than other comparisons at similar
distances. Such a large difference in the rates of IBD sharing between
regions, cannot
be explained by plausible differences in false
positive rates or power between populations, since this pattern holds even
at the longest length scales, where block identification is nearly perfect.

To better understand IBD within these groupings,
we show in figures \ref{fig:sharing_and_maps}G--I how average numbers of IBD blocks shared, 
in three different length categories,
depend on the geographic distance separating the two populations.
Even without taking into account regional variation,
mean numbers of shared IBD blocks decay exponentially with distance,
and further structure is revealed by breaking out populations by the regional groupings described above.
The exponential decays shown for each pair of groupings 
emphasize how the decay of IBD with distance becomes more rapid for longer blocks.
This is expected under models where migration is mostly local, since
as one looks further back in time,
the distribution of each individual's ancestors is less concentrated around the individual's location
(recall figure \ref{fig:conceptual}B).
Therefore, the expected number of ancestors shared by a pair of individuals decreases 
as the geographic distance between the pair increases;
and this decrease is faster for more recent ancestry.

This wider spread of older blocks can also explain why the decay of IBD with distance varies
significantly by region even if dispersal rates have been relatively constant.
For instance, the gradual decay of sharing with the Iberian and Italian peninsulas
could occur because these blocks are inherited from much longer ago
than blocks of similar lengths shared by individuals in other populations.

Conversely, there is a high level of sharing for ``E--E'' relationships over a broad range of distances.
This is especially true for our shortest (oldest) blocks:
individuals in our E grouping share on average more short blocks with individuals in distant E populations 
than do pairs of individuals in the same W population.
We argue below that this is because modern individuals in these locations
have a larger proportion of their ancestors 
in a relatively small population that subsequently expanded.

\begin{table}[!htp]
\begin{center}
\begin{tabular}{|c|rrrrr|}
  \hline
   IBD rate  & E & I & N & TC & W \\ 
  \hline
  E  & 2.57 & 0.44 & 0.99 & 0.62 & 0.53 \\ 
  I  & 0.44 & 0.80 & 0.43 & 0.41 & 0.45 \\ 
  N  & 0.99 & 0.43 & 2.62 & 0.33 & 0.86 \\ 
  TC & 0.62 & 0.41 & 0.33 & 1.43 & 0.25 \\ 
  W  & 0.53 & 0.45 & 0.86 & 0.25 & 0.93 \\ 
   \hline
\end{tabular}
\end{center}
\caption{
Rates of IBD within and between each geographic grouping given in table \ref{tab:ibd_summaries}.
\label{tab:group_rates}
}
\end{table}

Having seen the continent-wide patterns of IBD in figure~\ref{fig:sharing_and_maps},
it is natural to wonder if similar information is contained in
single-site summaries of relatedness,
such as mean Identity by State (IBS) values across European populations.
The mean IBS between populations $x$ and $y$ is defined as the probability that two randomly chosen alleles from $x$ and $y$
are identical (``By State''),
averaging over SNPs and individuals.
In the analogous plot of IBS against geographic distance (supplemental figure~\ref{sfig:ibs_by_dist}),
some of the patterns seen in figure~\ref{fig:sharing_and_maps} are present, and some are not.
For instance, there is a continuous decay with geographic distance (linear, not exponential),
and comparisons to the southern ``I'' group and to Cyprus/Turkey are even more well-separated below the others.
On the other hand, the ``E-E'' comparisons do not show higher IBS than the bulk of the remaining comparisons.

%%%%%%%%%%
\subsection{Ages and numbers of common ancestors}

Each block of genome shared IBD by a pair of individuals represents genetic material 
inherited from one of their genetic common ancestors. 
Since the distribution of lengths of IBD blocks differs depending on the age of the ancestors
-- e.g.\ older blocks tend to be shorter --
it is possible to use the distribution of lengths of IBD blocks
to infer numbers of most recent pairwise genetic common ancestors back
through time averaged across pairs of individuals.
For this inference, we restricted to blocks longer than 2cM, where we had good power to detect true IBD blocks.
We obtain dates in units of generations in the past,
and for ease of discussion convert these to years ago (ya)
by taking the mean human generation time to be 30 years \citep{fenner2005crosscultural}. 

\paragraph{Nature of the results on age inference}
There are two major difficulties to overcome, however.
First, detection is noisy: 
we do not detect all IBD segments (especially shorter ones),
and some of our IBD segments are false positives.
This problem can be overcome by 
careful estimation and modeling of error, described in section \ref{ss:error_model}.
The second problem is more serious and unavoidable:
as described in section \ref{ss:inversion_methods},
the inference problem 
is extremely ``ill conditioned'' \citep[in the sense of][]{petrov2005well},
meaning in this case that there are many possible histories of shared ancestry
that fit the data nearly equally well.
For this reason, there is a fairly large, unavoidable limit to the temporal resolution,
but we still obtain a good deal of useful information.

\begin{figure}[!htp]
  \begin{center}
    \wikiincludegraphics{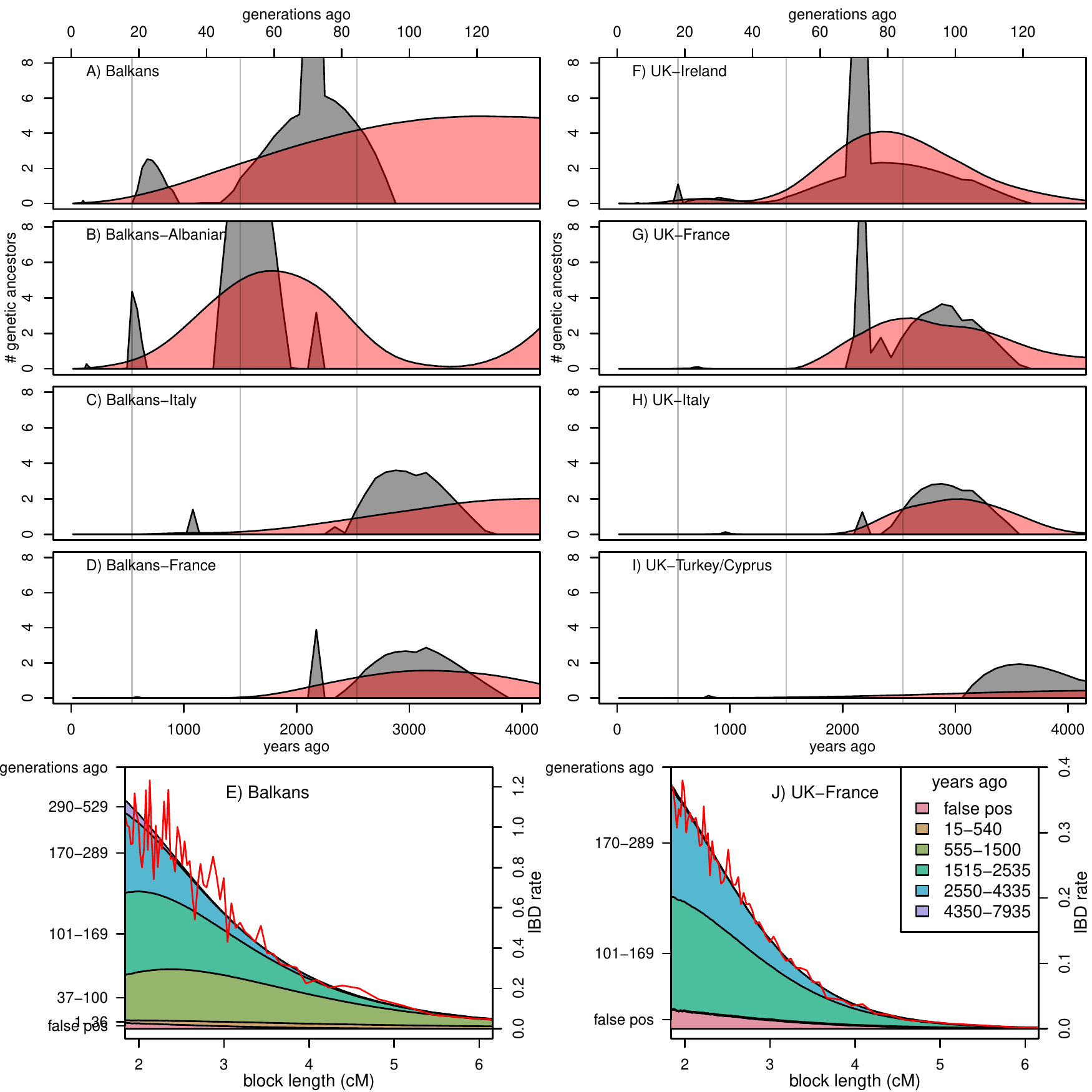}
    \caption{
    {\bf Estimated average number of most recent genetic common ancestors per generation back through time,}
    shared by {\bf (A)} pairs of individuals from ``the Balkans'' (former Yugoslavia, Bulgaria, Romania, Croatia, Bosnia, Montenegro, Macedonia, Serbia, and Slovenia,
    excluding Albanian speakers);
    and, shared by one individual from the Balkans with one individual from 
    {\bf (B)} Albanian speaking populations;
    {\bf (C)} Italy;
    or {\bf (D)} France.
    The black distribution is the maximum likelihood fit;
    shown in red is smoothest solution that still fits the data,
    as described in section \ref{ss:inversion_methods}.
    {\bf (E)} shows the observed IBD length distribution for pairs of individuals from the Balkans (red curve),
    along with the distribution predicted by the smooth (red) distribution in {(A)},
    as a stacked area plot partitioned by time period in which the common ancestor lived.
    The partitions with significant contribution are labeled on the left vertical axis (in generations ago), 
    and the legend in {(J)} gives the same partitions, in years ago; 
    the vertical scale is given on the right vertical axis.
    The second column of figures {\bf (F--J)} is similar,
    except that comparisons are relative to samples from the UK.
    \label{fig:inversion_distributions}
    }
  \end{center}
\end{figure}

\begin{figure}[!htp]
  \begin{center}
    \wikiincludegraphics{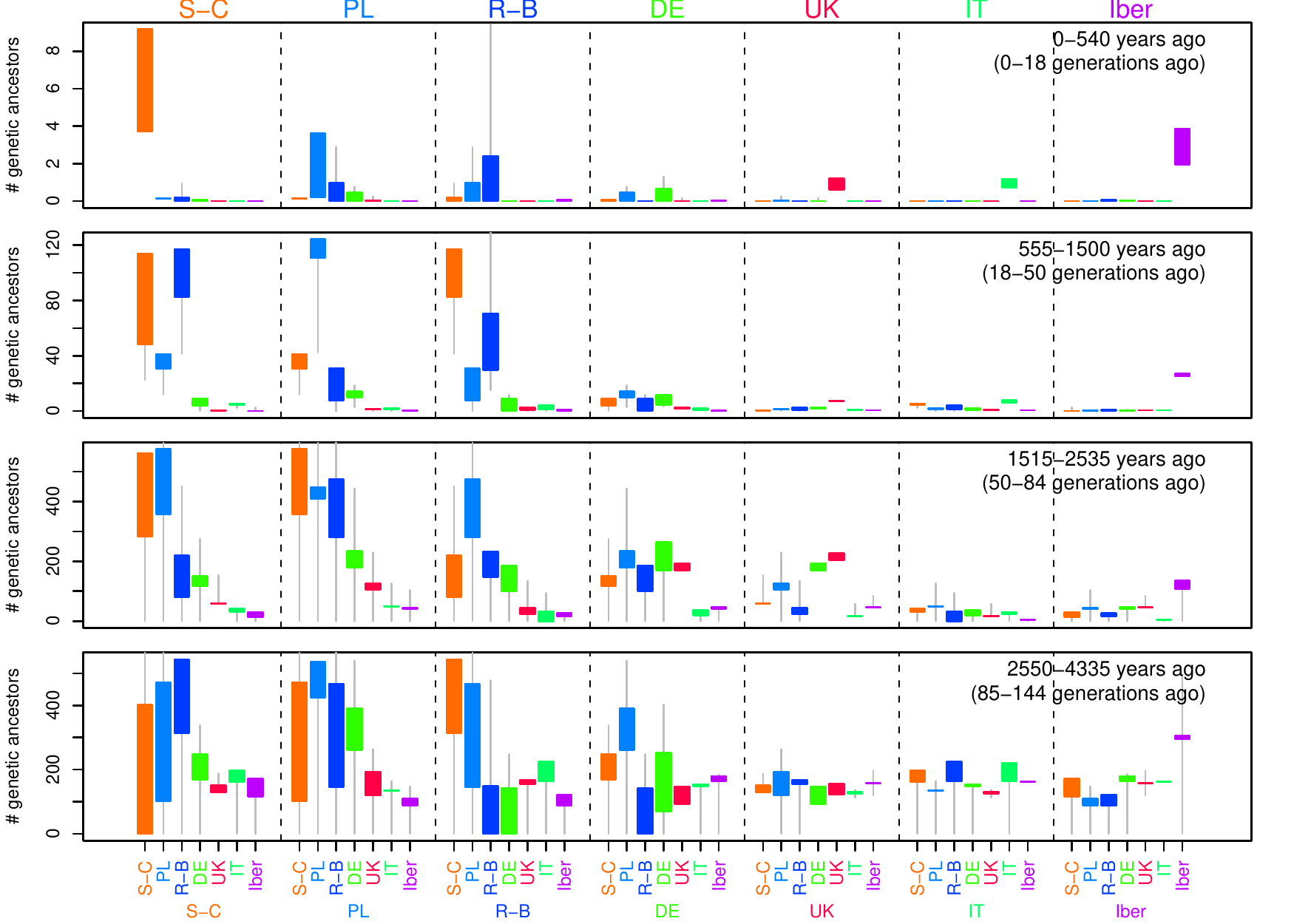}
    \caption{
    {\bf Estimated average total numbers of genetic common ancestors shared per pair of individuals in various pairs of populations,}
    in roughly the time periods 0--500ya, 500--1500ya, 1500--2500ya, and 2500--4300ya.
    We have combined some populations to obtain larger sample sizes:
    ``S-C'' denotes Serbo-Croatian speakers in former Yugoslavia,
    ``PL'' denotes Poland,
    ``R-B'' denotes Romania and Bulgaria,
    ``DE'' denotes Germany,
    ``UK'' denotes the United Kingdom,
    ``IT'' denotes Italy,
    and ``Iber'' denotes Spain and Portugal.
    For instance, the green bars in the leftmost panels tell us that 
    Serbo-Croatian speakers and Germans most likely share 0--0.25 most recent genetic common ancestor from the last 500 years,
    3--12 from the period 500--1500 years ago, 
    120--150 from 1500--2500 ya, and 170--250 from 2500--4400 ya.
    % The figure should be read as a boxplot,
    % as we are fairly confident the true number of genetic common ancestors 
    % lies in or near the colored bar in each case,
    % and almost certainly lies within the range demarcated by the vertical grey lines;
    % see section \ref{ss:inversion_methods} for the method.
    Although the lower bounds appear to extend to zero,
    they are significantly above zero in nearly all cases
    except for the most recent period 0--540ya.  \label{fig:inversion_boxplots}
    }
  \end{center}
\end{figure}

We deal with this uncertainty by describing the set of histories 
(i.e.\ historical numbers of common genetic ancestors)
that are consistent with the data,
summarized in two ways.
First, it is useful to look at individual consistent histories,
which gives a sense of recurrent patterns and possible historical signals.
Figure \ref{fig:inversion_distributions} shows for several populations both
the best-fitting history (in black) 
and the smoothest history that still fits the data (in red).
We can make general statements if they hold across all (or most) consistent histories.
Second, we can summarize the entire set of consistent histories 
by finding confidence intervals (bounds)  
for the total number of common ancestors aggregated in certain time periods.
These are shown in figure \ref{fig:inversion_boxplots},
giving estimates (colored bands) and bounds (vertical lines) 
for the total numbers of genetic common ancestors in each of three time periods,
roughly 0--500ya, 500--1500ya, and 1500--2500ya (``ya'' denotes ``years ago'').
Supplemental figures~\ref{sfig:inversion_boxplots_long} (and \ref{sfig:inversion_boxplots_long_coal})
is a version of figure~\ref{fig:inversion_boxplots} with more populations (in coalescent units, respectively),
and plots analogous to figure~\ref{fig:inversion_distributions} for all these histories
are shown in supplemental figure~\ref{sfig:tinyinversions}.
For a precise description of the problem and our methods, see section \ref{ss:inversion_methods}.
We validated the method through simulation (details in supplemental document), 
and found that it performed well to the temporal resolution discussed here. 
We note that in simulations where the population size changes smoothly, 
the maximum likelihood solution is often overly peaky, 
whereas the smoothed solution can smear out the signal of rapid change in population size. 
In light of that we encourage the reader to view truth as lying somewhere between these two solutions, 
and to not overinterpret specific peaks in the maximum likelihood,
which may occur due to numerical properties of the inference. 
That said, there are a number of sharp peaks in common ancestry shared across many population comparisons older than 2000 years ago, 
which may potentially indicate demographic events in a shared ancestral population. 
A more thorough investigation of these older shared signals would potentially need a more model-based approach, 
so we restrict ourselves here to talking about the broad differences between the distribution of common shared ancestors between regions.

The time periods we use for these bounds are quite large, but this is unavoidable,
because of a trade off between temporal resolution and uncertainty in numbers of common ancestors.
Also note that the lower bounds on numbers of common ancestors during each time
interval are often close to zero.
This is because one can (roughly speaking) obtain a history with equally good fit
by moving ancestors from that time interval into the neighboring ones, 
resulting in peaks on either side of the selected time interval (see figure \ref{sfig:example_inversion_bounds}),
even though these do not generally reflect realistic histories.
The reader should also bear in mind that 
we do not depict the dependence of uncertainty between intervals.

\paragraph{Results of age inference}
In figure \ref{fig:inversion_distributions} we show
how the age and number of shared pairwise genetic common ancestors changes as we move away from the Balkans (left column) and the UK (right column),
along with two examples of how the observed block length distribution 
is composed of ancestry from different depths.
(The average number of shared pairwise genetic common ancestors from generation $n$
is the probability that the most recent common ancestor of a pair at a single site lived in generation $n$ (i.e.~the coalescent rate)
multiplied by the expected number of segments that recombination has broken a pair of individuals' genomes into that many generations back, 
as shown in section \ref{ss:numbers_of_ancestors}.)
More plots of this form are shown in supplemental figure~\ref{sfig:tinyinversions}, and
coalescent rates between pairs of populations are shown in the (equivalent) supplemental figure \ref{sfig:inversion_distributions_coal}.

Most detectable recent common ancestors lived between 1500 and 2500 years ago,
and only a small proportion of blocks longer than 2cM are inherited from longer ago than 4000 years.
Obviously, there are a vast number of genetic common ancestors older than this,
but the blocks inherited from such common ancestors are sufficiently unlikely to be longer than 2cM
that we do not detect many.
For the most part, blocks longer than 4cM come from 500--1500 years ago,
and blocks longer than 10cM from the last 500 years.

In most cases, only pairs within the same population are likely to share genetic common ancestors within the last 500 years.
Exceptions are generally neighboring populations (e.g.\ UK and Ireland).
During the period 500--1500ya, individuals typically share tens to hundreds of genetic common ancestors
with others in the same or nearby populations,
although some distant populations have very low rates.
Longer ago than 1500ya, pairs of individuals from any part of Europe share hundreds of genetic ancestors in common,
and some share significantly more.

\paragraph{Regional variation: interesting cases}
We now examine some of the more striking patterns we see in more detail.

There is relatively little common ancestry shared between the Italian peninsula and other locations,
and what there is seems to derive mostly from longer ago than 2500ya.
An exception is that Italy and the neighboring Balkan populations share small but significant numbers of common ancestors in the last 1500 years,
as seen in supplemental figures~\ref{sfig:tinyinversions} or \ref{sfig:all_inversions}.
The rate of genetic common ancestry between pairs of Italian individuals seems to have been fairly constant for the past 2500 years,
which combined with significant structure within Italy suggests a constant exchange of migrants between coherent subpopulations.

Patterns for the Iberian peninsula are similar,
with both Spain and Portugal showing very few common ancestors with other populations over the last 2500 years.
However, the rate of IBD sharing within the peninsula is much higher than within Italy -- during the last 1500 years
the Iberian peninsula shares fewer than 2 genetic common ancestors with other populations,
compared to roughly 30 per pair within the peninsula;
Italians share on average only about 8 with each other during this period.

The higher rates of IBD between populations in the ``E'' grouping shown in figure \ref{fig:sharing_and_maps}
seem to derive mostly from ancestors living 1500--2500ya,
but also show increased numbers from 500--1500ya,
as shown in figure \ref{fig:inversion_boxplots} and supplemental figures \ref{sfig:all_inversions}.
For comparison, the IBD rate is high enough that
even geographically distant individuals in these eastern populations
share about as many common ancestors 
as do two Irish or two French-speaking Swiss.

By far the highest rates of IBD within any populations is found between Albanian speakers --
around 90 ancestors from 0--500ya, and around 600 ancestors from 500--1500ya
(so high that we left them out of figure~\ref{fig:inversion_boxplots}; see supplemental figure \ref{sfig:inversion_boxplots_long}).
Beyond 1500ya, the rates of IBD drop to levels typical for other populations in the eastern grouping.

There are clear differences in the number and timing of genetic common ancestors
shared by individuals from different parts of Europe,
These differences reflect the impact of major historical and demographic events, 
superimposed against a background of local migration and generally high genealogical relatedness across Europe.
We now turn to discuss possible causes and implications of these results.

\section{Discussion}

Genetic common ancestry within the last 2500 years across Europe
has been shaped by diverse demographic and historical events.
There are both continental trends, 
such as a decrease of shared ancestry with distance,
regional patterns,
such as higher IBD in eastern and northern populations,
and diverse outlying signals.
We have furthermore quantified numbers of genetic common ancestors that populations share with each other back through time,
albeit with a (unavoidably) coarse temporal resolution.
These numbers are intriguing not only because of the differences between populations,
which reflect historical events,
but the high degree of implied genealogical commonality between even geographically distant populations.

\paragraph{Ubiquity of common ancestry}
We have shown that typical pairs of individuals drawn from across
Europe have a good chance of sharing long stretches of identity by descent,
even when they are separated by thousands of kilometers. 
We can furthermore conclude that pairs of individuals across Europe 
are reasonably likely to share common genetic ancestors within the last 1000 years,
and are certain to share many within the last 2500 years.
From our numerical results, the average number of genetic common ancestors from the last 1000 years
shared by individuals living at least 2000km apart
is about 1/32 (and at least 1/80);
between 1000--2000ya they share about one;
and between 2000--3000ya they share above ten.
Since the chance is small that any genetic material has been transmitted along a particular
genealogical path from ancestor to descendent more than 8 generations deep
\citep{donnelly1983probability}
-- about .008 at 240ya, and $2.5\times10^{-7}$ at 480ya  -- % cbind(1:20,-expm1(-(2^(-2*(1:20))*(33*2*(1:20)+22))))
this implies, conservatively, 
thousands of shared genealogical ancestors in only the last 1000 years 
even between pairs of individuals separated by large geographic distances.
At first sight this result seems counterintuitive. 
However, 
as 1000 years is about 33 generations,
and $2^{33} \approx 10^{10}$ is far larger than the size of the European population,
so long as populations have mixed sufficiently,
by 1000 years ago everyone (who left descendants) 
would be an ancestor of every present day European.
% (See \citet{rohde2004modelling} for more rigorous discussion of this point.)
% Therefore, with even quite low migration rates,
% sharing of genealogical ancestry past this time should be ubiquitous across Europe. 
Our results are therefore one of the first genomic demonstrations of the counter-intuitive but
necessary fact that all Europeans are genealogically related over very short time periods,
and lends substantial support to models predicting 
close and ubiquitous common ancestry of all modern humans \citep{rohde2004modelling}.

The fact that most people alive today in Europe share nearly the same set of
(European, and possibly world-wide) ancestors from only 1000 years ago
seems to contradict the signals of long term, albeit subtle, population genetic structure within Europe
\citep[e.g.][]{novembre2008europe,lao2008correlation}. 
These two facts can be reconciled by the fact that even though the distribution of ancestors
(as cartooned in figure~\ref{fig:conceptual}B) 
has spread to cover the continent, 
there remain differences in degree of relatedness of modern individuals to these ancestral individuals.
For example, someone in Spain may be related to an ancestor in the Iberian peninsula 
through perhaps 1000 different routes back through the pedigree, 
but to an ancestor in the Baltic region by only 10 different routes, 
so that the probability that this Spanish individual inherited genetic material from the Iberian ancestor is roughly 100 times higher. 
This allows the amount of genetic material shared by pairs of extant individuals to vary
even if the set of ancestors is constant.

\paragraph{Relation to single-site summaries} 
Other work has studied fine-scale differentiation between populations within Europe
based on statistics such as $F_{ST}$, IBS \citep[e.g.][]{lao2008correlation,odushlaine2010population}, or PCA \citep{novembre2008europe},
that summarize in various ways single-marker correlations, averaged across loci.
Like rates of IBD, these measures of differentiation can be thought of as weighted averages of past coalescent rates 
\citep{malecot1970mathematics,slatkin1991inbreeding,rousset2002inbreeding,mcvean2009pca},
but take much of their information from much more distant times (tens thousands of generations).
As expected, we have seen both strong consistency between these measures and IBD (e.g.\ the decay with geographic distance),
as well as distinct patterns (e.g.\ higher sharing in eastern Europe).
These results highlight the fact that long segments of IBD contain information about much more recent events than do single-site summaries,
information that can be leveraged to learn about the timing of these events.

\paragraph{Limitations of Sampling}
A concern about our results is that the European individuals in the POPRES dataset
were all sampled in either Lausanne or London.
This might bias our results,
for instance, if an immigrant community originated
mostly from a particular small portion of their home population, 
thereby sharing a particularly high number of recent common ancestors with each other.
We see remarkably little evidence that this is the case:
there is a high degree of consistency in numbers of IBD blocks shared across samples from each population,
and between neighboring populations.
For instance, we could argue that the high degree of shared common ancestry among Albanian speakers
was because most of these sampled originated from a small area rather than uniformly across Albania and Kosovo.
However, this would not explain the high rate of IBD between Albanian speakers and neighboring populations.
Even populations from which we only have one or two samples,
which we at first assumed would be unusably noisy,
provide generally reliable, consistent patterns,
as evidenced by e.g.\ supplemental figure \ref{sfig:dotchart_long}.

Conversely, it might be a concern that individuals sampled in Lausanne or London
are more likely to have recent ancestors more widely dispersed than is typical for their population of origin.
This is a possibility we cannot discard, and if true,
would mean there is more structure within Europe than what we detect.
However, by the incredibly rapid spread of ancestry,
this is unlikely to have an effect over more than a few generations
and so does not pose a serious concern about our results about
the ubiquitous levels of common ancestry.
Fine-scale geographic sampling of Europe as a whole is needed to address these issues,
and these efforts are underway in a number of populations \citep[e.g.][]{price2009impact,jakkula2008genomewide,tylersmith2012british,winney2011people}.

Finally, we have necessarily have taken a narrow view of European
ancestry as we have restricted our sample to individuals who are not
outliers with respect to genetic ancestry, and
when possible to those having all four grandparents drawn from the same county. 
Clearly the ancestry of Europeans is far more diverse  
than those represented here, 
but such steps seemed necessary to make best initial use of this dataset.

\paragraph{Ages of particular common ancestors}
We have shown that the problem of inferring the average distribution of genetic common ancestors back through time
has a large degree of fundamental uncertainty.
The data effectively leave a large number of degrees of freedom unspecified,
so one must either describe the set of possible histories, as we do,
and/or use prior information to restrict these degrees of freedom.

A related but far more intractable problem is
to make a good guess of how long ago a {\em certain} shared genetic common ancestor lived,
as personal genome services would like to do, for instance:
if you and I share a 10cM block of genome IBD, 
when did our most recent common ancestor likely live?
Since the mean length of an IBD block inherited from 5 generations ago is 10cM,
we might expect the average age of the ancestor of a 10cM block to be from around 5 generations.
However, a direct calculation from our results says that the typical age of a 10cM block shared by two individuals from the UK is between 32 and 52 generations
(depending on the inferred distribution used). 
This discrepancy results from the fact that you are {\it a priori}
much more likely to share a common genetic ancestor further in the
past, and this acts to skew our answers away from the naive expectation
-- even though it is unlikely that a 10cM block is inherited from a particular shared ancestor from 40 generations ago,
there are a great number of such older shared ancestors. 
This also means that estimated ages must depend drastically on the populations' shared histories:
for instance, the age of such a block shared by someone from the UK with someone
from Italy is even older, usually from around 60 generations ago.
This may not apply to ancestors from the past very few (perhaps less than eight) generations,
from whom we expect to inherit multiple long blocks --
in this case we can hope
to infer a specific genealogical relationship with reasonable certainty
\citep[e.g.][]{huff2011maximumlikelihood,henn2012cryptic}, 
although even then care must be taken to exclude the
possibility that these multiple blocks have
not been inherited from distinct common ancestors.

Although the sharing of a long genomic segment can be an intriguing
sign of some recent shared ancestry,
the ubiquity of shared genealogical ancestry only tens of generations ago 
across Europe \citep[and likely the world,][]{rohde2004modelling}
makes such sharing unsurprising, and assignment to particular
genealogical relationships impossible.
What is informative about these chance sharing events from distant ancestors is that 
they provide a fine-scale view of an individual's distribution of ancestors
(e.g. figure \ref{fig:sharing_and_maps}),
and that in aggregate they can provide an unprecedented view 
into even small-scale human demographic history.

\paragraph{Where do your $n^\mathrm{th}$ cousins live?}
Our results also offer a way to understand the geographic location
of individuals of a given degree of relatedness. 
The values of figure~\ref{fig:inversion_boxplots} (and \ref{sfig:inversion_boxplots_long}) 
can be interpreted as the distribution of distant cousins for any reference population --
for instance, the set of bars for Poland (``PL'') in the top row shows
that a randomly chosen distant cousin of a Polish individual with the common ancestor
living in the past 500 years most likely lives in Poland  
but has reasonable chance of living in the Balkan peninsula or Germany.
Here ``randomly chosen'' means chosen randomly proportional to the paths through the pedigree --
concretely, take a random walk back through the pedigree to an ancestor in the appropriate time period,
and then take a random walk back down.
If one starts in Poland, then the chance of arriving in, say, Romania 
is proportional to the average number of (genetic) common ancestors shared by a pair from Poland and Germany,
which is exactly the number estimated in figure~\ref{fig:inversion_boxplots}.

\subsection*{The signal of history}
\label{ss:signalofhistory}

As we have shown, patterns of IBD provide ample but noisy geographic and temporal signals,
which can then be connected to historical events.
Rigorously making such connections is difficult, 
due to the complex recent history of Europe,
controversy about the demographic significance of many events,
and uncertainties in inferring the ages of common ancestors.
Nonetheless, our results can be plausibly connected to 
several historical and demographic events.

\paragraph{The migration period}
One of the striking patterns we see is the relatively high level of 
sharing of IBD between pairs of individuals across eastern Europe,
as high or higher than that observed within other, much smaller populations.
This is consistent with these individuals having a comparatively large proportion of ancestry 
drawn from a relatively small population 
that expanded over a large geographic area.
The ``smooth'' estimates of figure \ref{fig:inversion_distributions}
(and more generally figures \ref{fig:inversion_boxplots}, and \ref{sfig:all_inversions}), 
suggest that this increase in ancestry stems from around 1000--2000 years ago,
since during this time pairs of eastern individuals are expected to share a substantial number of common ancestors,
while this is only true of pairs of non-eastern individuals if they are from the same population.
For example, even individuals from widely separated eastern populations share about the same amount of IBD
as do two Irish individuals (see supplemental figure~\ref{sfig:dotchart_long}),
suggesting that this ancestral population may have been relatively small.

This evidence is consistent with the idea that these populations 
derive a substantial proportion of their ancestry from 
various groups that expanded during the ``migration period'' 
from the fourth through ninth centuries \citep{davies2010europe}.
This period begins with the Huns moving into eastern Europe towards the end of the fourth century,
establishing an empire including modern-day Hungary and Romania;
and continues in the fifth century as various Germanic groups moved into and ruled much of the western Roman empire.
This was followed by the expansion of the Slavic populations into regions of low population density beginning in the sixth century, 
reaching their maximum by the 10th century \citep{barford2001early}.
The eastern populations with high rates of IBD is highly coincident with the modern distribution of Slavic languages,
so it is natural to speculate that much of the higher rates were due to this expansion.
The inclusion of (non-Slavic speaking) Hungary and Romania in the group of eastern populations sharing high IBD
could indicate the effect of other groups (e.g.\ the Huns) on ancestry in these regions,
or because some of the same group of people who elsewhere are known as Slavs adopted different local cultures in those regions.
Greece and Albania are also part of this putative signal of expansion,
which could be because the Slavs settled in part of these areas (with unknown demographic effect),
or because of subsequent population exchange.
However, additional work and methods would be needed to verify this hypothesis.

The highest levels of IBD sharing are found in the Albanian-speaking individuals (from Albania and Kosovo),
an increase in common ancestry deriving from the last 1500 years.
This suggests that a reasonable proportion of the 
ancestors of modern-day Albanian speakers (at least those represented in POPRES)
are drawn from a relatively small, cohesive population
that has persisted for at least the last 1500 years.
These individuals share similar but slightly higher numbers of common ancestors with nearby populations
than do individuals in other parts of Europe (see figure~\ref{sfig:dotchart_long}),
implying that these Albanian speakers have not been a particularly isolated population so much as a small one.
Furthermore, our Greek and Macedonian samples
share much higher numbers of common ancestors with Albanian speakers than with other neighbors,
possibly a result of historical migrations, or else perhaps smaller effects of the Slavic expansion in these populations.
It is also interesting to note that the sampled Italians share nearly as much IBD with Albanian speakers as with each other.
The Albanian language is a Indo-European language without other close relatives \citep{hamp1966position}
that persisted through periods when neighboring languages were
strongly influenced by Latin or Greek, suggesting an intriguing link
between linguistic and genealogical history in this case.

\paragraph{Italy, Iberia, and France}
On the other hand,
we find that France and the Italian and Iberian peninsulas have the lowest rates of genetic common ancestry in the last 1500 years
(other than Turkey and Cyprus),
and are the regions of continental Europe thought to have been least affected by the Slavic and Hunnic migrations.
These regions were, however, moved into by Germanic tribes (e.g.\ the Goths, Ostrogoths, and Vandals),
which suggests that perhaps the Germanic migrations/invasions of these regions 
entailed a smaller degree of population replacement, 
than the Slavic and/or Hunnic,
or perhaps that the Germanic groups were less genealogically cohesive.
This is consistent with the argument that the Slavs moved into relatively depopulated areas,
while Gothic ``migrations'' may have been takeovers by small groups of extant populations
\citep{halsall2005barbarian,kobylinski2005slavs}.

In addition to the very few genetic common ancestors that Italians share both with each other and with other Europeans,
we have seen significant modern substructure within Italy (i.e.\ figure~\ref{fig:substructure}) that predates most of this common ancestry,
and estimate that most of the common ancestry shared between Italy and other populations
is older than about 2300 years (supplemental figure~\ref{sfig:tinyinversions}).
Also recall that most populations show no substructure with regards to the number of blocks shared with Italians,
implying that the common ancestors other populations share with Italy predate divisions within these other populations.
This suggests significant old substructure and large population sizes within Italy,
strong enough that different groups within Italy
share as little recent common ancestry as other distinct, modern-day countries,
substructure that was not homogenized during the migration period.
These patterns could also reflect in part geographic isolation within
Italy as well as a long history of settlement of Italy from diverse sources.

In contrast to Italy, the rate of sharing of IBD within the Iberian peninsula is similar to that within other populations in Europe.
There is furthermore much less evidence of substructure within our
Iberian samples than within the Italians, as shown in supplemental figure \ref{sfig:substructure_summaries}.
This suggests that the reduced rate of shared ancestry is due to geographic isolation (by distance and/or the Pyrenees)
rather than long-term stable substructure within the peninsula.

\paragraph{Future directions}

Our results show that patterns of recent identity by descent both
provide evidence of ubiquitous shared common ancestry and hold the 
potential to shed considerable light on the complex history of Europe.
However, these inferences also quickly run up against a fundamental limit 
to our ability to infer pairwise rates of recent common genetic ancestry.
In order to make a fuller model of European history, we will need to
make use of diverse sources of genomic information from large samples, 
including IBD segments and rare variants \citep{nelson2012abundance,tennessen2012evolution},
and develop methods that can more fully utilize this information across more than pairs of populations.
Another profound difficulty is that Europe -- 
and indeed any large continental region -- 
has such complex layers of history, 
through which ancestry has mixed so greatly,
that attempts to connect genetic signals in extant individuals 
to particular historical events 
requires the corroboration of other sources of information from many disciplines.
For example, the ability to isolate ancient autosomal DNA from individuals who lived during these time periods 
\citep[as do][]{skoglund2012origins,keller2012insights}
will help to overcome some of these these profound difficulties.
More generally, the quickly falling cost of sequencing, along with the development of new methods, 
will shed light on the recent demographic and genealogical history of populations of recombining organisms, 
human and otherwise.

\section{Materials and Methods}
\label{ss:methods}

\subsection*{Description of data and data cleaning}
\label{ss:thedata}

We used the two European subsets of the POPRES dataset --
the CoLaus subset, collected in Lausanne, Switzerland,
and the LOLIPOP subset, collected in London, England;
the dataset is described in \citet{nelson2008population}. 
Those collected in Lausanne reported parental and grandparental country of origin;
those collected in London did not.
We followed \citet{novembre2008europe} in 
assigning each sample to the common grandparental country of origin when available,
and discarding samples whose parents or grandparents were reported as originating in different countries.
We took further steps to restrict to individuals whose grandparents came from the same geographic region,
first performing principal components analysis on the data using SMARTPCA \citep{patterson2006population}, 
and excluding 41 individuals who clustered with populations outside Europe 
(the majority of such were already excluded by self-reported non-European grandparents). 
These individuals certainly represent an important part of the recent genetic ancestry of Europe,
but are excluded because we aim to study events stemming from older patterns of gene flow, and because we
do not model the whole-genome dependencies in recently admixed genomes.
We then used PLINK's inference of the fraction of single-marker IBD \citep[Z0, Z1, and Z2,][]{purcell2007plink} to identify very close relatives, 
finding 25 pairs that are first cousins or closer
(including duplicated samples), 
and excluded one individual from each pair.
We grouped samples into populations mostly by reported country,
but also used reported language in a few cases.
Because of the large Swiss samples, 
we split this group into three by language: French-speaking (CHf),
German-speaking (CHd),
or other (CH).
Many samples reported grandparents from Yugoslavia;
when possible we assigned these to a modern-day country by language,
and when this was ambiguous or missing we assigned these to ``Yugoslavia''.
Most samples from the United Kingdom reported this as their country of origin;
however, the few that reported ``England'' or ``Scotland'' were assigned this label.
This left us with 2257 individuals from 40 populations;
for sample sizes see table \ref{tab:ibd_summaries}.
Supplemental table \ref{stab:population_defns} further breaks this down,
and unambiguously gives the composition of each population.
Physical distances were converted to genetic distances using the hg36 map,
and the average human generation time was taken to be 30 years \citep{fenner2005crosscultural}.

All figures were produced in {\tt R} \cite{R},
with color palettes from packages {\tt colorspace} \cite{zeleis2009colorspace} and {\tt RColorBrewer} \cite{plate2011rsvgtipsdevice}.
Code implementing all methods described below is distributed along with IBD block data sufficient to reproduce the historical analyses
through \url{http://www.github.com/petrelharp/euroibd} and in the Dryad digital repository \cite{ourdryad}.

\subsection{Calling IBD blocks}

\begin{figure}[!htp]
  \begin{center}
    \wikiincludegraphics{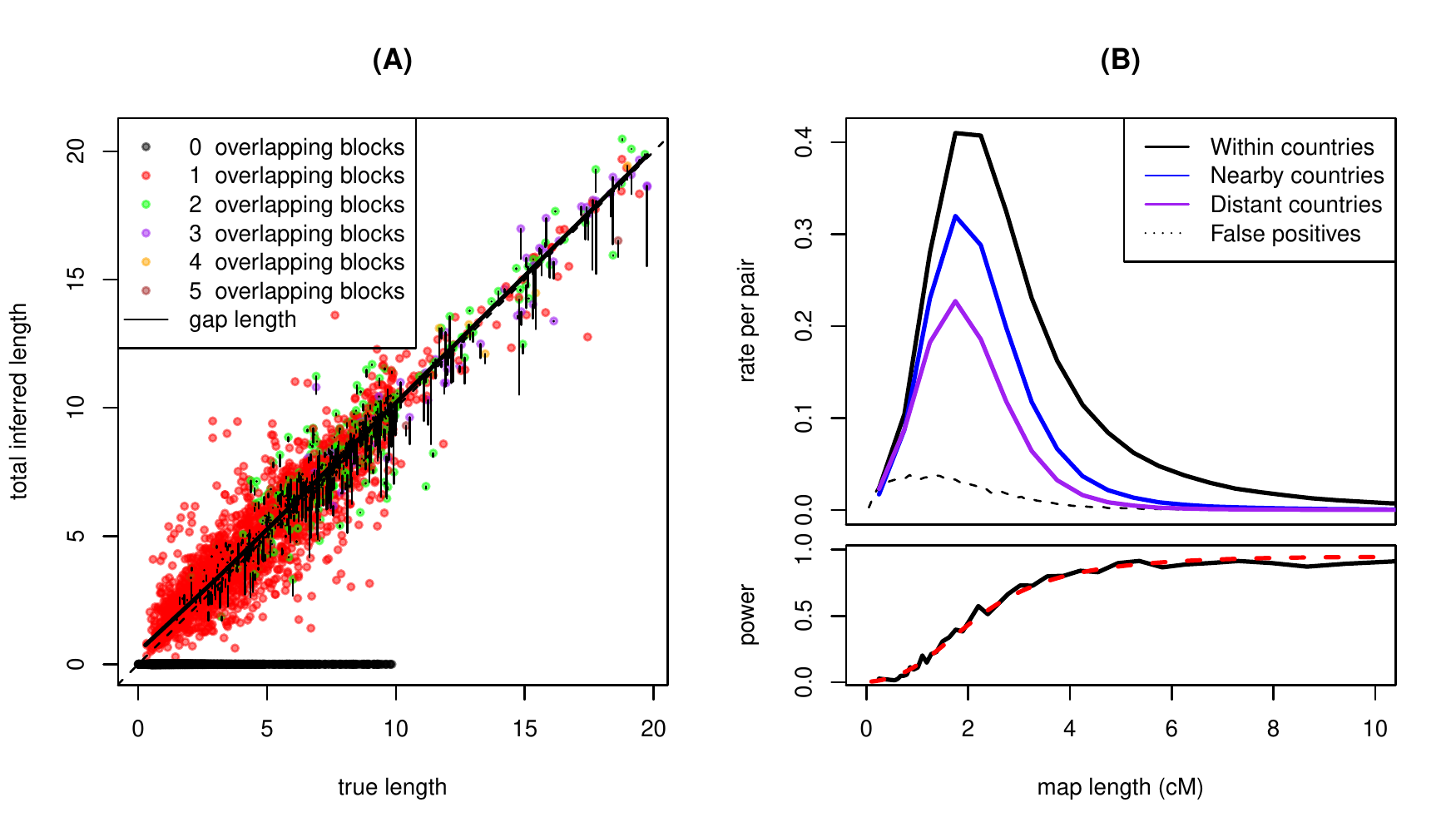}
    \caption{
    {\bf (A)} Bias in inferred length with lines $x=y$ (dotted) and a {\tt loess} fit (solid).
      Each point is a segment of true IBD (copied between individuals),
      showing its true length and inferred length after postprocessing.
      Color shows the number of distinct, nonoverlapping segments found by BEAGLE,
      and the length of the vertical line gives the total length of gaps between such segments
      that BEAGLE falsely inferred was not IBD (these gaps are corrected by our postprocessing).
    {\bf (B)}  Estimated false positive rate as a function of length.
Observed rates of IBD blocks, per pair and per cM, are also displayed for the purpose of comparison.
    ``Nearby'' and ``Distant'' means IBD between pairs of populations closer and farther away than 1000km, respectively.
    {\bf (C)} Below, the estimated power as a function of length (black line), together with the parametric fit $c(x)$ of equation \eqref{eqn:error_model} (red dotted curve).
      \label{fig:power_and_fp}
    }
  \end{center}
\end{figure}

To find blocks of IBD, we used {\tt fastIBD} \citep[implemented in BEAGLE,][]{browning2011powerful},
which records putative genomic segments shared IBD by pairs of individuals, along with a score indicating the strength of support.
As suggested by the authors, in all cases we ran the algorithm 10 times with different random seeds, 
and postprocessed the results to obtain IBD blocks.
Based on our power simulations described below, 
we modified the postprocessing procedure recommended by \citet{browning2011powerful} 
to deal with spurious gaps or breaks introduced into long blocks of IBD by low marker density or switch error, as follows:  
We called IBD segments by first removing any segments not overlapping a segment seen in at least one other run
(as suggested by \citet{browning2011powerful}, except with no score cutoff); 
then merging any two segments separated by a gap shorter than at least one of the segments and no more than 5cM long;
and finally discarding any merged segments that did not contain a subsegment with score below $10^{-9}$.
As shown in figure \ref{fig:power_and_fp}, 
this resulted in a false positive rate of between 8--15\% across length categories,
and a power of at least 70\% above 1cM, reaching 95\% by 4cM.
After post-processing, we were left with 1.9 million IBD blocks,
1 million of which were at least 2cM long (at which length we estimate 85\% power and a 10\% false positive rate).  

\subsection{Power and false positive simulations}
\label{ss:error_model}

All methods to identify haplotypic IBD rely on identifying long regions of near identical haplotypes 
between pairs of individuals (referred to as identical by state, IBS).  
However, long IBS haplotypes could potentially also result from the concatenation
of multiple shorter blocks of true IBD. 
While such runs can contain important information about deeper population history
\citep[e.g.][]{li2011inference, harris2012inferring}, 
we view them as a false positives as they do not represent single haplotypes shared without intervening recombination. 
The chance of such a false positive IBD segment decreases as the genetic length of shared haplotype increases. 
However, the density of informative markers also plays a role,
because such markers are necessary to infer regions of IBS.

\paragraph{Power}
If we are to have a reasonable false positive rate, we must accept
imperfect power. Power will also vary with the density and
informativeness of markers and length of segment considered. 
For example, it is intuitive that segments of genome containing many rare alleles are easier identify as IBD.
Conversely, rare immigrant segments from a population with different
allele frequencies may, if they are shared by multiple individuals
within the population, cause higher false positive rates.
For these reasons, when estimating statistical power and false positive rate,
it is important to use a dataset as similar to the one under consideration as possible.
Therefore, to determine appropriate postprocessing criteria and to estimate our statistical power,
we constructed a dataset similar to the POPRES with known shared IBD segments as follows:
we copied haploid segments randomly between 60 trio-phased individuals of European descent (using only one from each trio) 
from the HapMap dataset \citep[haplotypes from release \#21, 17/07/06][]{hapmap2007second},
reoriented alleles to match the strand orientation of POPRES,
substituted these for 60 individuals from Switzerland in the POPRES data,
and ran BEAGLE on the result as before.
These segments were copied between single chromosomes of randomly chosen individuals, for random lengths 0.5--20cM,
with gaps of at least 2cM between adjacent segments and without copying between the same two individuals twice in a row.
When copying, we furthermore introduced genotyping error by flipping alleles independently with probability .002
and marking the allele missing with probability .023 (error rates were
determined from duplicated individuals in the sample as given by \citet{nelson2008population}).
An important feature of the inferred data was that 
BEAGLE often reported true segments longer than about 5cM
as two or more shorter segments separated by a short gap, 
which led us to merge blocks as described above.

\paragraph{Length bias}
We also need a reasonably accurate assessment of our bias and false positive rates for our inference of numbers of genetic ancestors from the IBD length spectrum.
Although the estimated IBD lengths were approximately unbiased,
we fit a parametric model to the relationship between true and inferred lengths
after removing inferred blocks less than 1cM long.
A true IBD block of length $x$ is missed entirely with probability $1-c(x)$, and is otherwise inferred to have length $x+\epsilon$; 
with probability $\gamma(x)$ the error $\epsilon$ is positive; otherwise it is negative and conditioned to be less than $x$.
In either case, $\epsilon$ is exponentially distributed; if $\epsilon>0$ its mean is $1/\lambda_+(x)$, 
while if $\epsilon<0$ its (unconditional) mean is $1/\lambda_-(x)$.
The parametric forms were chosen by examination of the data;
these are, with final parameter values:
\begin{align} \label{eqn:error_model}
  \begin{split}
  c(x) &=  1 - 1/\left(1 + .077 x^2 \exp(.54 x)\right) \\
  \gamma(x) &= .34 \left( 1 - (1+ .51 (x-1)^+ \exp(.68 (x-1)^+) )^{-1} \right) \\
  \lambda_+(x) &=  1.40 \\
  \lambda_-(x) &=    \min( .40 + 1/(.18 x) , 12 ) 
\end{split}
\end{align}
where $z^+ = \max(z,0)$.
The parameters were found by maximum likelihood,
using constrained optimization as implemented in the {\tt R} package {\tt optim} \citep{R}
separately on three independent pieces:
the parameters in $c(x)$ and $\gamma(x)$;
the parameters in $\lambda_-$; 
and finally the parameters in $\lambda_+$;
the fit is shown in supplemental figure~\ref{sfig:fit_error_model}.

\paragraph{False positive rate}
To estimate the false positive rate, 
we randomly shuffled segments of diploid genome between individuals from the same population
(only those 12 populations with at least 19 samples)
so that any run of IBD longer than about 0.5cM would be broken up among many individuals.
Specifically, as we read along the genome we output diploid genotypes in random order;
we shuffled this order by exchanging the identity of each output individual with another
at independent increments chosen uniformly between 0.1 and 0.2cM.
This ensured that no output individual had a continuous run of length longer than 0.2cM copied from a single input individual,
while also preserving linkage on scales shorter than 0.1cM.
The results are shown in figure \ref{fig:power_and_fp}B;
from these we estimate that the mean density of false positives $x$ cM long per pair and per cM
is approximately 
\begin{equation} \label{eqn:fp}
  f(x) = \exp(-13-2x +4.3 \sqrt{x}), 
\end{equation}
a parametric form again chosen by examination of the data and fit by maximum likelihood.

We found that overall, the false positive rate was around $1/10^\mathrm{th}$ of the observed rate,
except for very long blocks (longer than 5cM or so, where it was close to zero),
and for very short blocks (less than 1cM, where it approached 0.4).
As {\tt fastIBD} depends on estimating underlying haplotype frequencies,
it is expected to have a higher false positive rate in populations that are more differentiated from the rest of the sample.
There was significant variation in false positive rate between different populations,
with Spain, Portugal, and Italy showing significantly higher false positive rates than the other populations we examined
-- see supplemental figure \ref{sfig:fp_rate_by_country} -- 
however, the variation was significant only for blocks shorter than 2cM
across all population pairs, with the exception of pairs of Portuguese individuals,
where the upwards bias may be significant as high as 4cM.

\paragraph{Differential sample sizes} 
Finally, one concern is that as 
{\tt fastIBD} calls IBD based on a model of haplotype frequencies in
the sample it may be unduly affected by the large-scale sample size
variation across the POPRES sample. In particular, the French-speaking Swiss
sample is very large, which could lead to systematic bias in
calling IBD in populations closely to the Swiss samples.  
%is also expected to have lower power to detect IBD segments in populations that are more differentiated from the rest of the sample.
%  Removed:
%      Austria        Belgium        England         France        Germany         Greece        Ireland          Italy         Kosovo    Netherlands         Poland 
%            1              2              2              6              2              1              2              8              1              1              2 
%     Portugal         Russia         Serbia          Spain   Swiss French   Swiss German         Turkey United Kingdom     Yugoslavia 
%            4              1              1              6            745              4              1             21              1 
To investigate this, we randomly discarded 745 French-speaking Swiss (all but 100 of these),
and a random sampling of the remaining populations (removing 812 in total, leaving 1445).
We then ran {\tt BEAGLE} on chromosome 1 of these individuals,
% discarding segments seen in only one of the ten {\tt BEAGLE} runs, and merging gaps as above
% (but without the final step of restricting to segments with score below $10^{-9}$).
postprocessing in the same way as for the full sample.
Reassuringly, there was high concordance between the two --
we found that 98\% (95\%) of the blocks longer than 2cM found in the analysis with the full dataset (respectively, with the subset) 
were found in both analyses.
Overall, more blocks were found by the analysis with the smaller dataset;
however, by adjusting the score cutoff by a fixed amount this difference could be removed,
leaving nearly identical length distributions between the two analyses.
This is a known attribute of the {\tt fastIBD} algorithm, and can alternatively be avoided by adjusting the model complexity
\citetext{S.~Browning, personal communication}.

We then tested the extent to which the effect of sample size varied by population,
for IBD blocks in several length categories (binning block lengths at 1, 2, 4, and 10cM).
Suppose that $F_{xy}$ is the number of IBD blocks found between populations $x$ and $y$ in the analysis of the full dataset,
and $S_{xy}$ is the number found in the analysis of the smaller dataset
(counted between the same individuals each time).
We then assume that $F_{xy}$ and $S_{xy}$ are Poisson with mean $\lambda^F_{xy}$ and $\lambda^S_{xy}$, respectively,
so that conditioned on $N_{xy} = F_{xy} + S_{xy}$ (the total number of blocks), 
$S_{xy}$ is binomial with parameters $N_{xy}$ and $p_{xy} = \lambda^S_{xy}/(\lambda^S_{xy}+\lambda^F_{xy})$.
We are looking for deviations from the null model under which the effect of smaller sample size affects all population pairs equally,
so that $\lambda^S_{xy} = C \; \lambda^F_{xy}$ for some constant $C$.
We therefore fit a binomial GLM \citep{mccullagh1989generalized} with a logit link,
with terms corresponding to each population -- in other words,
\begin{align*}
  p_{xy} = \left( 1 + \exp\left( - \alpha_0 - \alpha_x - \alpha_y \right) \right)^{-1} .
\end{align*}
We found statistically significant variation by population (i.e.\ several nonzero $\alpha_x$),
but all effect sizes were in the range of 0--4\%;
estimated parameters are listed in supplemental table \ref{stab:subset_glm_results}.
Notably, the coefficient corresponding to the French-speaking Swiss (the population with the largest change in sample size)
was fairly small.
We also fit the model not assuming additivity when $x=y$, i.e.\ adding coefficients $\alpha_{xx}$ to the formula for $p_{xx}$,
but these were not significant. These results suggest that sample size
variation across the POPRES data has only minor effects on the
distribution of IBD blocks shared across populations.

\subsection{IBD rates along the genome}
\label{ss:block_density}

To look for regions of unusual levels of IBD and to examine our assumption of uniformity,
we compared the density of IBD tracts of different lengths along the genome,

in supplemental figure~\ref{sfig:overlap_all}.
To do this, we first divided blocks up into nonoverlapping bins based on length,
with cutpoints at 1, 2.5, 4, 6, 8, and 10cM.
We then computed at each SNP the number of IBD blocks in each length bin that covered that site.
To control for the effect of nearby SNP density on the ability to detect IBD,
we then computed the residuals of a linear regression predicting number of overlapping IBD blocks
using the density of SNPs within 3cM.
To compare between bins, we then normalized these residuals,
subtracting the mean and dividing by the standard deviation;
these ``z-scores'' for each SNP are shown in figure \ref{sfig:overlap_all}.

\subsection{Correlations in IBD rates across populations}
\label{ss:grouping_methods}

We noted repeated patterns of IBD sharing across multiple populations (seen in supplemental figure~\ref{sfig:dotchart_long}),
in which certain sets of populations tended to show similar patterns of sharing.
To quantify this, we computed correlations between mean numbers of IBD blocks;
in supplemental figure \ref{sfig:correlations} we show correlations in numbers blocks of various lengths.
Specifically, if $I(x,y)$ is the mean number of IBD blocks of the given length 
shared by an individual from population $x$ with a (different) individual from population $y$,
there are $n$ populations,
and $\bar I(x) = (1/(n-1)) \sum_{y\neq x} I(x,y)$,
then figure \ref{sfig:correlations} shows for each $x$ and $y$,
\begin{equation}
  \frac{1}{n-2} \sum_{z \notin\{x,y\}} (I(x,z) -\bar I(x)) ( I(y,z) -\bar I(y) ) ,
\end{equation}
the (Pearson) correlation between $I(x,z)$ and $I(y,z)$ ranging across $z \notin \{x,y\}$.
Other choices of block lengths are similar, although shorter blocks show higher overall correlations
(due in part to false positives)
and longer blocks show lower overall correlations 
(as rates are noisier, and sharing is more restricted to nearby populations).
The geographic groupings of table \ref{tab:ibd_summaries} were then chosen by visual inspection.

\subsection{Substructure}

We assessed the overall degree of substructure within each population, 
by measuring, for each $x$ and $y$, the degree of inhomogeneity across individuals of population $x$ for shared ancestry with population $y$.
We measured inhomogeneity by the standard deviation in number of blocks shared with population $y$,
across individuals of population $x$.
We assessed the significance by a permutation test,
randomly reassigning each block shared between $x$ and $y$ to a individual chosen uniformly from population $x$,
and recomputing the standard deviation, 1000 times.
(If there are $k$ blocks shared between $x$ and $y$ and $m$ individuals in population $x$,
this is equivalent to putting $k$ balls in $m$ boxes, tallying how many balls are in each box,
and computing the sample standard deviation of the resulting list of numbers.)  
Note that some degree of inhomogeneity of shared ancestry is expected
even within randomly mating populations, due to randomness of the relationship between individuals in the pedigree.
These effects are likely to be small if the relationships are suitably deep, but
this is still an area of active research \citep{henn2012cryptic,carmi2012variance}. 
The resulting $p$-values are shown in supplemental figure~\ref{sfig:substructure_summaries}.
We did not analyze these in detail, particularly as we had limited power to detect substructure in populations with few samples,
but note that a large proportion (47\%) of the population pairs showed greater inhomogeneity than in all 1000 permuted samples (i.e.\ $p<.001$).
Some comparisons even with many samples in both populations (where we have considerable power to detect even subtle inhomogeneity)
showed no structure whatsoever --
in particular, the distribution of numbers of Italian IBD blocks shared by Swiss individuals is not distinguishable from Poisson,
indicating a high degree of homogeneity of Italian ancestry across Switzerland.

\subsection{Single-site summaries}
\label{ss:singlesite}
	
To assess the single marker measures of relatedness across the POPRES sample 
we calculated pairwise identity by state, 
the probability that two alleles sampled at random from a pair of individuals are identical,
averaged across SNPs. 
This was calculated for all pairs of individuals using the ``{\tt --genome}'' option in PLINK v1.07
\citep[][\url{http://pngu.mgh.harvard.edu/purcell/plink}]{purcell2007plink},
and is shown in Supplementary Figure \ref{sfig:ibs_by_dist} with points colored as in Figure \ref{fig:sharing_and_maps}.

We also calculated principal components of the POPRES genotype data
using the EIGENSOFT package v3.0 \citep{patterson2006population},
which were used in identifying outlying individuals
and in producing figures \ref{sfig:pca_map}, \ref{sfig:pca_ibd_it}, and \ref{sfig:pca_ibd_uk}.

\subsection{Inferring ages of common ancestors}
\label{ss:inversion_methods}

Here, our aim is to use the distribution of IBD block lengths
to infer how long ago the genetic common ancestors were alive from which these IBD blocks were inherited.
A pair of individuals who share a block of IBD of genetic length at least $x$
have each inherited contiguous regions of genome from a single common ancestor $n$ generations ago
that overlap for length at least $x$.
If we start with the population pedigree,
those ancestors from which the two individuals might have inherited IBD blocks
are those that can be connected to both by paths through the pedigree.
The distribution of possible IBD blocks is determined by the number of links
(i.e.\ the number of meioses) occurring along the two paths.

Throughout the article we informally often refer to ancestors living a certain ``number of generations in the past''
as if humans were semelparous with a fixed lifetime.
Keeping with this, it is natural to write the number of IBD blocks shared by a pair of individuals
as the sum over past generations of the number of IBD blocks inherited from that generation.
In other words, if $N(x)$ is the number of IBD blocks of genetic length at least $x$ shared by two individual chromosomes,
and $N_n(x)$ is the number of such IBD blocks inherited by the two along paths through the pedigree
having a total of $n$ meioses, then $N(x) = \sum_n N_n(x)$.
Therefore, averaging over possible choices of pairs of individuals,
the mean number of shared IBD blocks can be similarly partitioned as
\begin{equation} 
  \E[ N(x) ] = \sum_{n \ge 1} \E[ N_n(x) ] .
\end{equation}
In each successive generation in the past each chromosome is broken up into successively more pieces,
each of which has been inherited along a different path through the pedigree,
and any two such pieces of the two individual chromosomes
that overlap and are inherited from the same ancestral chromosome
contribute one block of IBD.
Therefore, the mean number of IBD blocks coming from $n/2$ generations ago
is the mean number of possible blocks
multiplied by the probability that a particular block
is actually inherited by both individuals from the same genealogical ancestor in generation $n/2$.
Allowing for overlapping generations,
the first part we denote by $K(n,x)$, 
the mean number of pieces of length at least $x$ obtained by cutting the chromosome at the recombination sites of $n$ meioses,
and the second part we denote by $\mu(n)$, 
the probability that the two chromosomes have inherited at a particular site along a path of total length $n$ meioses
(e.g.\ their common ancestor at that site lived $n/2$ generations ago).
Multiplying these and summing over possible paths, we have that
\begin{equation} \label{eqn:linear_relationship}
  \E[ N(x) ] = \sum_{n \ge 1} \mu(n) K(n,x) ,
\end{equation}
i.e.\ the mean rate of IBD is a linear function of the distribution of the time back to the most recent common ancestor averaged across sites.
The distribution $\mu(n)$ is more precisely known as the coalescent time distribution \citep{kingman1982coalescent,wakeley2005coalescent},
in its obvious adaptation to population pedigrees.
As a first application, note that the distribution of ages
of IBD blocks above a given length $x$ depends strongly on the demographic history -- 
a fraction $\mu(n) K(n,x)/\sum_m \mu(m) K(m,x)$ of these are from paths $n$ meioses long.

Furthermore, it is easy to calculate that for a chromosome of genetic length $G$, 
\begin{equation} \label{eqn:kernel_defn}
   K(n,x) = \left( n(G-x) + 1\right)\exp(-xn) ,
\end{equation}
assuming homogeneous Poisson recombination on the genetic map 
(as well as constancy of the map and ignoring the effect of interference, which is a reasonable for the range of $n$ we consider).
The mean number of IBD blocks of length at least $x$ shared by a pair of individuals
across the entire genome is then obtained by summing \eqref{eqn:linear_relationship} across all chromosomes,
and multiplying by four (for the four possible chromosome pairs).

Equations \eqref{eqn:linear_relationship} and \eqref{eqn:kernel_defn} give the relationship between lengths of shared IBD blocks 
and how long ago the ancestor lived from whom these blocks are inherited. 
Our goal is to invert this relationship to learn about $\mu(n)$,
and hence the ages of the common ancestors underlying our observed distribution of IBD block lengths.
To do this, 
we first need to account for sampling noise and estimation error.
Suppose we are looking at IBD blocks shared between any of a set of $n_p$ pairs of individuals,
and assume that $N(y)$, the number of {\em observed} IBD blocks shared between any of those pairs of length at least $y$, 
is Poisson distributed with mean $n_p M(y)$,
where
\begin{align} 
  M(y) &= \int_y^G f(z) + \sum_{n\ge1} \mu(n)  \left( \int_0^G c(x)R(x,z) dK(n,x) \right) dz , \quad \mbox{with}\label{eqn:error_relationship} \\
  R(x,y) &= \begin{cases} \gamma(x) \lambda_+(x) \exp(-\lambda_+(x) (y-x)) \quad &\mbox{for} \; y>x \\
    (1-\gamma(x)) \lambda_-\exp(-\lambda_-(x) (x-y)) / (1-\exp(-\lambda_-(x) x)) \quad &\mbox{for} \; y<x .  \end{cases} 
\end{align}
Here the false positive rate $f(z)$, power $c(x)$ and the components of the error kernel $R(x,y)$ are estimated as above,
with parametric forms given in equations \eqref{eqn:fp} and \eqref{eqn:error_model}.
The Poisson assumption has been examined elsewhere \citep[e.g.][]{fisher1954fuller,huff2011maximumlikelihood},
and is reasonable
because there is a very small chance of having inherited a block from each pair of shared genealogical ancestors;
there a great number of these, and if these events are sufficiently independent,
the Poisson distribution will be a good approximation \citep[see e.g.][]{grimmett2001probability}.
If this holds for each pair of individuals,
the total number of IBD blocks is also Poisson distributed,
with $M$ given by the mean of this number across all constituent pairs.
(Note that this does {\em not} assume that each pair of individuals
has the same mean number, so does not assume that our set of pairs
are a homogeneous population.)

We have therefore a likelihood model for the data,
with demographic history (parametrized by $\mu = \{\mu(n): n\ge1\}$) as free parameters.
Unfortunately, the problem of inferring $\mu$ is ill-conditioned \citep[unsuprising due to its similarity of the kernel \eqref{eqn:kernel_defn} to the Laplace transform, see][]{epstein2008badtruth},
which in this context means that the likelihood surface is flat in certain directions (``ridged''):
for each IBD block distribution $N(x)$
there is a large set of coalescent time distributions $\mu(n)$ that fit the data equally well.
A common problem in such problems is that the unconstrained maximum likelihood solution
is wildly oscillatory;
in our case, the unconstrained solution is not so obviously wrong,
since we are helped considerably by the knowledge that $\mu\ge0$.
For reviews of approaches to such ill-conditioned inverse problems, 
see e.g.\ \citet{petrov2005well} or \citet{stuart2010inverse};
the problem is also known as ``data unfolding'' in particle physics \citep{cowan1998statistical}.
If one is concerned with finding a point estimate of $\mu$,
most approaches add an additional penalty to the likelihood,
which is known as ``regularization'' \citep{tikhonov1977solutions} 
or ``ridge regression'' \citep{hoerl1970ridgeregression}.

However, our goal is parametric inference,
and so we must
describe the limits of the ``ridge'' in the likelihood surface in various directions,
(which can be seen as maximum {\it a posteriori} estimates under priors of various strengths).

To do this, we first discretize the data, so that $N_i$ is the number of IBD blocks 
shared by any of a total of $n_p$ distinct pairs of individuals
with inferred genetic lengths falling between $x_{i-1}$ and $x_i$.
We restrict to blocks having a minimum length of 2cM long, so that $x_0=2$.
To find a discretization so that each $N_i$ has roughly equal variance,
we choose $x_i$ by first dividing the range of block lengths into 100 bins with equal numbers of blocks falling in each;
discard any bins longer than 1cM; and divide the remainder of the range up into 1cM chunks. 
To further reduce computational time,
we also discretize time, effectively requiring $\mu_n$ to be constant on each interval $n_j \le n < n_{j+1}$,
with $n_{j+1}-n_j = \lfloor{j/10}\rfloor$, for $1 \le j \le 360$ -- so the resolution is finest for recent times,
and the maximum time depth considered is 6660 meioses, or 99900 years ago. 
(The discretization and upper bound on time depth were found to not affect our results.)
We then compute by numerical integration (using the function {\tt integrate} in {\tt R})
the matrix $L$ discretizing the kernel given in \eqref{eqn:error_relationship},
so that $L_{in} = \int_{x_{i-1}}^{x_i} \int_0^G c(x) R(x,z) dK(n,x) dz$
is the kernel that applied to $\mu$ gives the mean number of true IBD blocks per pair observed with lengths between $x_{i-1}$ and $x_i$,
and $f_i= \int_{x_{i-1}}^{x_i} f(z) dz$ is the mean number of false positives per pair with lengths in the same interval.
We then sum across chromosomes, as before.
The likelihood of the data is thus
\begin{equation}
    \exp\left( - n_p \sum_{i,n} L_{in} \mu_n + f_i \right) \prod_i \frac{ 1 }{ N_i ! } \left( n_p \sum_n L_{in} \mu_n + f_i \right)^{N_i} .
\end{equation}
To the (negative) log likelihood we add a penalization $\gamma$, 
after rescaling by the number of pairs $n_p$ (which does not affect the result but makes penalization strengths comparable between pairs of populations), 
and use numerical optimization \citep[the {\tt L-BFGS-B} method in {\tt optim},][]{R} to minimize the resulting functional
(which omits terms independent of $\mu$)
\begin{equation}
    \calL(\mu;\gamma,N) = \sum_{i,n} L_{in} \mu_n - \sum_i \frac{N_i}{n_p} \log\left( \sum_n L_{in} \mu_n + f_i \right) + \frac{\gamma(\mu)}{n_p} .
\end{equation}
Often we will fix the functional form of the penalization and vary its strength,
so that $\gamma(\mu) = \gamma_0 z(\mu)$,
in which case we will write $\calL(\mu;\gamma_0,N)$ for $\calL(\mu;\gamma_0\,z(\mu),N)$.

For instance, the leftmost panels in figure \ref{fig:inversion_distributions}
show the minimizing solutions $\mu$ for $\gamma(\mu)=0$ (no penalization)
and for $\gamma(\mu) = \gamma_0 \sum_n (\mu_{n+1}-\mu_n)^2$ (``roughness'' penalization).
Because our aim is to describe extremal reasonable estimates $\mu$,
in this and in other cases, we have chosen the strength of penalization $\gamma_0$
to be ``as large as is reasonable'', 
choosing the largest $\gamma_0$ such that the minimizing $\mu$ has log likelihood
differing by no more than 2 units from the unconstrained optimum.
This choice of cutoff can be justified as in \citet{edwards1984likelihood},
gave quite similar answers to other methods, and performed well
on simulated population histories (see supplemental document).

This can be thought of as taking the strongest prior that still gives us ``reasonable'' maximum {\it a posteriori} answers.
Note that the optimization is over {\em nonnegative} distributions $\mu$ 
also satisfying $\sum_n \mu(n) \le 1$ 
(although the latter condition does not enter in practice).

We would also like to determine bounds on total numbers of shared genetic ancestors who lived during particular time intervals,
by determining e.g.\ the minimum and maximum numbers of such ancestors that are consistent with the data.
Such bounds are shown in figure \ref{fig:inversion_boxplots}.
To obtain a lower bound for the time period between $n_1$ and $n_2$ generations,
we penalized the total amount of shared ancestry during this interval,
using the penalizations $\gamma_-(\mu) = \gamma_0^- \left( \sum_{n=n_1}^{n_2} \mu(n) \right)^2$,
and choosing $\gamma_0^-$ to give a drop of 2 log likelihood units, as described above.
The lower bound is then the total amount of coalescence $\sum_{n =n_1}^{n_2} \mu_-(n)$ for $\mu_-$ minimizing $\calL(\cdot;\gamma_-,N)$.
The upper bound is found by penalizing total shared ancestry {\em outside} this interval,
i.e.\ by applying the penalization $\gamma_+(\mu) = \gamma_0^+ \left( \sum_{n<n_1} \mu(n) + \sum_{n>n_2} \mu(n) \right)^2$.
It is almost always the case that lower bounds are zero,
since there is sufficient wiggle room in the likelihood surface to explain the observed block length distribution
using peaks just below $n_1$ and above $n_2$.
Examples are shown in supplemental figure~\ref{sfig:example_inversion_bounds}.
On the other hand, upper bounds seem fairly reliable.

In the above we have assumed that the minimizer of $\calL$ is unique,
thus glossing over e.g.\ finding appropriate starting points for the optimization.
In practice, we obtained good starting points by solving the natural approximating least-squares problem,
using {\tt quadprog} \citep{quadprog} in {\tt R}.
We then evaluated uniqueness of the minimizer by using different starting points,
and found that if necessary, adding only a very small penalization term was enough to ensure convergence to a unique solution.

\paragraph{Testing the method} 
To test this method, we implemented a whole-genome coalescent IBD simulator,
and applied our inference methods to the results under various demography scenarios.
We also used these simulations to evaluate the sensitivity of our method to misestimation of power or false positive rates.
The simulations, and the results, are described in Supplemental Document 1;
in all cases, the simulations showed that the method performed
well to the level of uncertainty discussed throughout the text and confirmed our understanding of the method described above.
We also found that misestimation of false positive rate 
only affects estimated numbers of common ancestors by a comparable amount,
and that misestimation of blocks less than 4cM long mostly affects estimates older than about 2000 years.
Therefore, if our false positive rates above 2cM are off by 10\% (the range that seems reasonable), 
which would change our estimated numbers of blocks by about 1\%,
this would only change our estimated numbers of shared ancestors by a few percent.

\paragraph{Extending to shorter blocks}
We only used blocks longer than 2cM to infer ages of common ancestors,
in part because the model we use does not seem to fit the data below this threshold.
Attempts to apply the methods to all blocks longer than 1cM 
reveals that there is no history of rates of common ancestry that, under this model,
produces a block length distribution reasonably close to the one observed --
small, but significant deviations occur below about 2cM.
This occurs probably in part because our estimate of false positive rate is expected to be less accurate
at these short lengths.
Furthermore, our model does not explicitly model the overlap of multiple short IBD segments 
to create on long segment deriving from different ancestors,
which could start to have a significant effect at short lengths.
(The effect on long blocks we model as error in length estimation.)
This could be incorporated into a model \citep[in a way analogous to][]{li2011inference},
but consideration of when several contiguous blocks of IBD might have few enough differences
to be detected as a long IBD block
quickly runs into the need for a model of IBD detection,
which we here treat as a black box.
Use of these shorter blocks, which would allow inference of older ancestry,
will need different methods, and probably sequencing rather than
genotyping data.

\subsection{Numbers of common ancestors}
\label{ss:numbers_of_ancestors}

Estimated numbers of genetic common ancestors can be found by simply solving for $N(0)$ using an estimate of $\mu(n)$ in 
equations \eqref{eqn:linear_relationship} and \eqref{eqn:kernel_defn}
(still restricting to genetic ancestors on the autosomes).
These tell us that given the distribution $\mu(n)$,
the mean number of genetic common ancestors coming from generation $n/2$ --
i.e.\ the mean number of IBD blocks of {\em any} length inherited from such common ancestors --
is $N(0) = \mu(n) \sum_{k=1}^{22} \left( n G_k + 1 \right)$, where
$G_k$ is the total sex-averaged genetic length of the $k^\mathrm{th}$ human chromosome.
Since the total sex-averaged map length of the human autosomes is about 32 Morgans, this is about $\mu(n) ( 32 n + 22 )$.
This procedure has been used in figures \ref{fig:inversion_distributions} and \ref{fig:inversion_boxplots}.

Converting shared IBD blocks to numbers of shared {\em genealogical} common ancestors is more problematic.
Suppose that modern-day individuals $a$ and $b$ both have $c$ as a grand$^{n-1}$parent.
Using equation \eqref{eqn:kernel_defn} at $x=0$, 
we know that the mean number of blocks that $a$ and $b$ both inherit from $c$
is $r(2n)$, with $r(n):=2^{-n}(32n+22)$, since each block has chance $2^{-2n}$ of being inherited across $2n$ meioses.
First treat the endpoints of each distinct path of length $n$ back through the pedigree as a grand$^{n-1}$parent,
so that everyone has exactly $2^n$ grand$^{n-1}$parents,
and some ancestors will be grand$^{n-1}$parents many times over.
Then if $a$ and $b$ share $m$ genetic grand$^{n-1}$parents,
a moment estimator for the number of genealogical grand$^{n-1}$parents is $m/r(n)$.
However, the geometric growth of $r(n)$ means that small uncertainties in $n$ 
have large effects on the estimated numbers of genealogical common ancestors
-- and we have large uncertainties in $n$.

Despite these difficulties, we can still get some order-of-magnitude estimates.
For instance, we estimate that someone from Hungary shares on average about 5 genetic common ancestors with someone from the UK
between 18 and 50 generations ago.
Since $1/r(36)=5.8\times10^7$, we would conservatively estimate that for every genetic common ancestor
there are tens of millions of genealogical common ancestors.
Most of these ancestors must be genealogical common ancestors many times over, 
but these must still represent at least thousands of distinct individuals.

\section*{Acknowledgements}
Thanks to Razib Khan, Sharon Browning, and Don Conrad for several useful discussions, 
and to Jeremy Berg, Ewan Birney, Yaniv Brandvain, Joe Pickrell, Jonathan Pritchard, Alisa Sedghifar, and Joel Smith for useful comments on earlier drafts.  
We also thank the four anonymous reviewers, as well as Amy Williams 
(at \href{http://haldanessieve.org/2012/10/05/our-paper-the-geography-of-recent-genetic-ancestry-across-europe/comment-page-1/}{Haldane's sieve}),
for their helpful suggestions.

\bibliographystyle{abbrvnat}

\clearpage

\section*{Supplemental material}

\renewcommand{\thefigure}{S\arabic{figure}}
\setcounter{figure}{0}
\renewcommand{\thetable}{S\arabic{table}}
\setcounter{table}{0}

\begin{figure}[!htp]
  \begin{center}
    \includegraphics[height=.9\textheight]{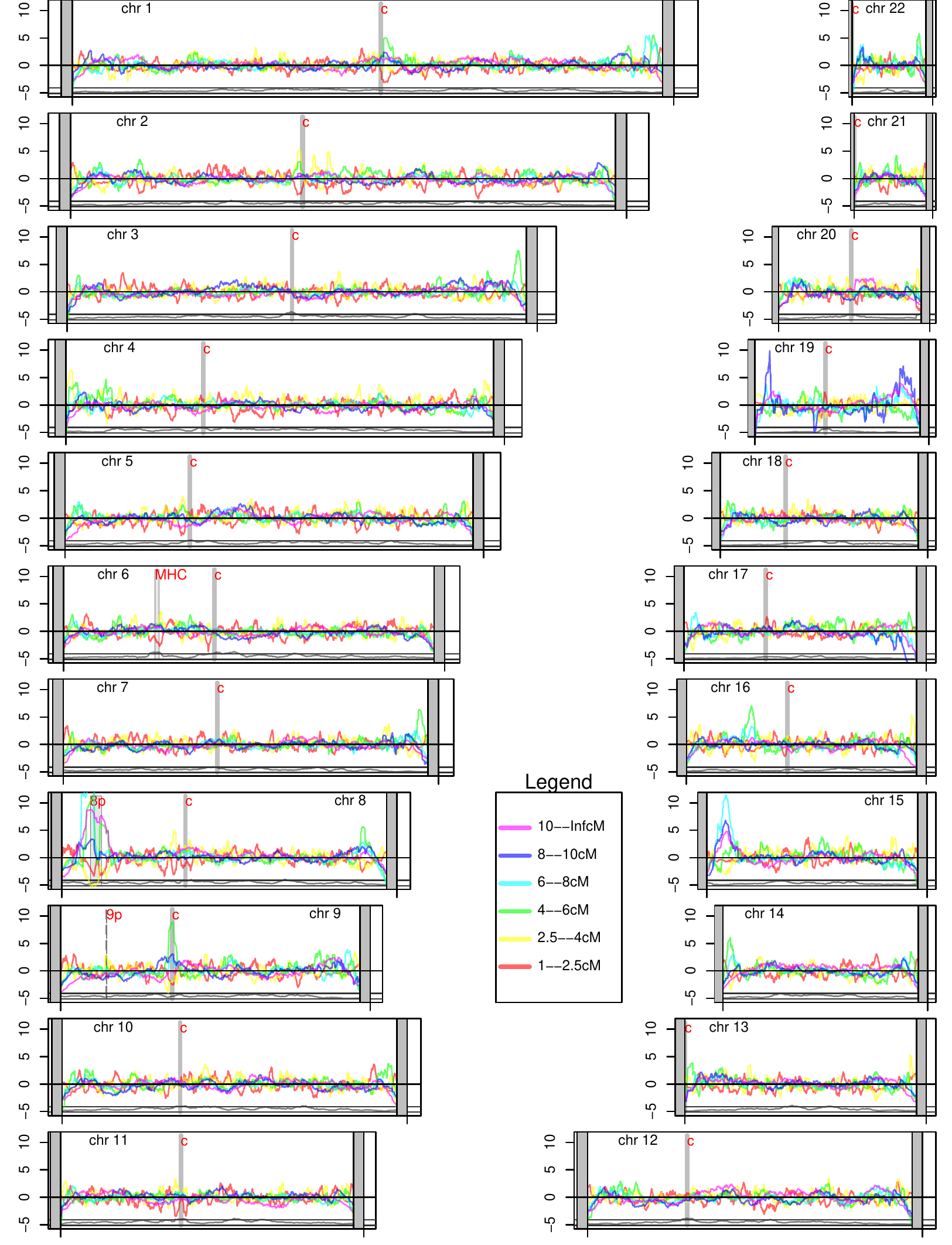}
    \caption{
       Normalized density of IBD blocks of different lengths, corrected for SNP density, across all autosomes
       (see section \ref{ss:block_density} for details).
       Marked with a grey bar and ``c'' are the centromeres; 
       and marked with ``8p'' is a large, segregating inversion \citep{giglio2001olfactory}.
       The grey curve along the bottom shows normalized SNP density. 
      \label{sfig:overlap_all}
    }
  \end{center}
\end{figure}

\begin{figure}[!htp]
  \begin{center}
    \includegraphics[height=.7\textheight]{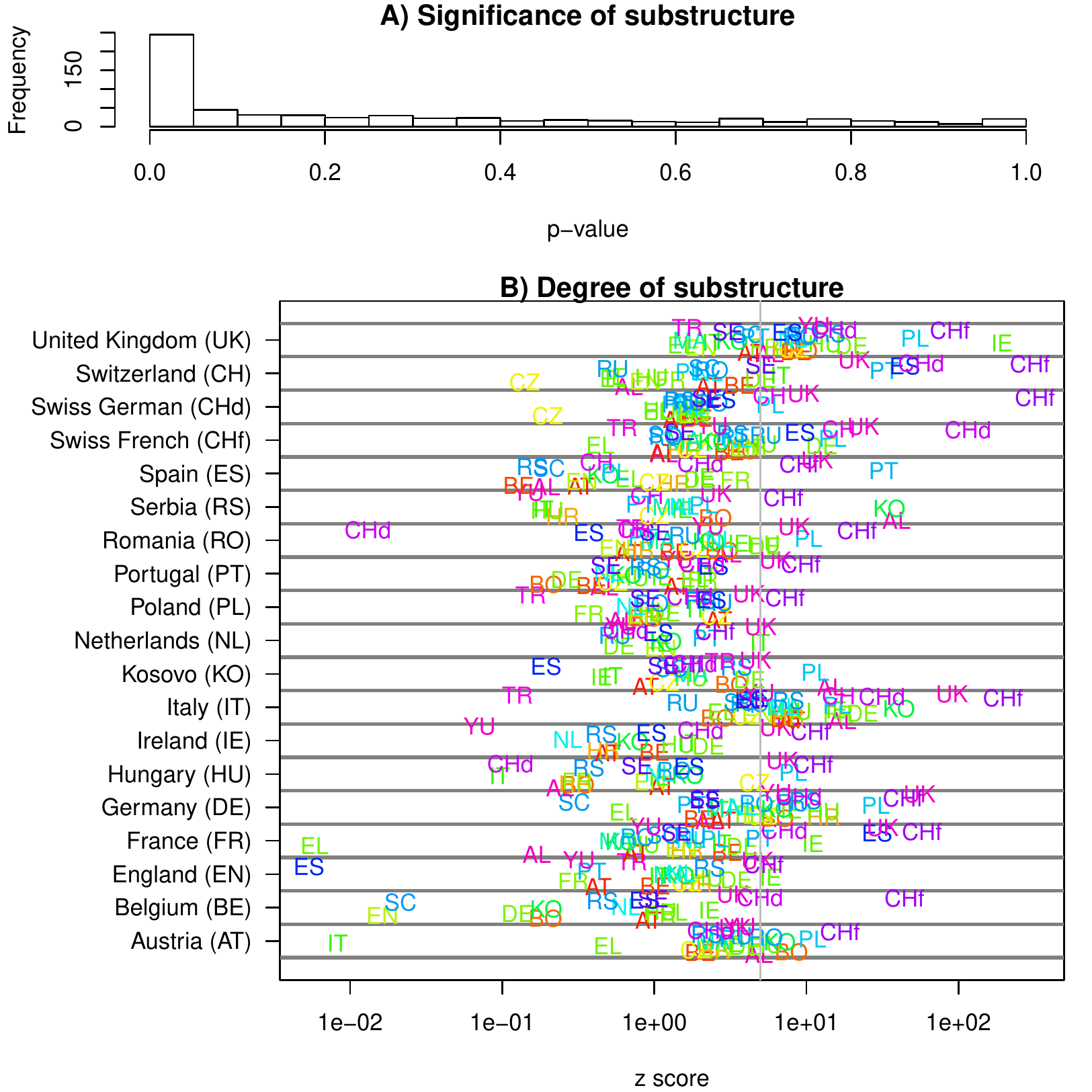}
    \caption{
    Two measures of overdispersal of block numbers across individuals (i.e.\ substructure):
    Suppose we have $n$ individuals from population $x$,
    and $N_{iy}$ is the number of IBD blocks of length at least 1cM
    that individual $i$ shares with anyone from population $y$.
    Our statistic of substructure within $x$ with respect to $y$ is the variance of these numbers,
    $s_{xy} = \frac{1}{n-1} \left( \sum_i N_{iy}^2 - \frac{1}{n} \left( \sum_i N_{iy} \right)^2 \right)$.
    We obtained a ``null'' distribution for this statistic by randomly reassigning all blocks shared between $x$ and $y$
    to an individual from $x$, and used this to evaluate the strength and the statistical significance of this substructure.
    {\bf (A)} Histogram of the ``$p$-value'', of the proportion of 1000 replicates that showed a variance greater than or equal to the observed variance $s_{xy}$,
    for all pairs of populations $x$ and $y$ with at least 10 individuals in population $y$.
    {\bf (B)} The ``$z$ score'', which is observed value $s_{xy}$ minus mean value divided by standard deviation,
    estimated using 1000 replicates.
    The population $x$ is shown on the vertical axis,
    with text labels giving $y$,
    so for instance, Italians show much more substructure with most other populations than do Irish.
    Note that sample size still has a large effect -- it is easier to see substructure with respect to the Swiss French ($x=$CHf)
    because the large number of Swiss French samples allows greater resolution.
    A vertical line is shown at $z=5$.
    Only pairs of populations with at least 3 samples in country $x$ and 10 samples in country $y$ are shown.
    Because of the log scale, only pairs with a positive $z$ score are shown, but no comparisons had $z < -2.5$,
    and only three had $z<-2$.
      \label{sfig:substructure_summaries}
    }
  \end{center}
\end{figure}

\begin{figure}[!htp]
  \begin{center}
    
    \vspace{2em}
    \begin{center}
      \includegraphics[height=.9\textheight]{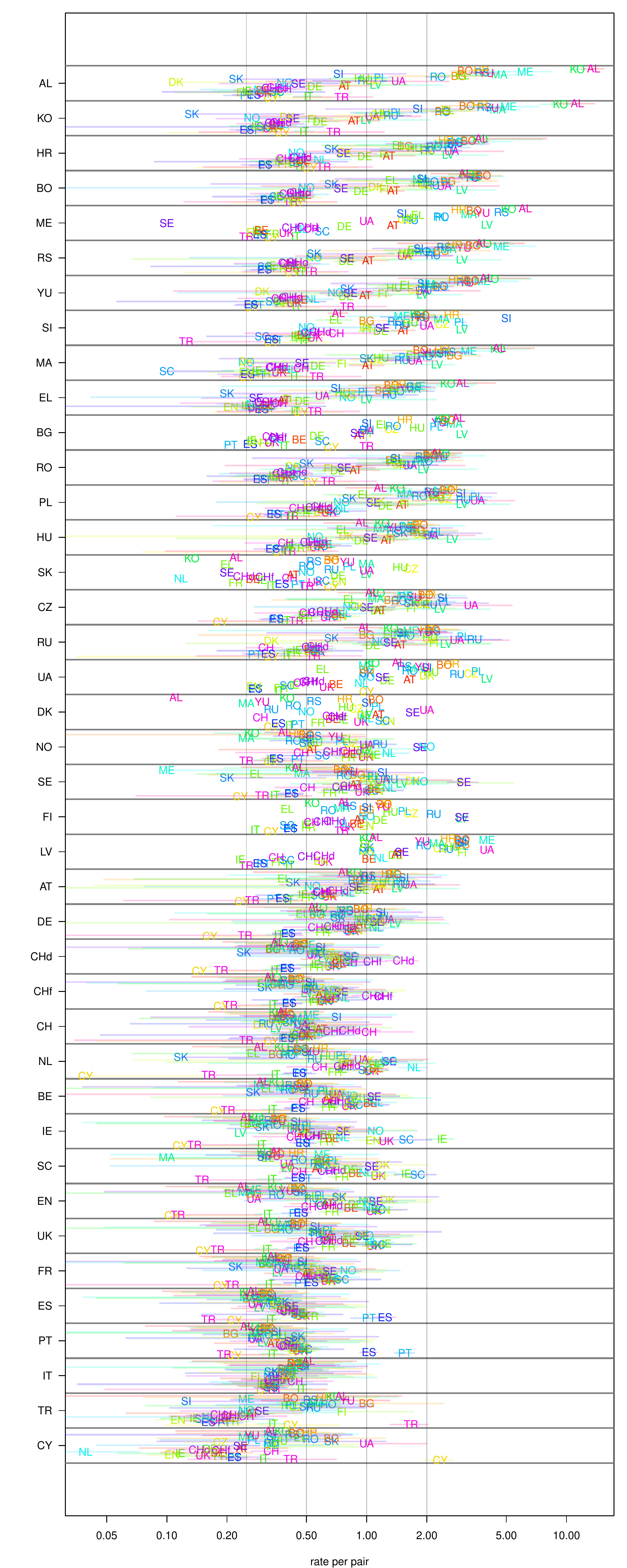}
    \end{center}
    \vspace{2em}
    \caption{
    {\bf (A)} Mean numbers of IBD blocks of length at least 1cM per pair of individuals,
    shown as a modified Cleveland dotchart,
    with $\pm$2 standard deviations shown as horizontal lines.
    For instance, on the bottom row we see that someone from the UK shares on average about one IBD block with someone else from the UK
    and slightly less than 0.2 blocks with someone from Turkey.
    Note that in most cases, the distribution of block numbers is fairly concentrated,
    and that nearby populations show quite similar patterns.
    \label{sfig:dotchart_long}
    }
  \end{center}
\end{figure}

\begin{figure}[!htp]
  \begin{center}
    \includegraphics{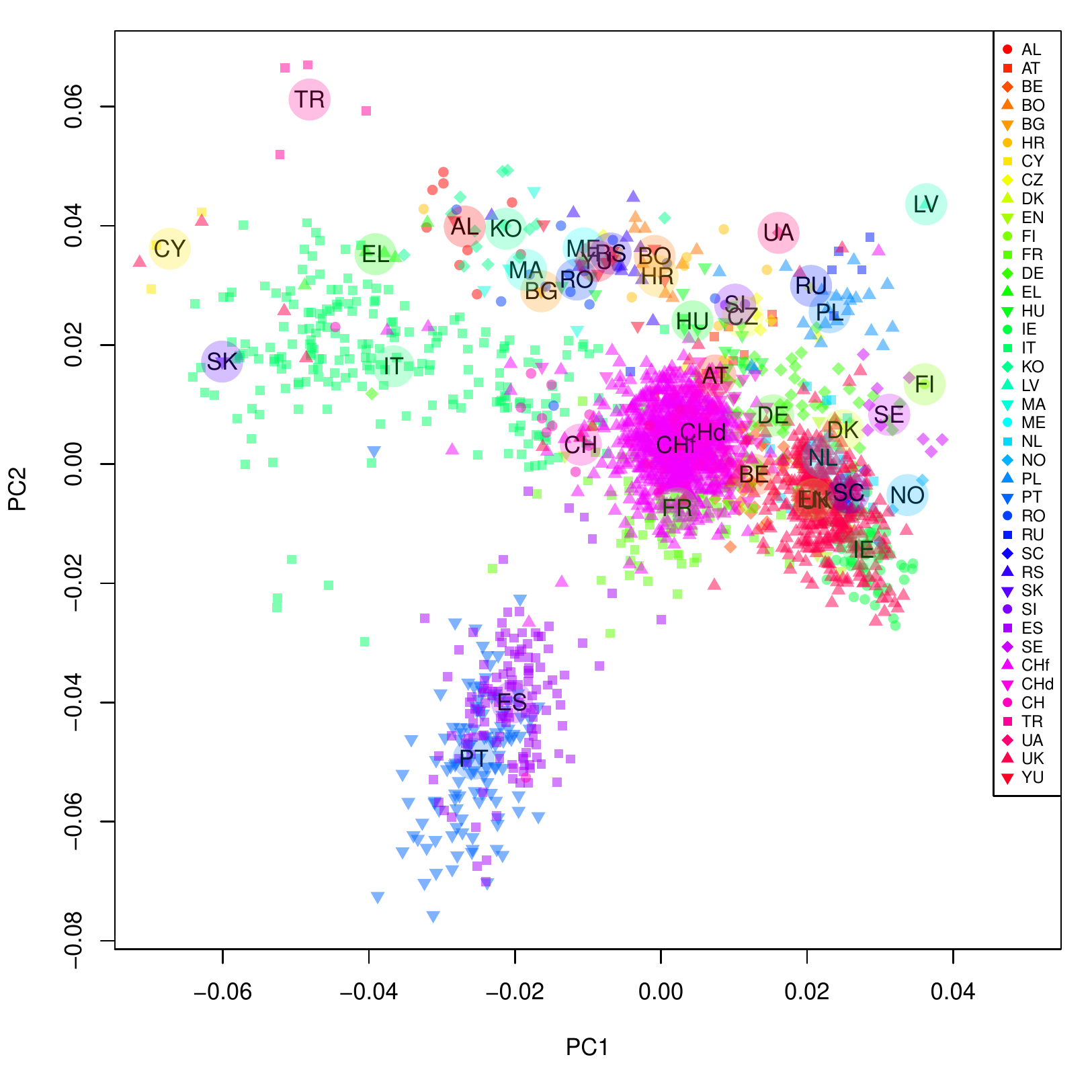}
    \caption{
    The positions of our sample of the first two principal components of the genotype matrix,
    as produced by EIGENSTRAT \citep{price2006eigenstrat}.
    Population centroids are marked by text and a transparent circle.
    Note the correspondence to a map of Europe, after a rotation and flip.
    \label{sfig:pca_map}
    }
  \end{center}
\end{figure}

\begin{figure}[!htp]
  \begin{center}
    \includegraphics{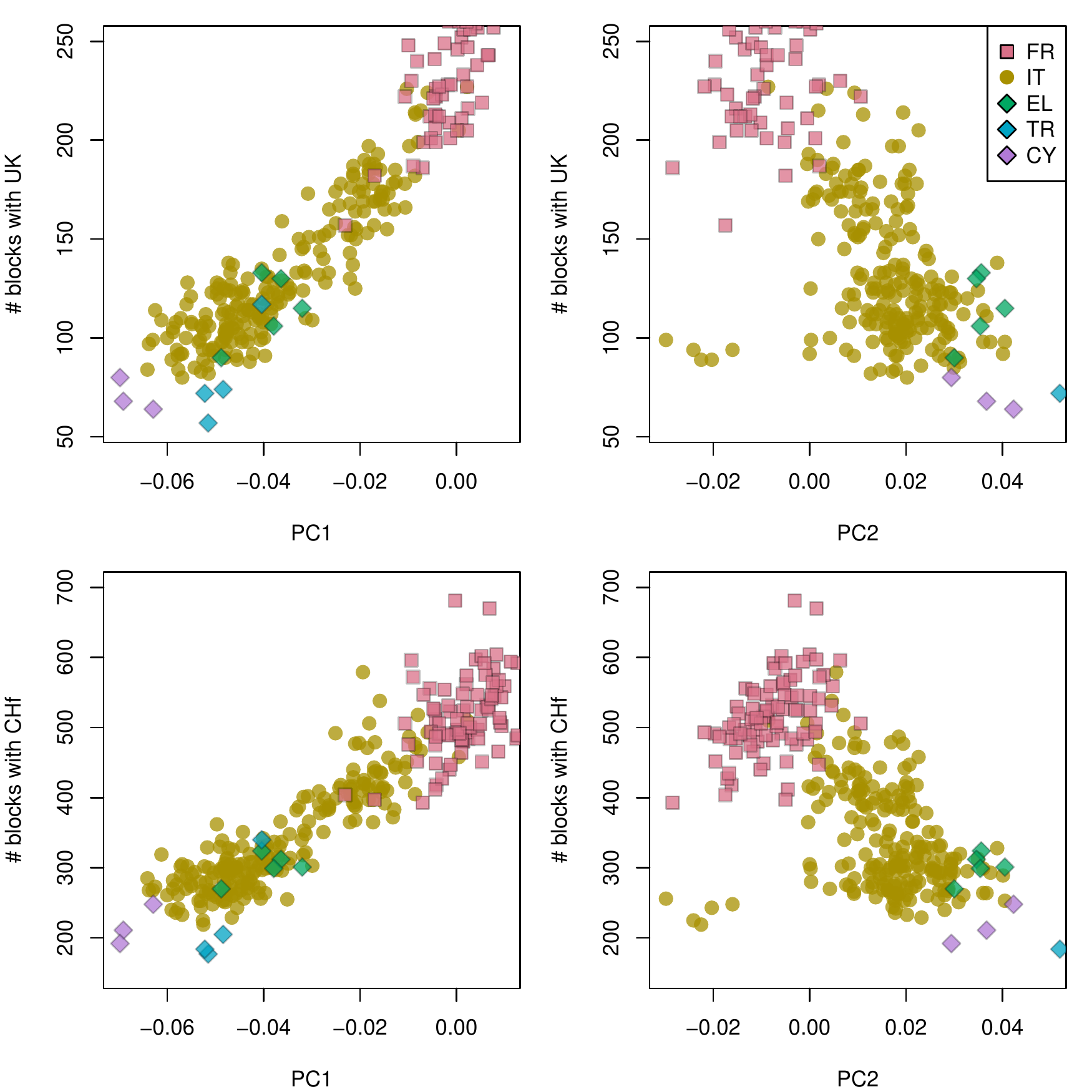}
    \caption{
    Comparison of figure \ref{fig:substructure}A in the main text to figure \ref{sfig:pca_map} --
    the axes are self-explanatory; the colors and symbols are the same as in figure \ref{fig:substructure}A.
    \label{sfig:pca_ibd_it}
    }
  \end{center}
\end{figure}

\begin{figure}[!htp]
  \begin{center}
    \includegraphics{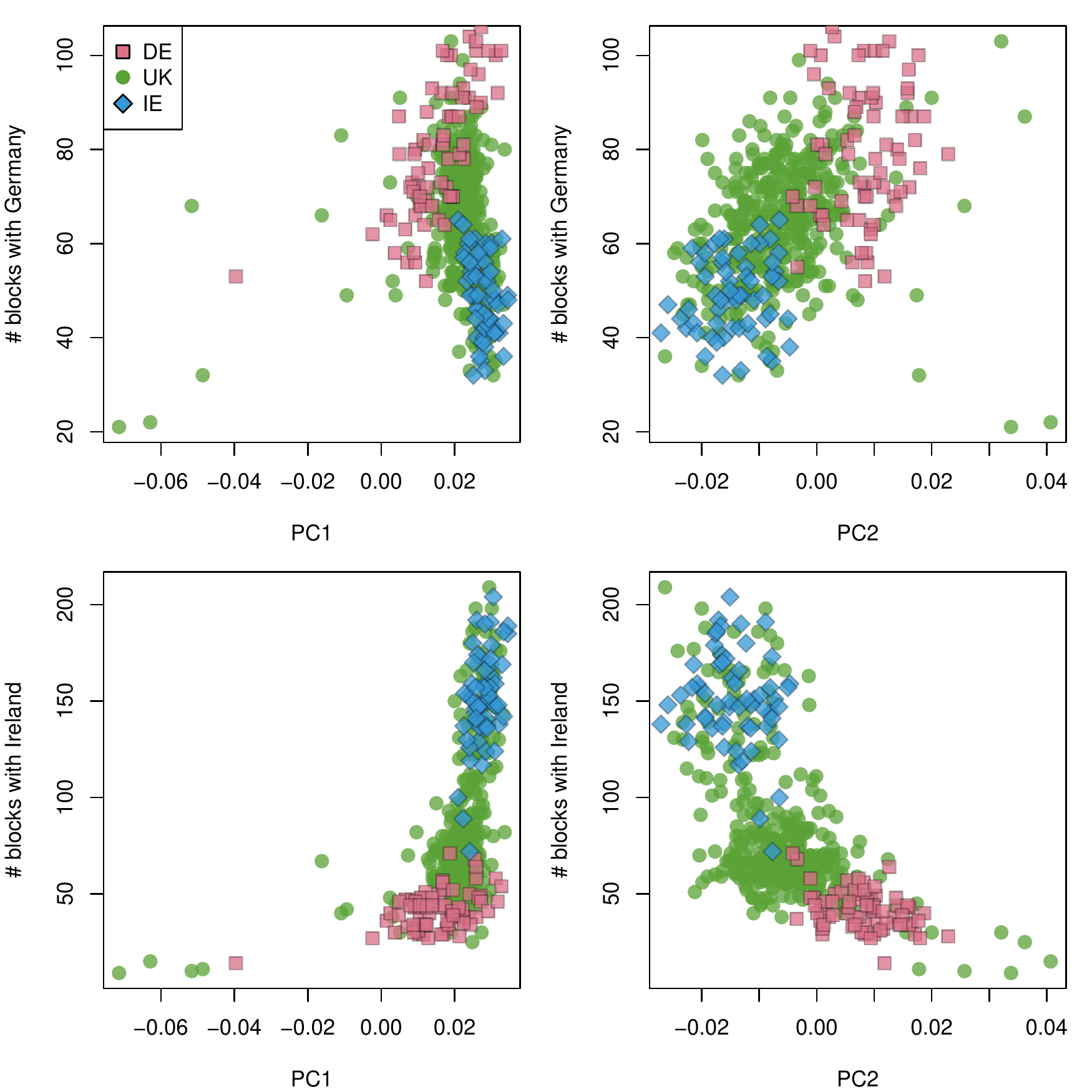}
    \caption{
    Comparison of figure \ref{fig:substructure}B in the main text to figure \ref{sfig:pca_map} --
    the axes are self-explanatory; the colors and symbols are the same as in figure \ref{fig:substructure}B.
    The four outlying UK individuals are, as in figure \ref{fig:substructure}B, 
    three who share a very high number of IBD blocks with Italians,
    and one who shares a very high number with the Slovakian sample.
    \label{sfig:pca_ibd_uk}
    }
  \end{center}
\end{figure}

\begin{figure}[!htp]
  \begin{center}

    \vspace{2em}
    \begin{center}
      \includegraphics[width=\textwidth]{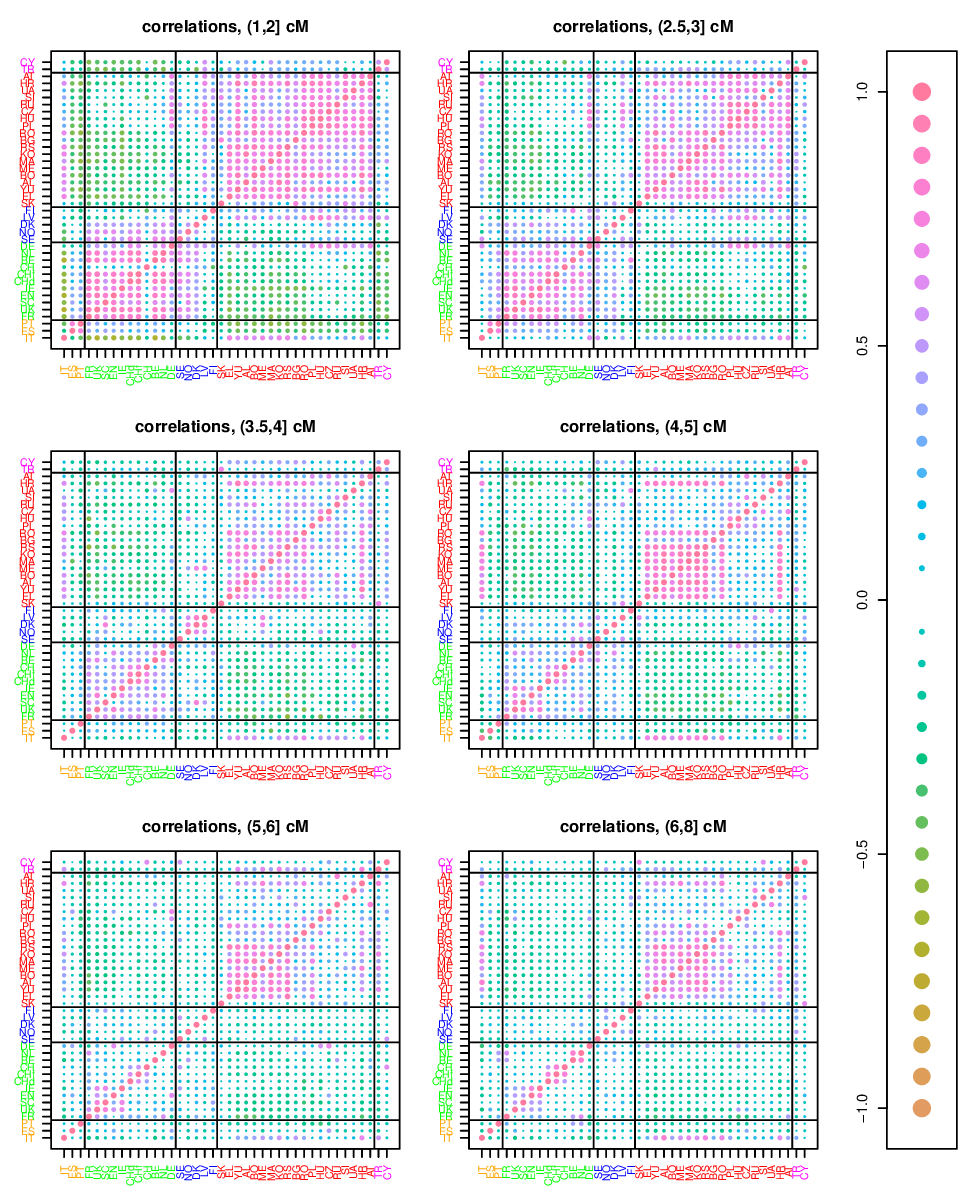}
    \end{center}
    \vspace{2em}
    \caption{
     Correlations in IBD rates, for six different length windows
     (omitted length windows are similar).
     If there are $n$ populations,
     $I(x,y)$ is the mean number of blocks in the given length range shared by a pair from populations $x$ and $y$,
     and $\bar I(x) = (1/(n-1))\sum_{z \neq x} I(x,z)$,
     shown is $(1/(n-2)) \sum_{z\notin\{x,y\}} (I(x,z)-\bar I(x))(I(y,z)-\bar I(y))$.
    \label{sfig:correlations}
    }
  \end{center}
\end{figure}

\begin{figure}[!htp]
  \begin{center}
    \vspace{2em}
    \begin{center}
    {\tt http://www.eve.ucdavis.edu/plralph/ibd/sharing-rates.svg}
    \end{center}
    \vspace{2em}
    \caption{
    The same plot as figure \ref{fig:sharing_and_maps}G--I, but rendered as an SVG figure
    with tooltips that allow identification of individual points \citep[using {\tt R}][]{plate2011rsvgtipsdevice} --
    open the file in a reasonably compliant browser (e.g.\ Firefox, Opera) or SVG browser (e.g. squiggle)
    and hover the mouse over a point of interest to see the label.
    \label{sfig:sharing_rates_svg}
    }
  \end{center}
\end{figure}

\begin{figure}[!htp]
  \begin{center}
    \includegraphics{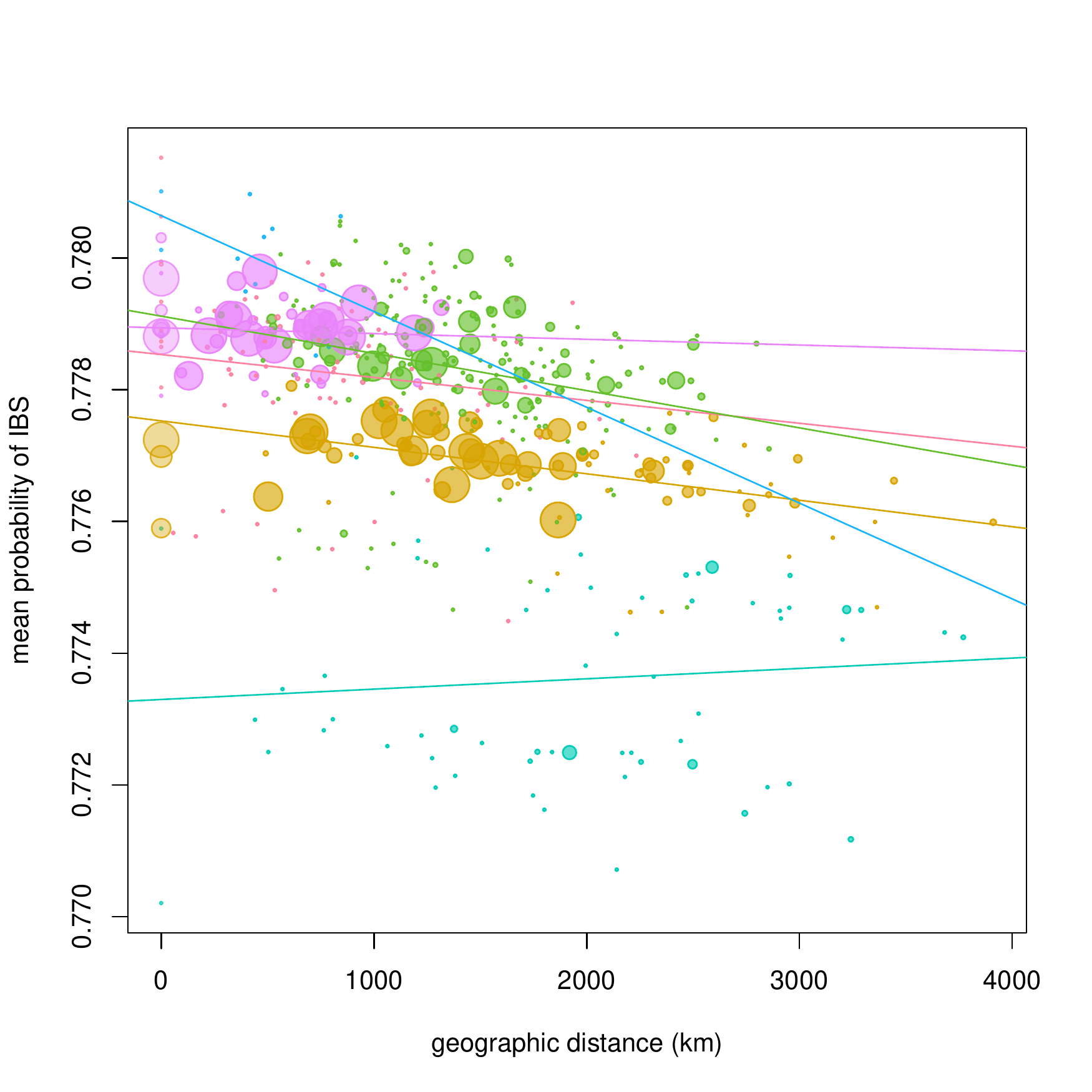}
    \caption{
    Mean IBS (``Identity by State'') against geographic distance,
    calculated using {\tt plink} \citep{purcell2007plink} as described in the main text,
    using the same groups and fitting the same curves as in figure \ref{fig:sharing_and_maps} of the main text.
    The lowest set of points, roughly following a line, are mean IBS with Turkey;
    unlike with IBD, mean IBS with Cyprus was significantly higher.
    In fact, the other rough line of points (between the comparisons to Turkey and the orange points)
    is almost entirely mean IBS with Cyprus, as well as mean IBS to Slovakia.
    Since Slovakia is only represented by a single individual in the dataset, we cannot reach further conclusions.
    \label{sfig:ibs_by_dist}
    }
  \end{center}
\end{figure}

\begin{figure}[!htp]
  \begin{center}
    \includegraphics[width=\textwidth]{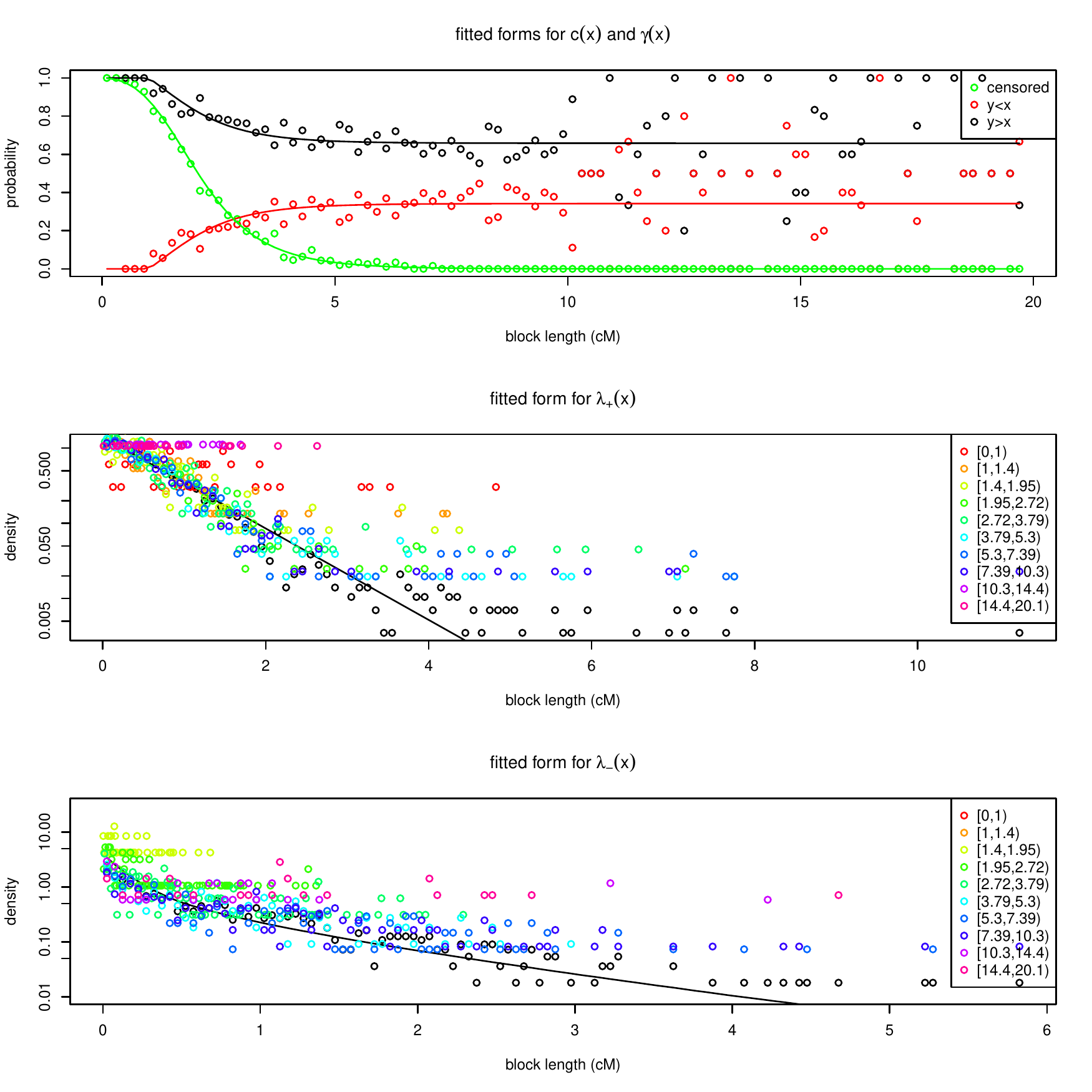}
    \caption{
    Goodness-of-fit for our estimated error distribution --
    points show data from simulations (described in the text),
    and lines show the parametric forms of equations \eqref{eqn:error_model}.
    Each simulated IBD block of length $x$ was either found by BEAGLE (and passed our filters) or was not;
    and if it was found, it had inferred length $y = x+\epsilon$, i.e.\ with length error $\epsilon$.
    The top figure shows the probability that a segment of a given length is missed entirely (and $1-c(x)$) in green,
    the probability that $\epsilon >0$ given the segment was found (and $\gamma(x)$) in black, 
    and the probability that $\epsilon \le 0$ given the segment was found (and $1-\gamma(x)$) in red.
    The second figure shows the probability density of all positive $\epsilon$ (in black, with $\lambda_+(x)$),
    and probability densities of positive $\epsilon$ for various categories of true length $x$ (colors).
    The third figure is similar to the second, except that it shows negative $\epsilon$.
    Note that blocks with inferred length $y<1$ were omitted.
    \label{sfig:fit_error_model}
    }
  \end{center}
\end{figure}

\begin{figure}[!htp]
  \begin{center}
    \includegraphics[width=\textwidth]{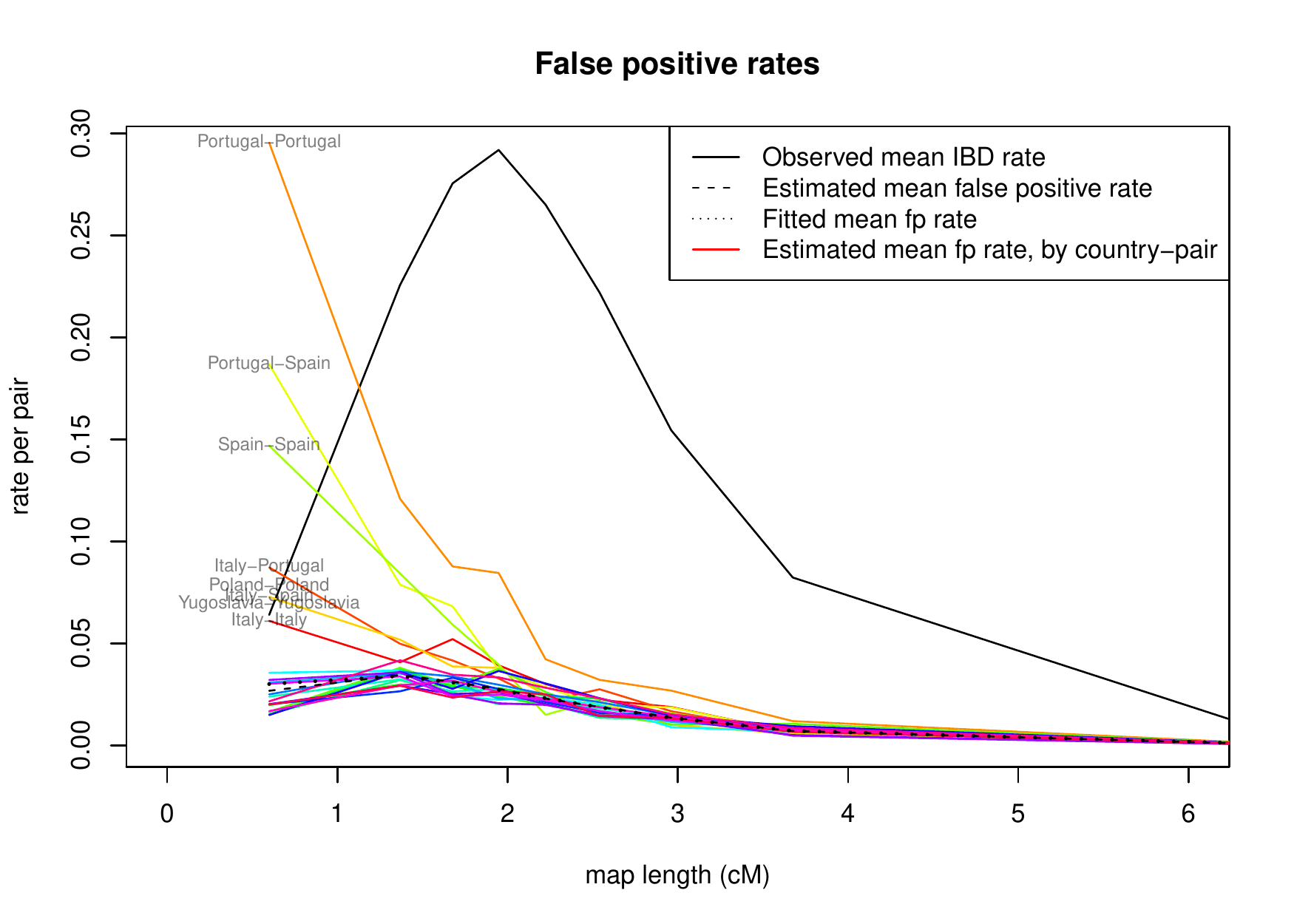}
    \caption{
    Estimated false positive rates per pair, compared to the observed rate, as a function of block length.
    The black dotted curves show the mean number of IBD blocks per pair observed in the false positive simulations (see section \ref{ss:error_model} of the main text),
    per centiMorgan, binned at 0, 1.2, 1.5, 1.8, 2.1, 2.4, 2.7, 3.2, 4.5, and 7.5cM,
    and the parametric fit described in the text.
    The colored curves show the same quantity, separately for each pair of country comparisons, 
    with the extreme values labeled.
    No comparisons other than Portugal--Portugal show any significant deviations from the parametric fit above 2cM.
    For comparison, the black solid curve shows the mean observed IBD rate across the same set of individuals;
    note that e.g.\ the false positive rate pairs of Portuguese individuals is higher than this at short lengths
    because the observed IBD rate between Portuguese at short block lengths is much higher than the overall mean.
    \label{sfig:fp_rate_by_country}
    }
  \end{center}
\end{figure}

\begin{figure}[!htp]
  \begin{center}
    
    \vspace{2em}
    \begin{center}
      \includegraphics[angle=90,height=.9\textheight]{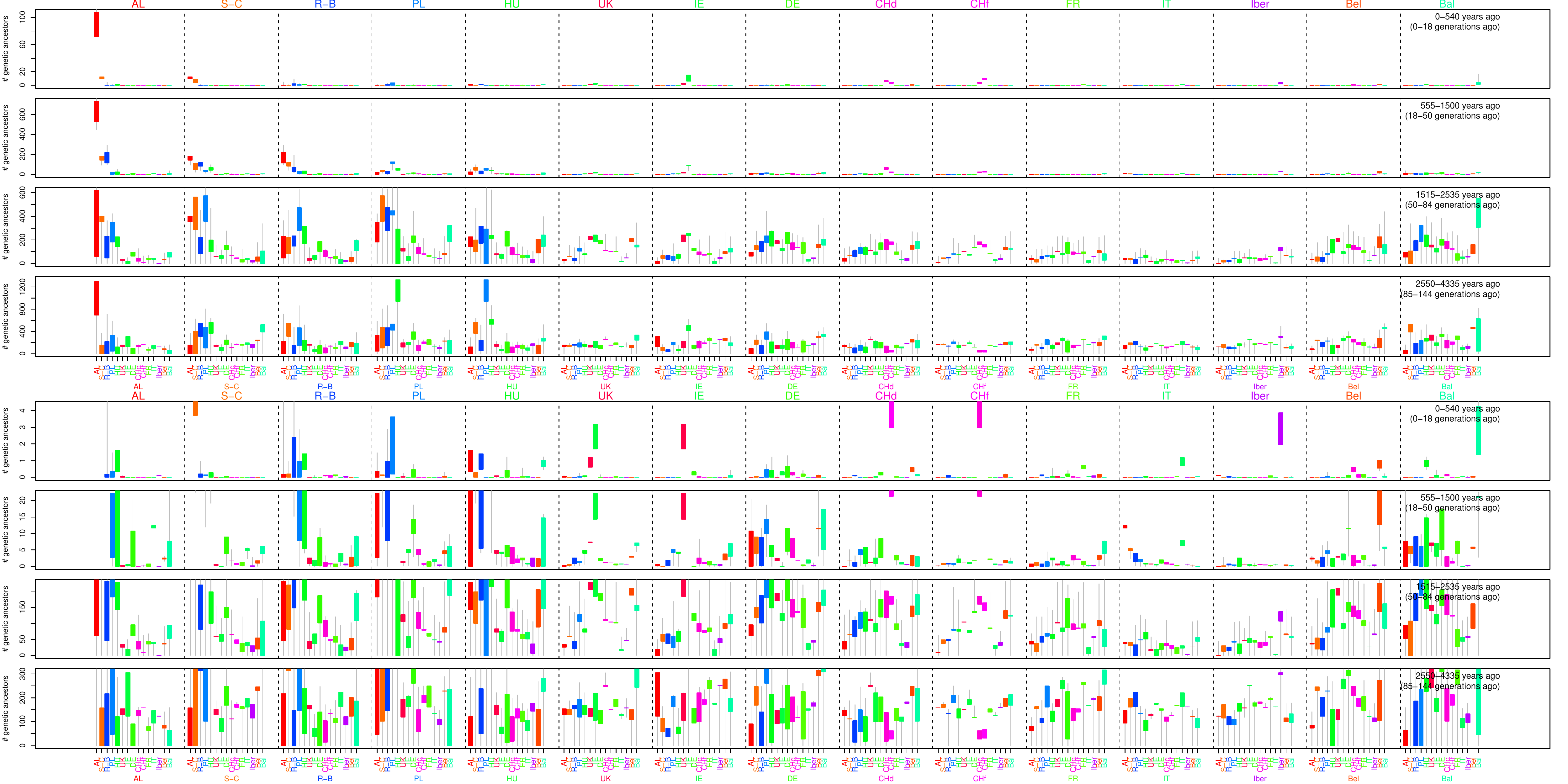}
    \end{center}
    \vspace{2em}
    \caption{
    Estimated total numbers of genetic common ancestors shared by various pairs of populations,
    in roughly the time periods 0--500ya, 500--1500ya, 1500--2500ya, and 2500--4300ya.
    The population groupings are:
    ``AL'', Albanian speakers (Albania and Kosovo);
    ``S-C'', Serbo-Croatian speakers in Bosnia, Croatia, Serbia, Montenegro, and Yugoslavia;
    ``R-B'', Romania and Bulgaria;
    ``UK'', United Kingdom, England, Scotland, Wales;
    ``Iber'', Spain and Portugal;
    ``Bel'', Belgium and the Netherlands;
    ``Bal'', Latvia, Finland, Sweden, Norway, and Denmark;
    and denotes a single population with the same abbreviations as in table \ref{tab:ibd_summaries} otherwise.
    \label{sfig:inversion_boxplots_long}
    }
  \end{center}
\end{figure}

\begin{figure}[!htp]
  \begin{center}
    
    \vspace{2em}
    \begin{center}
      \includegraphics[angle=90,height=.9\textheight]{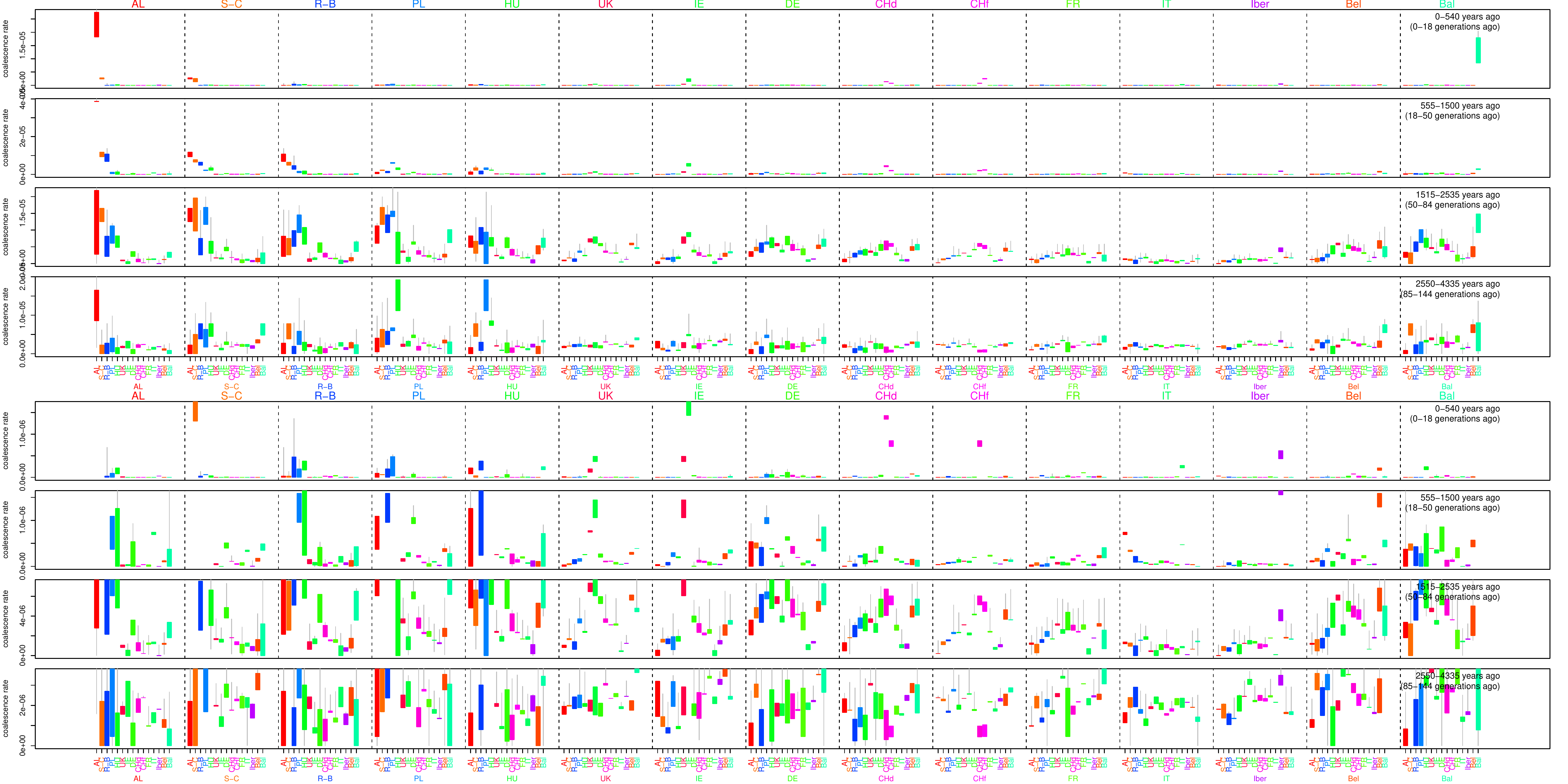}
    \end{center}
    \vspace{2em}
    \caption{
    For those who are used to thinking in effective population sizes,
    the equivalent figure to figure~\ref{sfig:inversion_boxplots_long},
    except with coalescent rate on the vertical axis, 
    rather than numbers of most recent genetic common ancestors.
    \label{sfig:inversion_boxplots_long_coal}
    }
  \end{center}
\end{figure}

\begin{figure}[!htp]
  \begin{center}
    \wikiincludegraphics{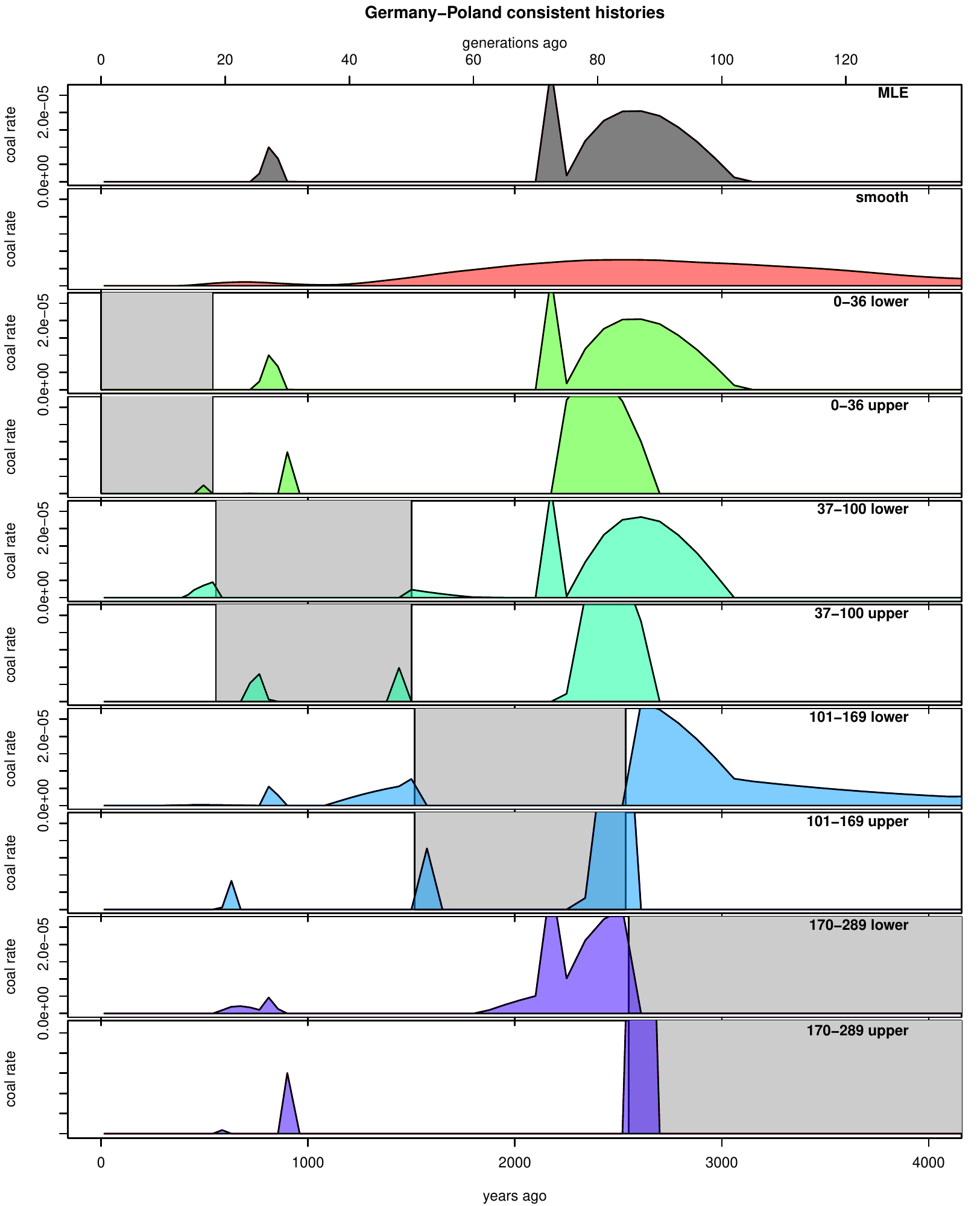}
    \caption{
    An example of the set of consistent histories (as coalescent distributions $\mu(n)$)
    used to find upper and lower bounds
    in figures~\ref{sfig:inversion_boxplots_long} and \ref{fig:inversion_boxplots}.
    The example shown is Poland--Germany;
    ``MLE'' is the maximum likelihood history;
    ``smooth'' is the smoothest consistent history;
    and the remaining plots show the histories giving lower and upper bounds
    for the referenced time intervals (in numbers of generations).
    In each case, the segment of time on which we are looking for a bound is shaded.
    \label{sfig:example_inversion_bounds}
    }
  \end{center}
\end{figure}

\begin{figure}[!htp]
  \begin{center}
    \wikiincludegraphics{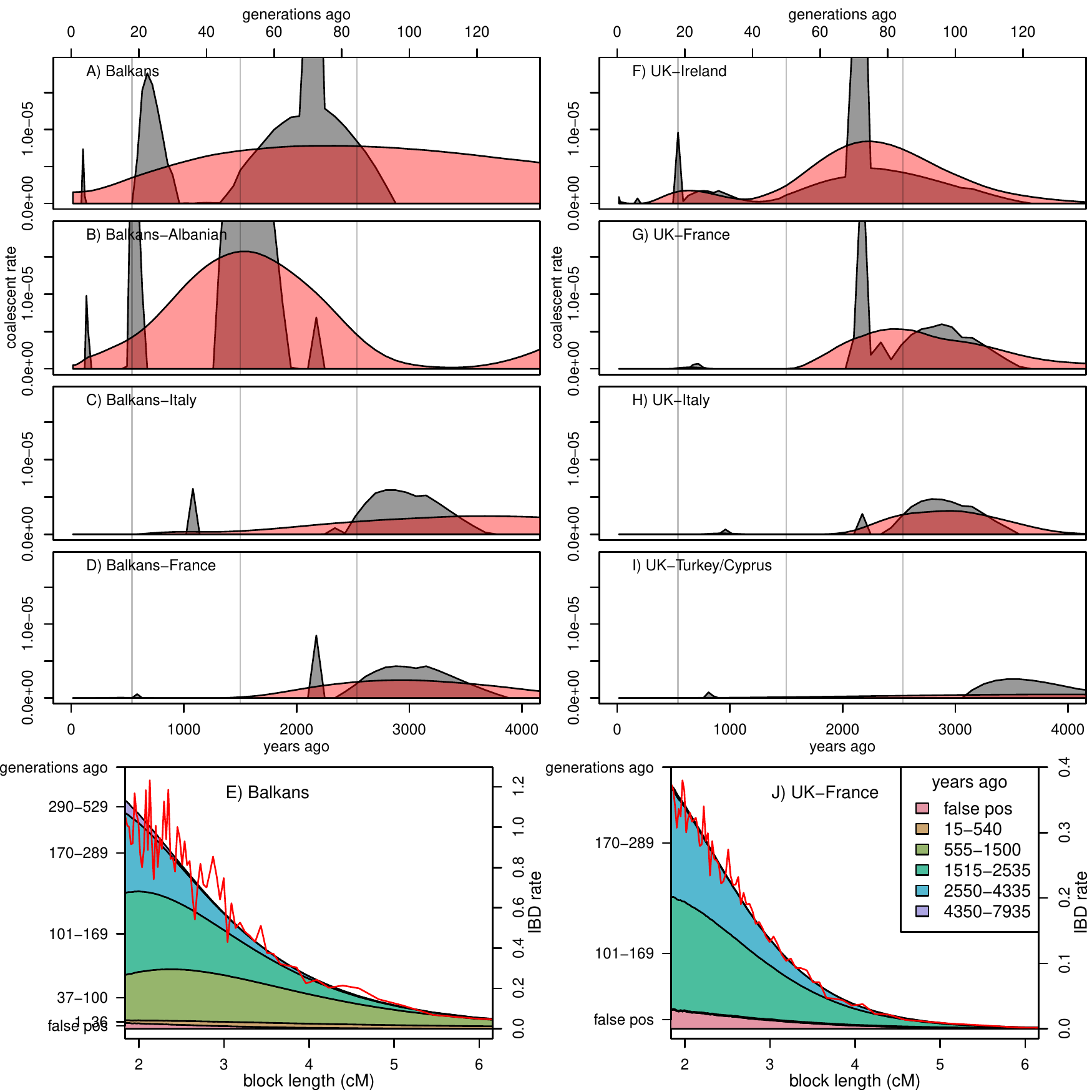}
    \caption{
    For those who are used to thinking in effective population sizes,
    the equivalent figure to figure~\ref{fig:inversion_distributions},
    except with coalescent rate on the vertical axis, 
    rather than numbers of most recent genetic common ancestors.
    \label{sfig:inversion_distributions_coal}
    }
  \end{center}
\end{figure}

\begin{figure}[!htp]
  \begin{center}
    \wikiincludegraphics{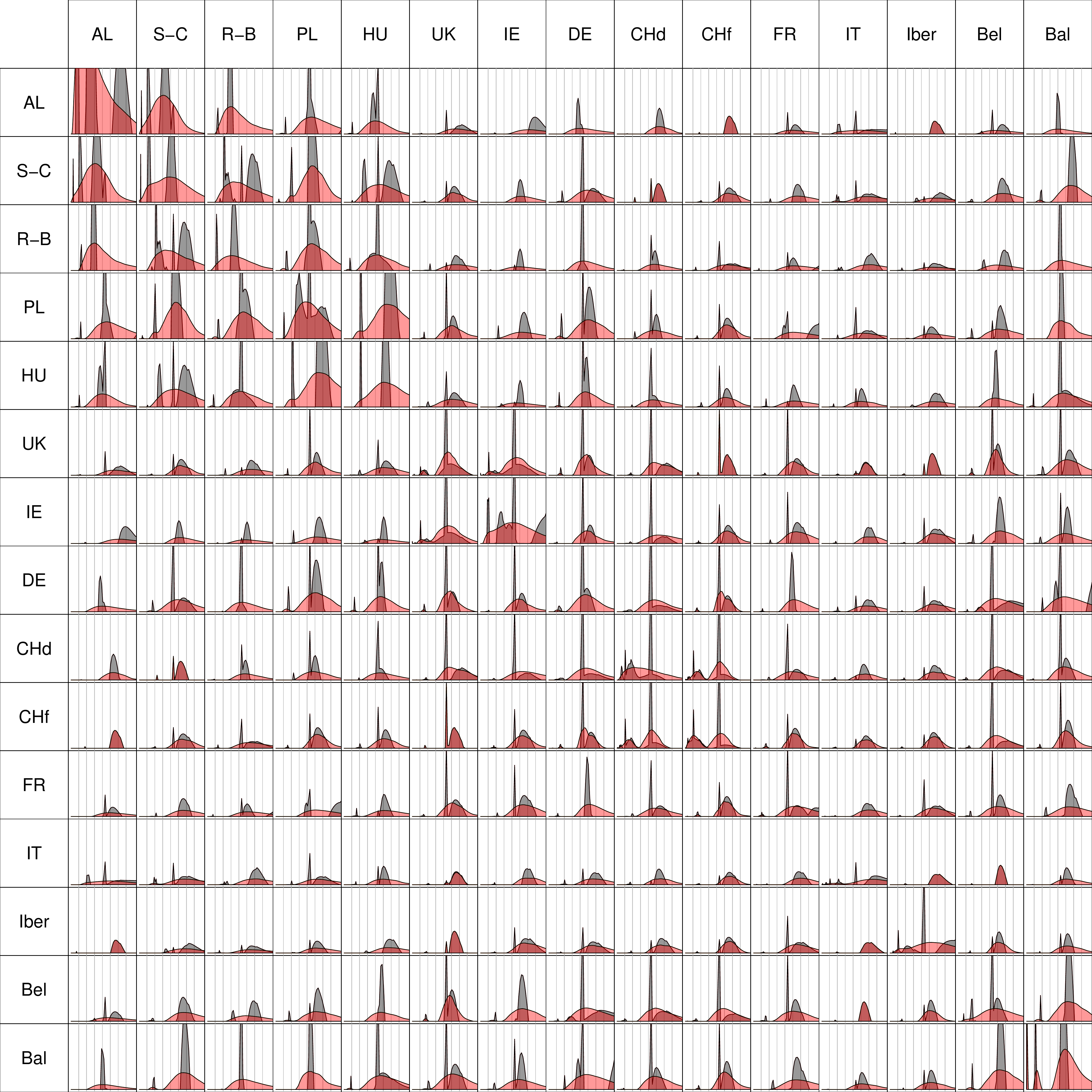}
    \caption{
    The maximum likelihood history (grey) and smoothest consistent history (red)
    for all pairs of population groupings of figure~\ref{sfig:inversion_boxplots_long} 
    (including those of figure~\ref{fig:inversion_boxplots}).
    Each panel is analogous to a panel of figure~\ref{fig:inversion_distributions};
    time scale is given by vertical grey lines every 500 years.
    For these plots on a larger scale see supplemental figure \ref{sfig:all_inversions}.
    \label{sfig:tinyinversions}
    }
  \end{center}
\end{figure}

\begin{figure}[!htp]
  \begin{center}
    
    \vspace{2em}
    \begin{center}
      {\tt http://www.eve.ucdavis.edu/$\sim$plralph/ibd/boxplotted-inversions.pdf }
    \end{center}
    \vspace{2em}
    \caption{
    All inversions shown in \ref{sfig:tinyinversions}, one per page (225 pages total).
    There is one page per pair of comparisons used in figure \ref{fig:inversion_boxplots}.
    On each page, there is one large plot, showing 10 distinct consistent histories (numbers of genetic ancestors back through time),
    and below are 10 histograms of IBD block length, one for each consistent history,
    showing both the observed distribution
    and the partitioning of blocks into age categories predicted by that history.
    The names of the two groupings are shown in the upper right:
    ``pointy'' is the unconstrained maximum likelihood solution;
    ``smooth'' is the smoothest consistent history;
    ``$a$--$b$ lower'' is the history used to find the lower bound for the time period $a$--$b$ generations ago in figure \ref{fig:inversion_boxplots};
    and ``$a$--$b$ upper'' is the history used to find the corresponding upper bound.
    Each of these are described in more detail in the Methods.
    \label{sfig:all_inversions}
    }
  \end{center}
\end{figure}

\clearpage

\begin{table}[!htp]
\begin{center}
\begin{tabular}{|r|llll|}
  \hline
 & 0-1 cM & 1-2 cM & 2-4 cM & 4-10 cM \\ 
  \hline
(Intercept) & ~0.08313    & ~0.1436 ** & ~0.07034    & -0.0193    \\ 
  Albania & -0.00097    & ~0.0063    & -0.04232    & -0.0323    \\ 
  Austria & ~0.36424    & ~0.0365    & ~0.05874    & -0.1399    \\ 
  Belgium & -0.02009    & ~0.0863    & ~0.02874    & ~0.0374    \\ 
  Bosnia & -0.13914    & -0.0327    & ~0.02277    & ~0.1032    \\ 
  Bulgaria & -0.14857    & ~0.0467    & -0.21836    & ~0.2427    \\ 
  Croatia & ~0.08645    & -0.0223    & ~0.02045    & -0.0371    \\ 
  Cyprus & -0.15556    & -0.0775    & ~0.08306    & ~0.2158    \\ 
  Czech Republic & -0.14301    & ~0.1013    & ~0.02656    & ~0.1607    \\ 
  Denmark & ~0.66101    & ~0.1265    & ~0.03675    & ~0.3095    \\ 
  England & ~0.12131    & ~0.0854    & ~0.03784    & ~0.0199    \\ 
  Finland & ~0.24617    & ~0.1491    & -0.09775    & ~0.9455    \\ 
  France & ~0.05049    & ~0.0533    & ~0.09670 ** & ~0.1655    \\ 
  Germany & ~0.06837    & ~0.0454    & ~0.08617 ** & ~0.1288    \\ 
  Greece & ~0.06752    & -0.2266    & ~0.00086    & ~0.3340    \\ 
  Hungary & ~0.04491    & -0.0070    & ~0.11499 *  & ~0.1179    \\ 
  Ireland & ~0.07873    & ~0.0676    & ~0.04292    & ~0.0466    \\ 
  Italy & ~0.02728    & ~0.0218    & ~0.06694 *  & ~0.1607    \\ 
  Kosovo & ~0.17256    & ~0.0384    & -0.02022    & ~0.0109    \\ 
  Latvia & ~0.76499    & ~0.1805    & -0.07177    & -0.0787    \\ 
  Macedonia & -0.12898    & ~0.1414    & -0.04692    & -0.0093    \\ 
  Montenegro & ~0.50084    & -0.0845    & -0.01400    & ~0.2746    \\ 
  Netherlands & ~0.10448    & ~0.0842    & ~0.11875 *  & ~0.0423    \\ 
  Norway & -0.46112    & -0.0176    & ~0.07972    & ~0.4042    \\ 
  Poland & ~0.21095    & ~0.0021    & ~0.05353    & ~0.1314    \\ 
  Portugal & ~0.03987    & -0.0019    & ~0.06771 *  & ~0.0664    \\ 
  Romania & -0.24251    & ~0.0545    & -0.05095    & -0.0845    \\ 
  Russia & ~0.12047    & -0.0094    & ~0.00484    & ~0.1568    \\ 
  Scotland & ~0.25426    & ~0.0183    & ~0.03205    & -0.1172    \\ 
  Serbia & ~0.06965    & ~0.0384    & -0.04688    & ~0.1397    \\ 
  Slovakia & -0.00988    & -0.0458    & ~0.21389    & -0.3070    \\ 
  Slovenia & -0.19993    & ~0.1456    & -0.00855    & ~0.4673    \\ 
  Spain & ~0.02470    & ~0.0268    & ~0.08928 ** & ~0.1084    \\ 
  Sweden & ~0.02043    & -0.0076    & ~0.20759 ** & ~0.2666    \\ 
  Swiss French & ~0.11099    & ~0.1394 ** & ~0.07782 *  & ~0.0525    \\ 
  Swiss German & ~0.14555    & ~0.0943 ** & ~0.08075 ** & ~0.0564    \\ 
  Switzerland & ~0.13339    & ~0.0102    & ~0.04665    & ~0.1135    \\ 
  Turkey & -0.01107    & ~0.0762    & ~0.13784    & -0.1003    \\ 
  Ukraine & -0.33535    & -0.0543    & ~0.03820    & -1.2068    \\ 
  United Kingdom & ~0.10379    & ~0.0868 ** & ~0.11605 ** & ~0.1309    \\ 
  Yugoslavia & -0.10508    & ~0.0436    & -0.00218    & ~0.0018    \\ 
   \hline
\end{tabular}
\end{center}
\caption{
Estimated coefficients describing the effect of changing population sample size,
as described in the text (section \ref{ss:error_model}, ``Differential sample sizes'').
Stars denote statistical significance: ``*'' corresponds to $p<.05$ and ``**'' corresponds to $p<.01$.
The coefficients are from a binomial GLM with a logit link function,
applied to the number of IBD segments detected in the same set of individuals run with and without an additional 812 individuals.
For instance, the top three entries in the lefthand column tell us that if $F$ is the number of segments greater than 1cM found 
between Albanian and Austrian individuals
in analysis with the full dataset,
and $S$ is the corresponding number in the analysis with only the subset,
that the model predicts that $S/(S+F) \approx (1+\exp(-0.08313+0.00097-0.36424))^{-1} = 0.61$ (plus binomial sampling noise).
Note that coefficients producing effect sizes larger than 4\% (e.g.~Austria for 0--1cM)
all correspond to populations with small sample sizes, and are not significant.
\label{stab:subset_glm_results}
}
\end{table}

\begin{table}[!htp]
\begin{center}
\begin{tabular}{|l|l|l||l|r|}
  \hline
COUNTRY\_SELF &  COUNTRY\_GFOLX & PRIMARY\_LANGUAGE &   Population & $n$ \\
  \hline
         Albania &        Albania &         Albanian &        Albania &    3 \\
      Yugoslavia &         Serbia &         Albanian &        Albania &    1 \\
      Yugoslavia &     Yugoslavia &         Albanian &        Albania &    5 \\
         Austria &                &           German &        Austria &    3 \\
         Austria &        Austria &           German &        Austria &   10 \\
           Spain &        Austria &           German &        Austria &    1 \\
         Belgium &        Belgium &            Dutch &        Belgium &    4 \\
         Belgium &        Belgium &          Flemish &        Belgium &    3 \\
         Belgium &        Belgium &           French &        Belgium &   28 \\
         Germany &        Belgium &           French &        Belgium &    1 \\
     Switzerland &        Belgium &           French &        Belgium &    1 \\
          Bosnia &         Bosnia &          Bosnian &         Bosnia &    4 \\
          Bosnia &         Bosnia &          Serbian &         Bosnia &    1 \\
          Bosnia &         Bosnia &   Serbo-Croatian &         Bosnia &    4 \\
        Bulgaria &       Bulgaria &        Bulgarian &       Bulgaria &    1 \\
         Croatia &                &         Croatian &        Croatia &    1 \\
         Croatia &        Croatia &         Croatian &        Croatia &    6 \\
      Yugoslavia &     Yugoslavia &         Croatian &        Croatia &    1 \\
         Croatia &        Croatia &   Serbo-Croatian &        Croatia &    1 \\
          Cyprus &                &          English &         Cyprus &    1 \\
          Cyprus &                &            Greek &         Cyprus &    1 \\
          Cyprus &         Cyprus &            Greek &         Cyprus &    1 \\
  Czech Republic & Czech Republic &            Czech & Czech Republic &    9 \\
         Denmark &                &           Danish &        Denmark &    1 \\
         England &        England &          English &        England &   18 \\
          Turkey &        England &          English &        England &    1 \\
  United Kingdom &        England &          English &        England &    3 \\
         Finland &        Finland &          Finnish &        Finland &    1 \\
          France &                &           French &         France &    2 \\
          France &         France &           French &         France &   82 \\
         Germany &         France &           French &         France &    1 \\
     Switzerland &         France &           French &         France &    1 \\
         Germany &                &                  &        Germany &    1 \\
         Germany &                &          English &        Germany &    2 \\
         Germany &        Germany &           French &        Germany &    1 \\
         Germany &                &           German &        Germany &    1 \\
         Germany &        Germany &           German &        Germany &   63 \\
     Switzerland &        Germany &           German &        Germany &    1 \\
         Hungary &        Germany &        Hungarian &        Germany &    1 \\
         Germany &        Germany &           Polish &        Germany &    1 \\
     Switzerland &         Greece &           French &         Greece &    1 \\
          Greece &         Greece &            Greek &         Greece &    4 \\
        \hline
          \multicolumn{5}{|c|}{(continued on next page)} \\
      \hline
\end{tabular}
\end{center}
\caption{
The composition of our populations.
``COUNTRY\_SELF'' is the reported country of origin;
``COUNTRY\_GFOLX'' is the country of origin of all reported grandparents
(individuals with reported grandparents from different countries were removed);
``PRIMARY\_LANGUAGE'' is the reported primary language;
``Population'' is our population label;
and $n$ gives the number of individuals falling in this category.
\label{stab:population_defns}
}
\end{table}

\begin{table}[!htp]
\begin{center}
\begin{tabular}{|l|l|l||l|r|}
  \hline
COUNTRY\_SELF &  COUNTRY\_GFOLX & PRIMARY\_LANGUAGE &   Population & $n$ \\
  \hline
         Hungary &        Hungary &           French &        Hungary &    1 \\
         Hungary &        Hungary &        Hungarian &        Hungary &   17 \\
         Hungary &        Hungary &          Russian &        Hungary &    1 \\
         Ireland &                &                  &        Ireland &   19 \\
         Ireland &                &          English &        Ireland &   38 \\
         England &        Ireland &          English &        Ireland &    1 \\
         Ireland &        Ireland &          English &        Ireland &    1 \\
         Ireland &        Ireland &           French &        Ireland &    1 \\
           Italy &                &                  &          Italy &    1 \\
          France &          Italy &           French &          Italy &    1 \\
           Italy &          Italy &           French &          Italy &    8 \\
     Switzerland &          Italy &           French &          Italy &    9 \\
           Italy &          Italy &           German &          Italy &    1 \\
           Italy &                &          Italian &          Italy &    3 \\
          France &          Italy &          Italian &          Italy &    1 \\
           Italy &          Italy &          Italian &          Italy &  170 \\
         Romania &          Italy &          Italian &          Italy &    1 \\
          Sweden &          Italy &          Italian &          Italy &    1 \\
     Switzerland &          Italy &          Italian &          Italy &   17 \\
          Kosovo &                &                  &         Kosovo &    1 \\
      Yugoslavia &         Kosovo &         Albanian &         Kosovo &   10 \\
      Yugoslavia &         Kosovo &          Kosovan &         Kosovo &    2 \\
      Yugoslavia &         Kosovo &   Serbo-Croatian &         Kosovo &    2 \\
          Latvia &         Latvia &          Latvian &         Latvia &    1 \\
       Macedonia &      Macedonia &       Macedonian &      Macedonia &    4 \\
      Yugoslavia &     Montenegro &          Serbian &     Montenegro &    1 \\
     Netherlands &    Netherlands &            Dutch &    Netherlands &   15 \\
         Holland &                &          English &    Netherlands &    1 \\
     Netherlands &    Netherlands &           French &    Netherlands &    1 \\
          Norway &         Norway &        Norwegian &         Norway &    2 \\
          France &         Poland &           French &         Poland &    1 \\
          Poland &                &           Polish &         Poland &    4 \\
          France &         Poland &           Polish &         Poland &    2 \\
          Poland &         Poland &           Polish &         Poland &   15 \\
          France &       Portugal &       Portuguese &       Portugal &    1 \\
        Portugal &       Portugal &       Portuguese &       Portugal &  114 \\
         Romania &        Romania &         Romanian &        Romania &   14 \\
         Romania &         Russia &         Romanian &         Russia &    1 \\
          Russia &         Russia &          Russian &         Russia &    5 \\
        Scotland &                &          English &       Scotland &    3 \\
        Scotland &       Scotland &          English &       Scotland &    2 \\
      Yugoslavia &         Serbia &        Hungarian &         Serbia &    1 \\
          Serbia &         Serbia &          Serbian &         Serbia &    1 \\
      Yugoslavia &         Serbia &          Serbian &         Serbia &    4 \\
      Yugoslavia &     Yugoslavia &          Serbian &         Serbia &    2 \\
         Croatia &         Serbia &   Serbo-Croatian &         Serbia &    1 \\
      Yugoslavia &         Serbia &   Serbo-Croatian &         Serbia &    2 \\
        \hline
          \multicolumn{5}{|c|}{(continued on next page)} \\
      \hline
\end{tabular}
\end{center}
\caption{
Continuation of table \ref{stab:population_defns}.
\label{stab:population_defnsB}
}
\end{table}

\begin{table}[!htp]
\begin{center}
\begin{tabular}{|l|l|l||l|r|}
  \hline
COUNTRY\_SELF &  COUNTRY\_GFOLX & PRIMARY\_LANGUAGE &   Population & $n$ \\
  \hline
        Slovakia &       Slovakia &        Slovakian &       Slovakia &    1 \\
           Italy &       Slovenia &          Slovene &       Slovenia &    1 \\
        Slovenia &       Slovenia &          Slovene &       Slovenia &    1 \\
           Spain &          Spain &         Columbia &          Spain &    2 \\
     Switzerland &          Spain &         Columbia &          Spain &    2 \\
           Spain &          Spain &           French &          Spain &    5 \\
     Switzerland &          Spain &           French &          Spain &    2 \\
           Spain &          Spain &         Galician &          Spain &    2 \\
           Spain &                &          Spanish &          Spain &    4 \\
           Spain &          Spain &          Spanish &          Spain &  106 \\
     Switzerland &          Spain &          Spanish &          Spain &    7 \\
          Sweden &                &                  &         Sweden &    1 \\
          Sweden &         Sweden &          Swedish &         Sweden &    9 \\
     Switzerland &                &           French &   Swiss French &    1 \\
         Belgium &    Switzerland &           French &   Swiss French &    1 \\
  Czech Republic &    Switzerland &           French &   Swiss French &    1 \\
          France &    Switzerland &           French &   Swiss French &    7 \\
          Poland &    Switzerland &           French &   Swiss French &    1 \\
        Portugal &    Switzerland &           French &   Swiss French &    1 \\
           Spain &    Switzerland &           French &   Swiss French &    1 \\
     Switzerland &    Switzerland &           French &   Swiss French &  826 \\
     Switzerland &    Switzerland &           German &   Swiss German &  103 \\
           Italy &    Switzerland &          Italian &    Switzerland &    2 \\
     Switzerland &    Switzerland &          Italian &    Switzerland &   12 \\
     Switzerland &    Switzerland &           Patois &    Switzerland &    1 \\
     Switzerland &    Switzerland &         Romansch &    Switzerland &    1 \\
           Spain &    Switzerland &          Spanish &    Switzerland &    1 \\
          Turkey &         Turkey &          Turkish &         Turkey &    4 \\
         Ukraine &        Ukraine &         Ukranian &        Ukraine &    1 \\
  United Kingdom &                &                  & United Kingdom &   87 \\
  United Kingdom &                &          English & United Kingdom &  270 \\
  United Kingdom & United Kingdom &          English & United Kingdom &    1 \\
      Yugoslavia &                &                  &     Yugoslavia &    1 \\
      Yugoslavia &     Yugoslavia &           French &     Yugoslavia &    1 \\
      Yugoslavia &     Yugoslavia &         Romanian &     Yugoslavia &    1 \\
      Yugoslavia &     Yugoslavia &   Serbo-Croatian &     Yugoslavia &    3 \\
      Yugoslavia &     Yugoslavia &      Yugoslavian &     Yugoslavia &    4 \\
      \hline
\end{tabular}
\end{center}
\caption{
Continuation of table \ref{stab:population_defnsB}.
\label{stab:population_defnsC}
}
\end{table}

\clearpage

\section*{Supplemental material}

\renewcommand{\thefigure}{S\arabic{figure}}
\setcounter{figure}{0}
\renewcommand{\thetable}{S\arabic{table}}
\setcounter{table}{0}
\renewcommand{\thesection}{S\arabic{section}}
\setcounter{section}{0}
\renewcommand{\thesubsection}{S\arabic{subsection}}
\setcounter{subsection}{0}

\section{Simulation methods}

To test our inference procedure, we implemented a simple whole-genome pedigree simulator,
which we briefly describe here.
We simulate in reverse time, 
keeping track of those parts of the pedigree along which genomic material have actually passed;
effectively constructing the ancestral recombination graph \citep{griffiths1997ancestral} in its entireity.
This is computationally taxing, but it is still feasible to generate all relationships between 100 sampled humans
(with the actual chromosome numbers and lengths) 
going back 300 generations in an hour or so on a modern machine with only 8GB of RAM;
or on the same machine, 1000 sampled humans going back 150 generations, overnight.
(Since memory use and time to process a generation scale linearly with the number of generations and the number of samples,
a machine with more RAM could produce longer or larger simulations.)
The demographic scenarios are restricted to (arbitrary) discrete population models
with changing population sizes and migration rates.
The code (python and R) is freely available at \url{http://github.org/petrelharp}.

The process we want to simulate is as follows:
we have $n$ sampled diploid individuals in the present day,
and at some time $T$ in the past, wish to know which of the $2n$ sampled haplotypes
inherited which portions of genome from the same ancestral haplotypes (i.e.\ are IBD by time $T$).
We work with diploid individuals, always resolved into maternal and paternal haplotypes, and only work with the autosomes.
We treat all chromosomes in a common coordinate system by laying them down end-to-end,
with the chromosomal endpoints $g_1 < \cdots < g_c$ playing a special role.
The algorithm iterates through previous generations,
and works as follows to produce the state at $t+1$ generations ago from the state at $t$ generations ago.
Each sampled haplotype can be divided into segments inheriting from distinct ancestral haplotypes from $t$ generations ago.
For each of the sampled haplotypes, indexed by $1 \le i \le 2n$,
we record the sequence of genomic locations separating these segments as $b(i,t) = (b_1(i,t), \ldots, b_{B(i,t)}(i,t))$, 
and the identities of the corresponding ancestors from $t$ generations ago as $a(i,t) = (a_1(i,t), \ldots, a_{B(i,t)}(i,t))$,
where $B(i,t)$ is the total number of segments the $i^\mathrm{th}$ sampled haplotype is divided into $t$ generations ago.
The first genomic location $b_1(i,t)$ is always 0, and for notational convenience, let $b_{B(i,t)+1}(i,t) = g_c$ (the total genome length).
The meaning of $a(i,t)$ is that if two samples $i$ and $j$ match on overlapping segments,
i.e.\ for some $k$ and $\ell$, $a_k(i,t) = a_\ell(i,t)$ and $[x,y] = [b_k(i,t),b_{k+1}(i,t)] \cap [b_{\ell}{j,t},b_{\ell+1}{j,t}]$, 
then both have inherited the genomic segment $[x,y]$ from the same ancestral haplotype $a_k(i,t)$ alive at time $t$,
and are thus IBD on that segment from sometime in the past $t$ generations.

As parameters, we are given $N_t(u)$, the effective population size in subpopulation $u$ at time $t$ in the past.
% and $m_t(u,v)$, the (reverse-time) migration rate from $u$ to $v$ at time $t$ in the past.

\begin{figure}[!htp]
  \begin{center}
    \includegraphics[width=.35\textwidth]{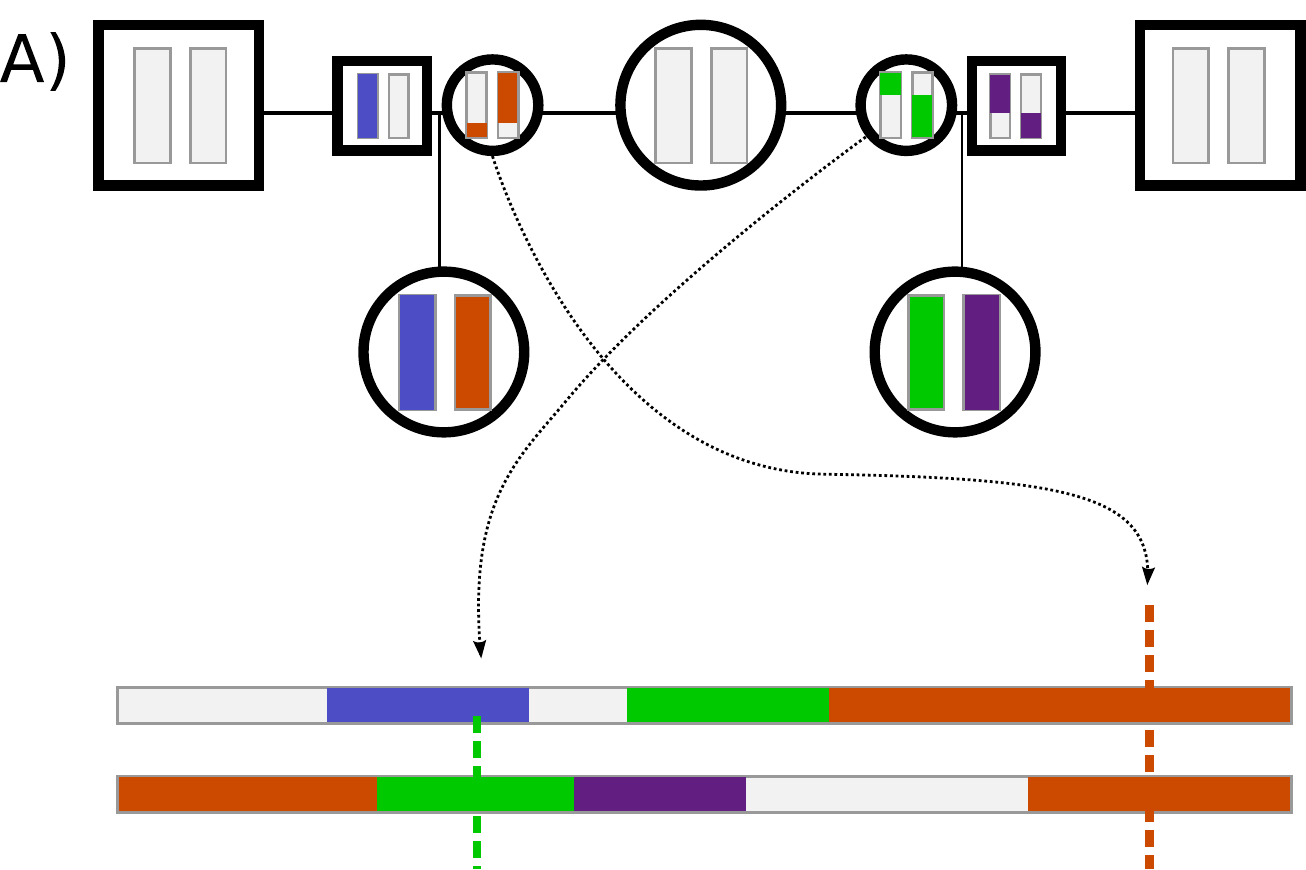}
\hspace{.1\textwidth}
    \includegraphics[width=.35\textwidth]{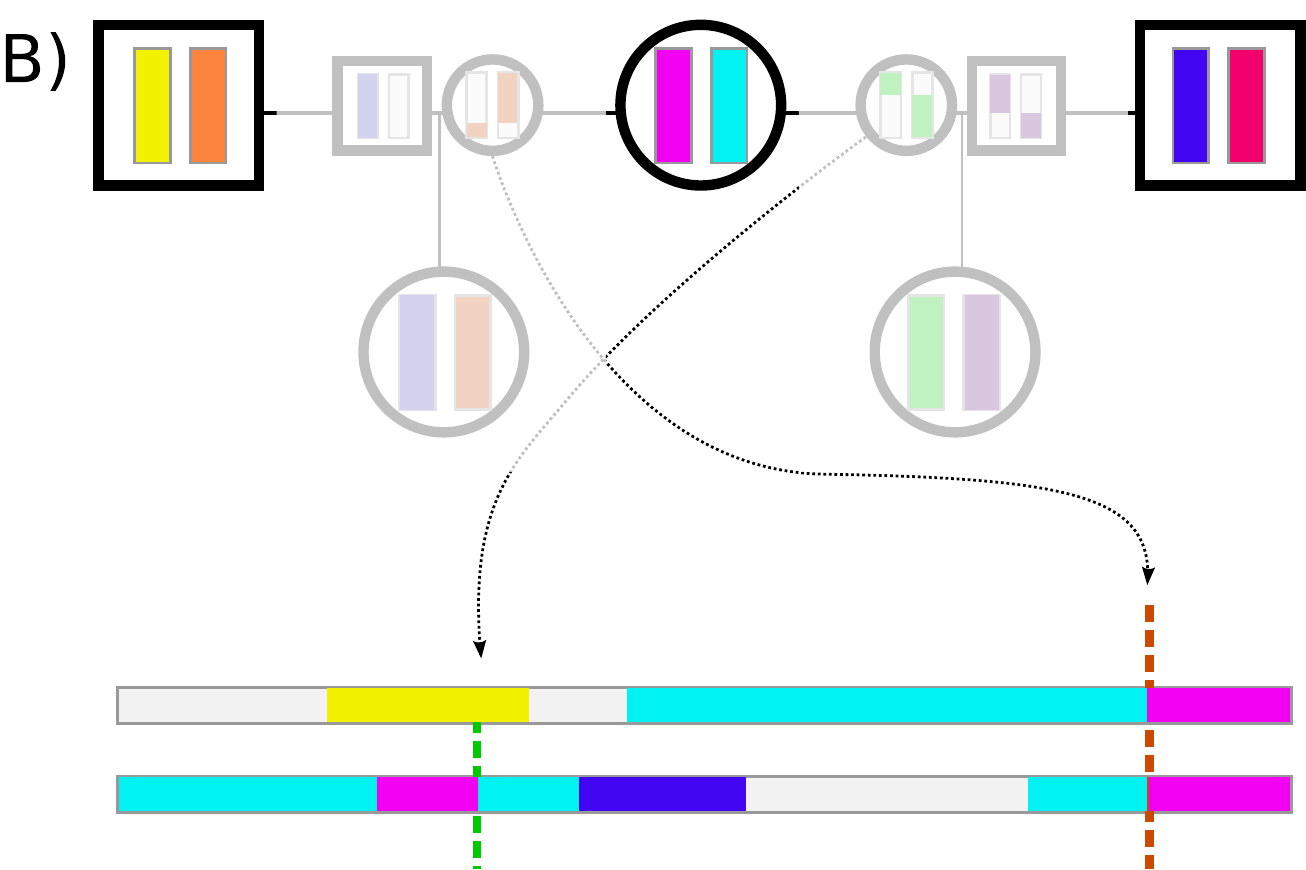}
    \caption{
        \label{fig:updates}
        An illustration of the update procedure, moving to the previous generation.
At the bottom of {\bf (A)} is the current state in generation $t$, with haplotype segments
colored by which generation-$t$ ancestral haplotype they derive from (e.g.~$a(i,t)$),
only showing colors for segments inheriting from the depicted two ancestors
(all segments in fact have labels).
In {\bf (B)} these have been updated to be colored corresponding to which of the generation-$(t+1)$ haplotypes they derive from.
The two depicted ancestors are half sibs.
The smaller symbols with partially-colored chromosomes depict the location of recombination breakpoints,
which are located on the sampled haplotypes by arrows and vertical dotted lines.
    }
  \end{center}
\end{figure}

To update from $t$ to $t+1$, we need to pick parents for each generation-$t$ ancestor,
choose the recombination breakpoints for the meiosis leading to each generation-$t$ haplotype
-- so, each unique value of $a(\cdot,t)$ has a corresponding (diploid) parent and a set of recombination breakpoints.
These steps are performed iteratively along each haplotype,
checking if parents and recombination breakpoints have been chosen already for each ancestor,
and randomly generating these if not.
Recombinations are generated as a Poisson process of unit rate along the genome (so, lengths are in Morgans);
to this set each chromosomal endpoint is added independently with probability $1/2$ each.
To choose a parent, 
% first a subpopulation $v$ is chosen from the migration kernel,
% so that if $a_k(i,t)$ is in subpopulation $u$, the parent is chosen from subpopulation $v$ with probability $m_{t+1}(u,v)$;
% then the parent is chosen uniformly from the $N_{t+1}(v)$ members of that subpopulation.
% Then, 
stretches of genome between alternating recombination breakpoints are assigned to the two haplotypes of the parent.
For instance, if the recombination breakpoints of a generation-$t$ haplotype labeled $a$ are at $r_1 < \cdots < r_R$,
and the maternal and paternal haplotypes of the parent of $a$ are labeled $h_m$ and $h_p$ respectively,
any $a_k(i,t)$ with $b_k(i,t)<r_1$ would be changed to $h_m$, while those with $r_1 \le b_k(i,t)<r_2$ would be changed to $h_p$,
and new segments are added when breakpoints fall inside of an existing segment ($a_k(i,t)<r_1\le a_{k+1}(i,t)$).

The algorithm is run for a given number of generations, 
after which an algorithm iterates along all sampled haplotypes in parallel, 
writing out any pairwise blocks of IBD longer than a given threshold.
An IBD block here is a contiguous piece of a single chromosome
over which both sampled chromosomes share the same state.

%%%%%
\section{Test of inference methods}

\begin{figure}[htp!]
\begin{center}
\includegraphics{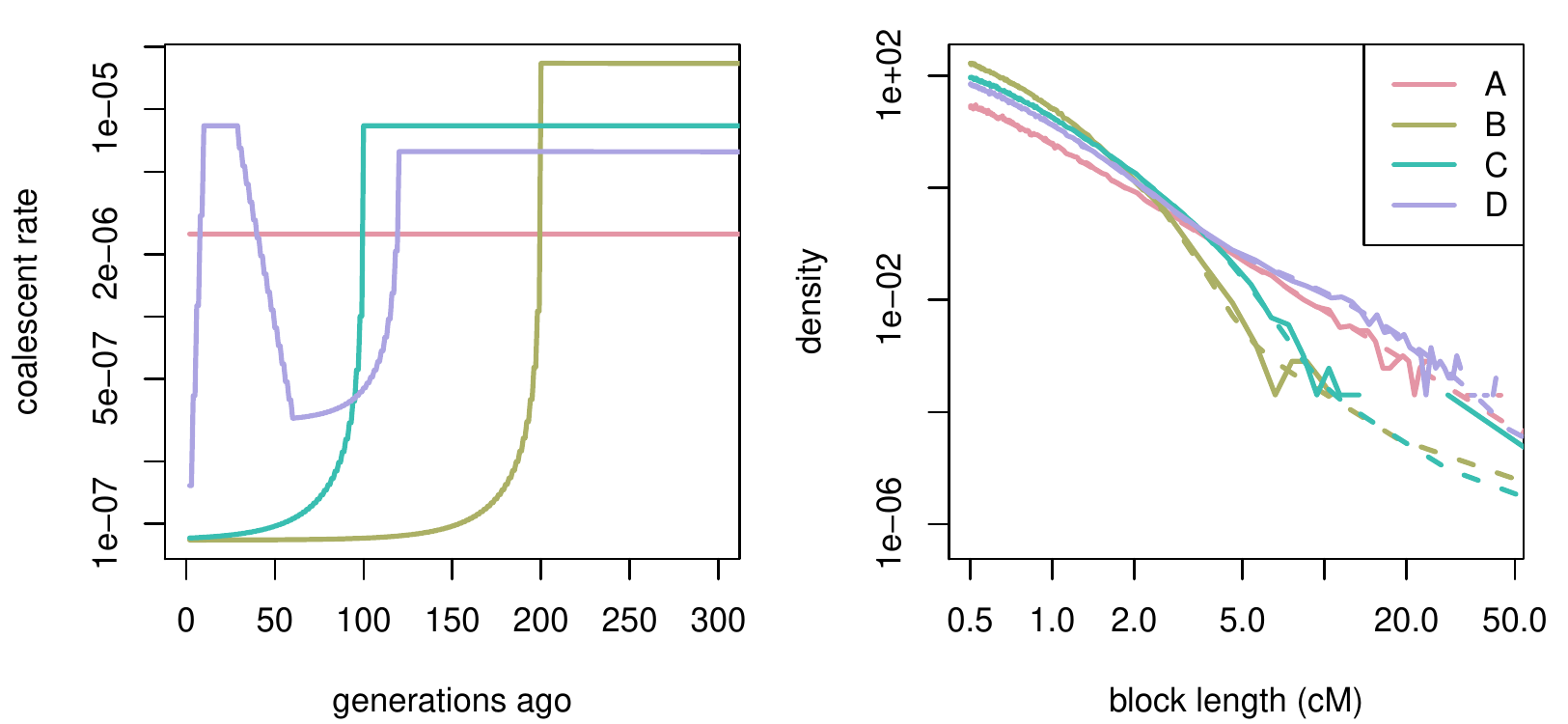}
\caption{
\label{fig:spectra_comparisons}
Coalescent rate {\bf (left)} and IBD length spectra {\bf (right)} for the three scenarios.
For the length distributions, the value given is observed blocks per pair and per centiMorgan;
for each, the dotted line gives the theoretical value predicted from the theoretical coalescent time distribution,
and the solid line is the observed distribution.
}
\end{center}
\end{figure}

We simulated from three simple demographic scenarios,
with parameters chosen to roughly match the mean number of IBD blocks per pair longer than 2cM that we see in the data.
The scenarios are as follows:
\begin{itemize}
    \item[(A)] Constant effective population size $10^5$ -- average 0.79 IBD blocks longer than 2cM per pair.
    \item[(B)] Exponential growth, starting from (constant) effective population size $1.5 \times 10^4$ prior to 100 generations ago,
        and approaching $3\times 10^6$ exponentially, as $N_e(t) = 3 \times 10^6 - (3 \times 10^6 - 1.5 \times 10^4) \exp(-0.077(100-t))$
        -- average 0.51 IBD blocks longer than 2cM per pair.
    \item[(C)] Exponential growth as in B, except expanding only 50 generations ago, and beginning with an effective population size of $3 \times 10^4$
      -- average 1.11 IBD blocks longer than 2cM per pair.
    \item[(D)] A more complex scenario: constant size $4\times10^4$ older than 60 generations ago;
      growing logistically to $8\times10^5$ between 60 and 30 generations ago;
      decreasing logistically to $3\times10^4$ between 30 and 15 generations ago;
      constant between 5 and 15 generations ago;
      and growing again to $4\times10^6$ until the present -- average 1.09 IBD blocks longer than 2cM per pair.
      (Imagine a population that grows large, has a small group split off gradually, which then grows in the present day;
      not motivated by any specific history, but chosen to test the methods when the true history is more "bumpy".)
\end{itemize}
For computational convenience, we simulated only up to 300 generations ago,
and retained only blocks longer than 0.5cM (but often restricted analysis to those longer than 2cM);
as in the paper we merged any blocks separated by a gap that was shorter than at least one adjacent block and shorter than 5cM.
Coalescent rates and block length distributions are shown in figure \ref{fig:spectra_comparisons}.
Even though we have not modeled gap removal, the results still closely match theory,
since very few blocks fell so close to each other.

For each scenario, we applied the inversion procedure described in the text to the full, error-free set of blocks
as well as to various subsets and modifications of it.
The inversion procedure we followed for each was as follows.
We chose a discretization for block lengths as described in the text,
by starting with the percentiles of the distribution,
and refining further so that the largest bin length was 1cM.
We then computed the matrix $L$ as described in the text,
except with no error distribution or false positive rate, 
so that if the $i^\mathrm{th}$ length bin is $[x_i,x_{i+1})$,
then $L_{in} = \sum_{g=1}^{22} K_g(n,\min(x_{i+1},G_g)) - K(n,\min(x_i,G_g))$,
with $G_g$ the length of the $g^\mathrm{th}$ chromosome and $K_g(n,x) = (n(G_g-x)+1)\exp(-nx)$.
For most simulations, we did not discretize time any further,
but allowed $n$ to range from 1 up to 300 generations.
We then used constrained optimization as implemented in the {\tt R} package {\tt optim} \citep[L-BFGS-B method,][]{R}
to maximize the penalized likelihoods described in the text,
beginning at the solution to the natural approximating least-squares problem \citep[using {\tt quadprog},][]{quadprog}.
For each case, we show the ``maximum likelihood solution''
(estimated by adding a small amount of smoothness penalty to ensure numerical uniqueness, allowing the algorithm to converge),
and the ``smoothest consistent solution'' 
-- the largest $\gamma$ so that the solution has decreased in log-likelihood by no more than 2 units.

In each case, we also show the exact coalescent distribution,
as well as the block length spectrum predicted by theory from 
the true coalescent distribution and each coalescent distribution found by penalized maximum likelihood.

Note that we could have incorporated false positives, missed blocks, or length misestimation into the simulations,
and subsequently modified the kernel $L$ to incorporate these rates,
but this would only add additional layers of simulation code, and would not make the task of inference more difficult,
since we account for these effects analytically.
The sensitivity of the methods to {\em misestimation} of these rates is a concern,
but this amounts to misestimation of the kernel $L$,
which we investigate below.

\begin{figure}[htp!]
\begin{center}
\includegraphics{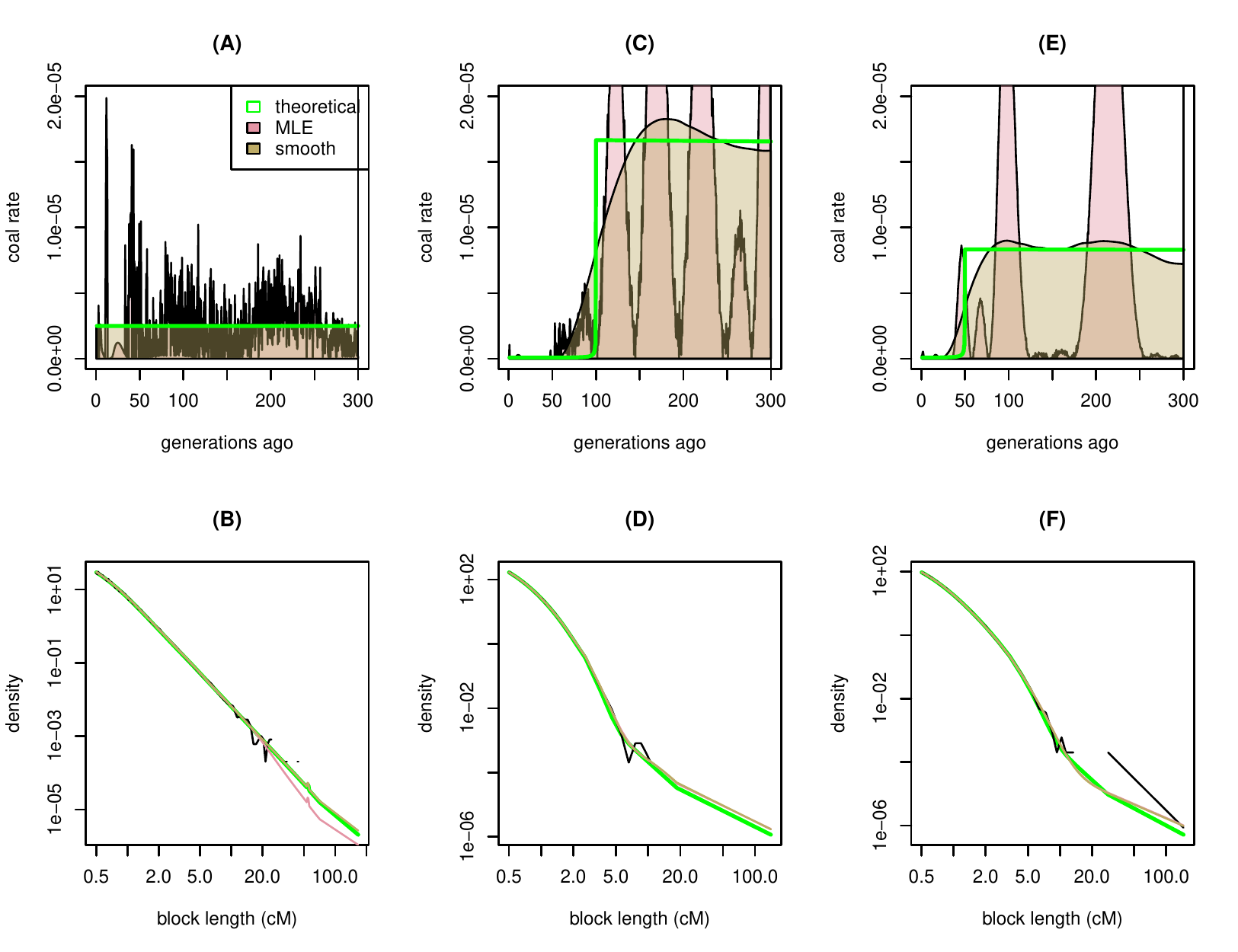}
\caption{
\label{fig:full_inversions}
Results of the inference procedure applied to all data (all blocks at least 0.5cM) with 300 generations as the upper limit.
Above are true (green) and inferred (shaded) coalescent rates;
below are block length distributions (density per pair), observed (black) and predicted by the inferred coalescent distributions in the respective upper panel.
{\bf (A--B)}, scenario A;
{\bf (C--D)}, scenario B; and
{\bf (E--F)}, scenario C.
The dangling line at the end of several plots is due to a few rare long blocks
and is not a significant deviation from the expectation.
}
\end{center}
\end{figure}

\begin{figure}[htp!]
\begin{center}
\includegraphics{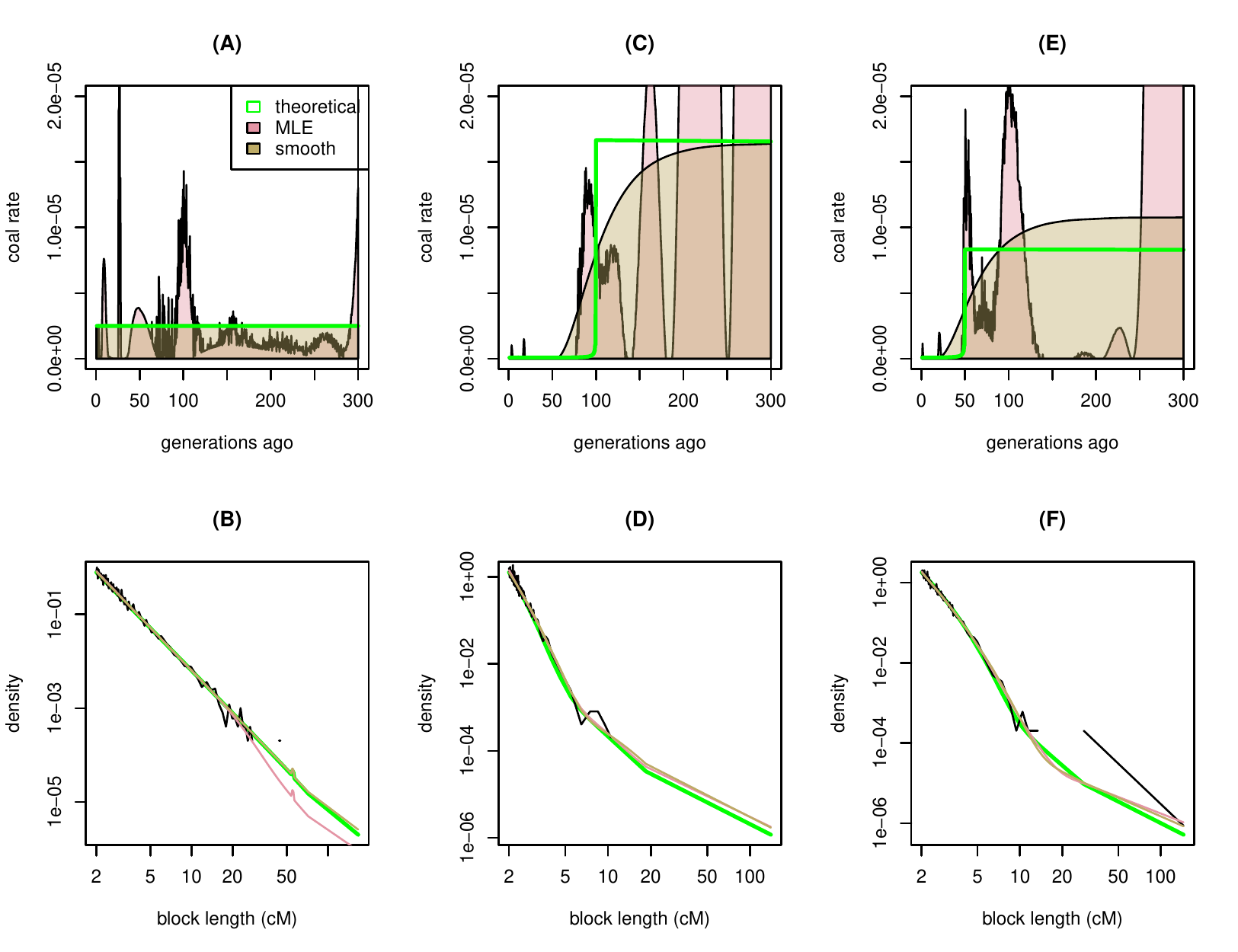}
\caption{
\label{fig:long_inversions}
{\bf Longer blocks only} -- as in figure \ref{fig:full_inversions}, except in each case we have only used blocks above at least 2cM.
}
\end{center}
\end{figure}

\begin{figure}[htp!]
\begin{center}
\includegraphics{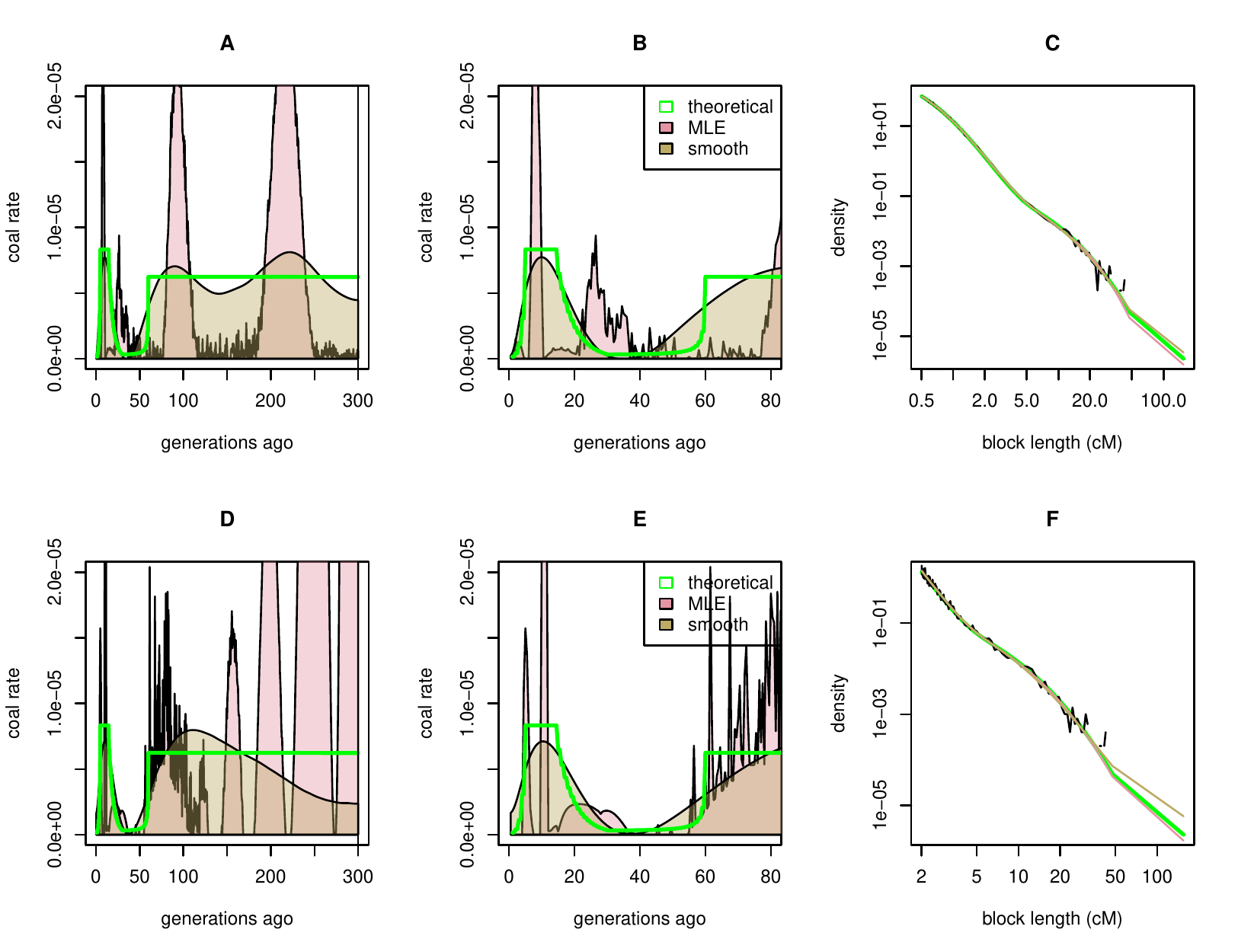}
\caption{
\label{fig:complex_inversions}
{\bf Scenario D:} Here we show results for the more complex demographic scenario.
{\bf (A--C)} use all blocks (down to 0.5cM), with (A) and (C) the same as in figure \ref{fig:full_inversions},
and (B) the same as (A) except zooming in on more recent times.
{\bf (D--F)} is as (A--C), except using only blocks longer than 2cM.
}
\end{center}
\end{figure}

In figure \ref{fig:full_inversions} we show the results of the inference procedure applied to the full set of blocks longer than 0.5cM;
figure \ref{fig:long_inversions} is the same, except using only blocks longer than 2cM,
and figure \ref{fig:complex_inversions} shows the results for scenario D separately.
Comparing these, we see that the short blocks 0.5--2cM does not significantly improve the resolution in recent times,
but does allow better estimation of coalescent rates longer ago in time.
Using blocks longer than 2cM gives us good resolution on the time scale we consider (the past 100 generations),
and including those down to 0.5cM does not make the likelihood much less ridged (as expected from theory).

\begin{figure}[htp!]
\begin{center}
  \includegraphics{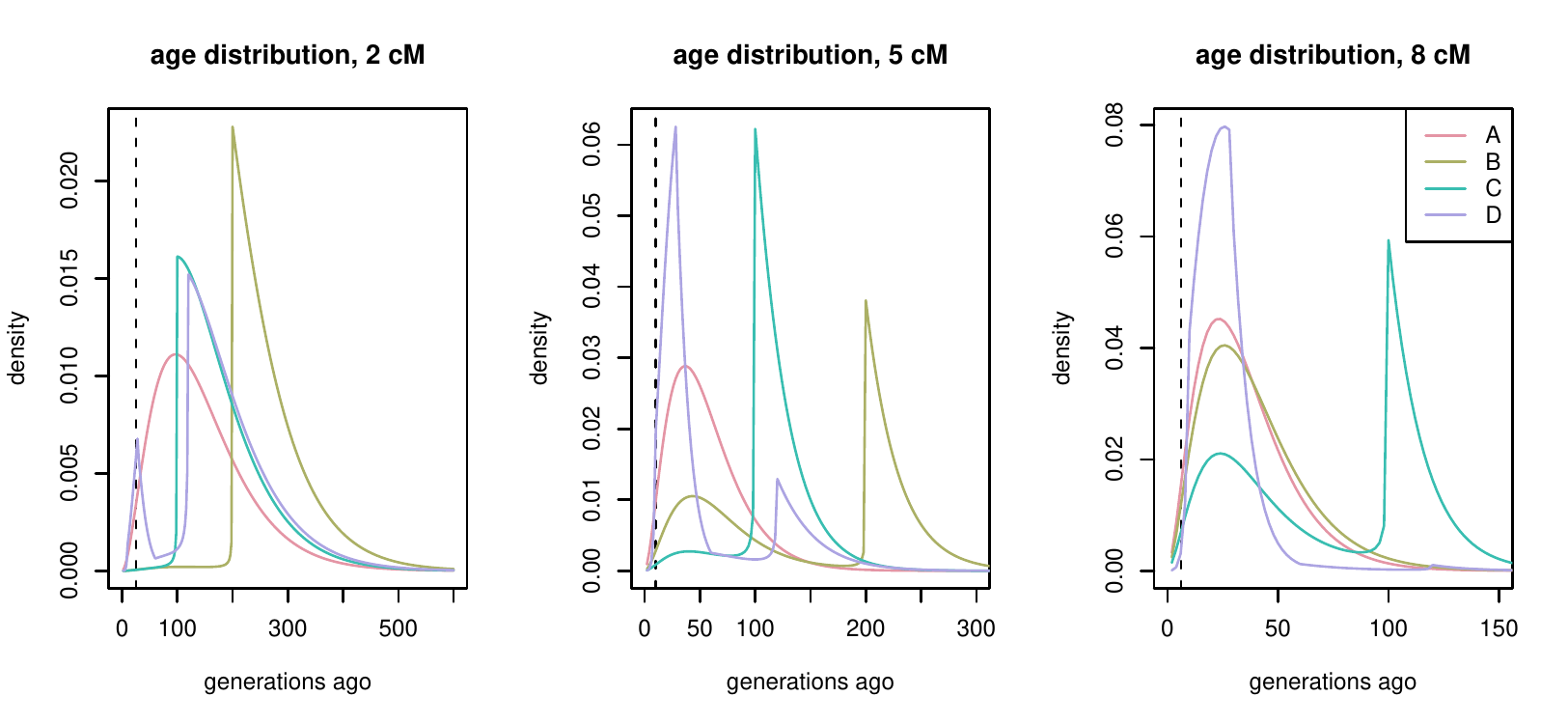}
\caption{
\label{fig:age_distributions}
Age distributions of blocks under each of the four scenarios --
each curve shows the probability distribution for the age of a 2cM, 5cM, and 8cM block
under each of the four scenarios.
For the age distribution of blocks $x$ cM long, the vertical dotted line is at $50/x$ generations,
the naive expectation for the typical age of such blocks.
}
\end{center}
\end{figure}

One counter-intuitive result we obtained was that the coalescent history could have a dramatic effect 
on the estimation of ages of blocks given their lengths.
In figure \ref{fig:age_distributions} we show the probability distribution of the ages of blocks of various lengths under the four scenarios,
i.e.\ how many generations ago the ancestors lived from whom the samples inherited blocks of that length.
These are counter-intuitive because a block inherited from $n$ generations ago has mean length $50/n$ cM,
but the age distributions of blocks in practice show that the converse is not true --
blocks $x$ cM long are usually much older than $50/n$ generations.
This is computed simply as follows:
the mean number of IBD blocks of length $x$ per unit of coalescence from $n/2$ generations ago (from paths of $n$ meioses)
is $K(n,x) = \sum_{i=1}^22 n(n(G_i-x)+1)\exp(-nx)$;
so the probability that a block of length $x$ came from $n/2$ generations ago is $K(n,x)/\sum_m K(m,x)$.

\section{Sensitivity analysis}

\begin{figure}[htp!]
\begin{center}
\includegraphics{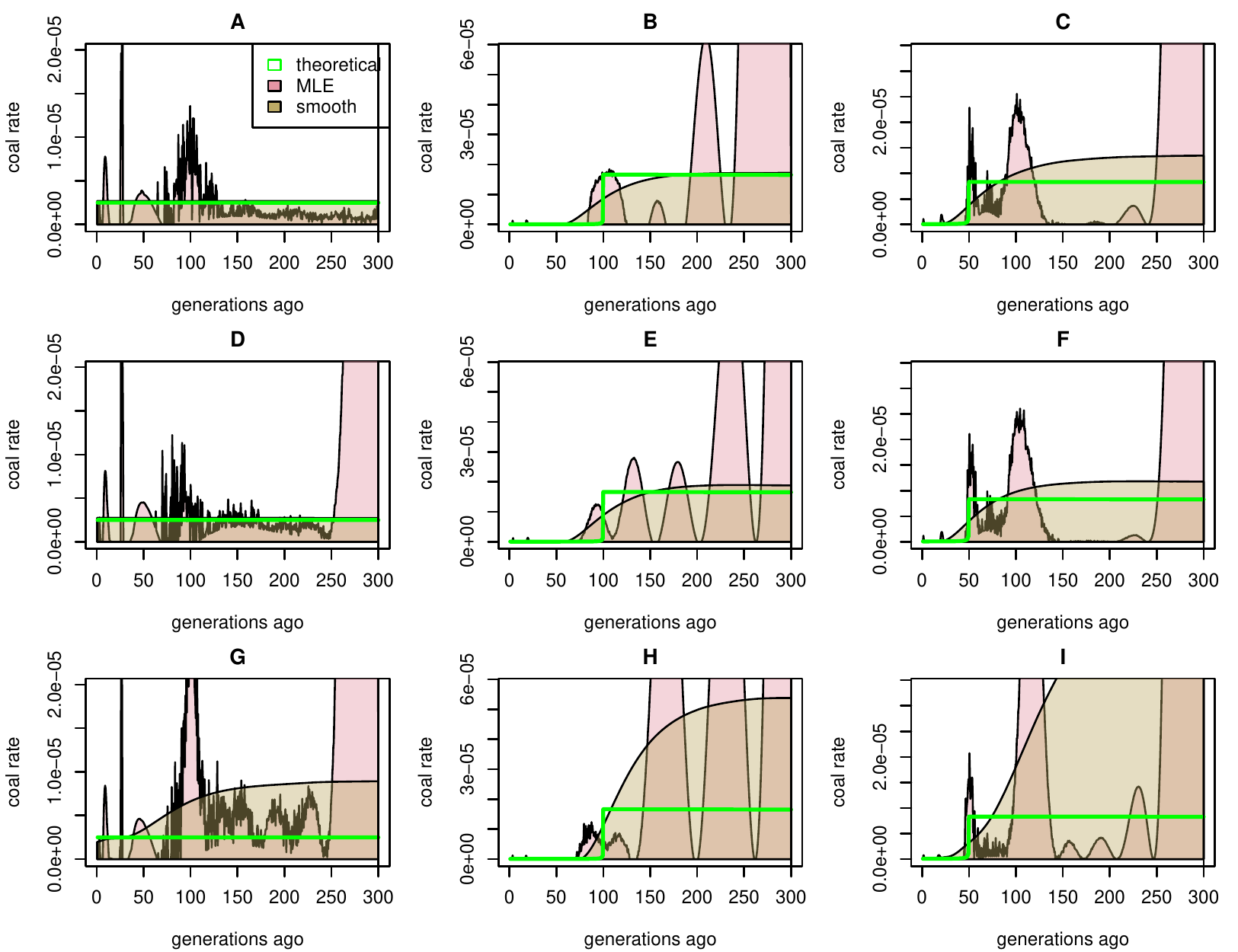}
\caption{
\label{fig:fp_inversions}
{\bf False positives} -- each row shows inference results from the three scenarios with different amounts of spurious false positives added on.
If the predicted number of blocks in the bin with length midpoint $x$cM is $m(x)$,
we added a Poisson number of blocks with mean $h(x)m(x)$ to the number observed, with $h$ varying.
In the first row {\bf(A--C)}, $h(x) = 0.1\exp(2-x)$,
in the second row {\bf(D--F)}, $h(x) = 0.1\exp((2-x)/4)$,
and in the third row {\bf (G--I)}, $h(x) = \exp(2-x)$.
The third scenario is not thought to be realistic,
but demonstrates that misestimation at short lengths only affects inference at older times.
}
\end{center}
\end{figure}

We also used these simulations to evaluate our sensitivity to error.
Of particular concern is error arising due to misestimation of the false positive rate --
we have seen that false positive rate at short lengths can vary somewhat by population -- 
we estimate as much as 10\% around 2cM.
To evaluate the effect on inference, we added to the numbers of blocks observed in each category
some number of ``false positives'', but applied the inference methods without accounting for these (so we still have $f=0$).
The numbers of false positives added to each length bin are Poisson with mean equal to the theoretical mean predicted for that bin,
multiplied by a factor that depends on the length and decreases (so there is an artifical inflation of short blocks).
The results for three different false positive rates are shown in figure \ref{fig:fp_inversions}.
From these, we see that if IBD rate is only increased by a maximum of 10\% 
-- even if the effect extends out to 6 or 8cM --
the effect on the inferred coalescent distribution is minor.
It is also useful to add a unrealistically high level of unaccounted-for false positives,
as it is natural to suspect that an excess of short blocks will only increase the coalescent rate at relatively older time periods.
This is indeed the case -- doubling the distribution at the short end (about 2--4cM)
only affects inferred coalescent rates beyond about 100 generations,
because this is when the bulk of the 2--4cM blocks have come from.

\end{document}